\pdfoutput=1
\documentclass[11pt]{article}

\usepackage[T1]{fontenc}
\usepackage[utf8]{inputenc}
\usepackage{textcomp}
\usepackage[a4paper,margin=1in]{geometry}
\usepackage{setspace}
\usepackage{microtype}

\usepackage{amsmath,amssymb,amsthm}
\usepackage{upgreek}

\usepackage{graphicx}
\usepackage{epstopdf}
\usepackage{float}
\usepackage{booktabs}
\usepackage{array}
\usepackage{tabularx}
\usepackage{multirow}
\usepackage{caption}
\usepackage{xcolor}
\usepackage{colortbl}
\usepackage{soul}
\usepackage{tikz}
\usepackage{seqsplit}
\usepackage{enumitem}
\usepackage{longtable}

\usepackage[round,authoryear]{natbib}
\usepackage{url}
\usepackage[hidelinks]{hyperref}
\usepackage[nameinlink,noabbrev]{cleveref}

\newcommand{\orcidA}{\textsuperscript{\href{https://orcid.org/0009-0008-6090-1827}{ORCID}}}
\newcommand{\orcidB}{\textsuperscript{\href{https://orcid.org/0009-0008-6913-1371}{ORCID}}}
\newcommand{\orcidC}{\textsuperscript{\href{https://orcid.org/0000-0003-0272-5622}{ORCID}}}
\providecommand{\keywords}[1]{\par\vspace{0.5em}\noindent\textbf{Keywords:} #1\par}

\title{Data-Driven Duration Management: Term Structure Forecasting Using Machine Learning}
\author{
Tobias Lausser\orcidA{} \and
Joao Eduardo Vuolo\orcidB{} \and
Rudi Zagst\orcidC{}\\[0.75em]
\normalsize Department of Mathematics, School of Computation, Information and Technology\\
\normalsize Technical University of Munich, Parkring 11, Garching bei M\"unchen, Bavaria, Germany\\
\normalsize \texttt{tobias.lausser@tum.de}
}
\date{Preprint.}

\begin{document}
\maketitle

\begin{abstract}
    This paper compares different methods for forecasting the term structure of U.S. and European zero-coupon government bonds using both traditional econometric and Machine Learning (ML) approaches. We compare classical models (e.g., Dynamic Nelson-Siegel (DNS) and Principal Component Analysis (PCA)) with different Neural Network (NN) architectures, including those inspired by the classical models, on the U.S. Treasury market and bonds issued by the European Central Bank (ECB). To enhance predictive performance, macroeconomic variables are incorporated. The findings for both markets are separately analyzed and compared. To this end, we propose a robust model evaluation framework combining statistical accuracy metrics—such as RMSE, MAE, and directional accuracy—with the economic relevance of a quantitative bond trading strategy. Results show that NNs consistently outperform traditional models in both forecasting accuracy and portfolio performance. For the U.S., the most effective approach is a direct-forecasting NN that incorporates DNS factors to reduce the dimensionality of zero-rate data and an Autoencoder (AE) to extract macroeconomic features, while for Europe, the optimal model is a factor-based NN using PCA-derived zero-rate factors without the integration of macroeconomic variables. Overall, the paper demonstrates how combining traditional modeling approaches with modern ML techniques and evaluation can improve yield curve forecasts and support applications in fixed-income portfolio construction.
\end{abstract}


\section{Introduction}

The bond market is a key pillar of the global financial system, reflecting market expectations regarding interest rates, inflation, and economic growth. As of March 2025, U.S. gross national debt is about \$37 trillion, compared to a stock market capitalization of \$49.8 trillion. In the EU, government debt stood at roughly €13.3 trillion (81\% of GDP) by end-2024, while stock market capitalization is projected at \$13.39 trillion in 2025.

The primary objective of our investigations is the term structure of interest rates, also known as the yield curve. It represents the relationship between bond yields and time to maturity \citep{fabozzi2012fixed}. Institutional investors, including pension funds and insurance companies, typically hold large proportions of fixed-income securities \citep{blake2001pension}, making yield curve forecasting both academically and practically significant. Moreover, interest rate changes directly impact the valuation of all financial assets \citep{diebold2006forecasting, cochrane2005bond}.

One reason for the complexity of term structure forecasting is its high dimensionality \citep{cochrane2005bond, diebold2006forecasting}. Prominent approaches to modeling the term structure involve compressing its information to a few latent factors.
A foundational example is the Nelson-Siegel (NS) model from \citet{nelson1987parsimonious}, later extended into a dynamic framework by \citet{diebold2006forecasting}.
This model parsimoniously describes the zero-rate curve using three intuitive factors interpreted as level, slope, and curvature.
The forecasting process is then two-staged: first, the latent factors are extracted from the term structure at each point in time,
and second, the dynamics of those three factors are modeled with Autoregressive (AR) and Vector Autoregressive (VAR) time-series models.
Building on this, \citet{christensen2011affine} developed the Arbitrage-Free Nelson-Siegel (AFNS) model, which imposes no-arbitrage conditions to ensure theoretical consistency, albeit at the cost of increased complexity requiring estimation via state-space methods like the Kalman Filter (KF).
In a parallel line of research, \citet{ang2003no} demonstrated the importance of macroeconomic variables, incorporating them alongside latent factors within a no-arbitrage VAR framework to model the zero-rate curve.

More recently, \citet{salachas2024term} examined the predictive power of the zero-rate curve during the COVID-19 pandemic by comparing traditional term structure models with versions augmented by pandemic-related information. Their study applied DNS and AFNS, as well as VAR models, on alternative latent factors. The authors incorporated COVID-19 indicators—such as infection rates, mobility measures, and government stringency indices—into the models to capture the macro-financial disruptions of this period. They provide evidence that these enhanced-information models significantly improve short- and medium-term interest rate forecasts, highlighting the predictive power of such exogenous variables.

These traditional models typically rely on linearity assumptions and may fall short in capturing the nonlinear and dynamic nature of zero-rate curve movements, particularly during periods of market stress, structural breaks, or unconventional monetary policy regimes. As financial markets have become increasingly complex and volatile, there is growing interest in exploring more flexible modeling techniques that can better account for nonlinearities and higher-order interactions among variables, an area where Machine Learning (ML) methods have demonstrated promising results \citep{vela2013forecasting,nunes2019comparison}. The most prominent example of ML architecture are Neural Networks (NN).

Early research into NNs for zero-rate curve modeling focused on forecasting the latent factors of traditional parametric models. For instance, \citet{vela2013forecasting} compared different NN architectures for predicting the factors of the NS model against AR and VAR. While finding that NNs performed best on U.S. data, the authors noted inconsistent results across different markets, concluding that there was insufficient evidence to declare a universally superior forecasting method. 

Shifting the focus from factor models to direct zero-rate prediction, subsequent research underscored the importance of incorporating macroeconomic data. In their work on the European zero-rate curve, \citet{nunes2019comparison} provided a comprehensive comparison between two NN architectures: Single-Task Learning (STL), where an independent model is trained for each maturity, and Multi-Task Learning (MTL), which forecasts all maturities jointly using a shared NN. Their results showed that NNs, particularly the MTL approach, consistently outperformed traditional linear benchmarks, such as Ordinary Least Squares (OLS). A key contribution of their work was the rigorous feature selection process, which began with a broad set of macroeconomic variables and systematically narrowed it down to the most informative predictors. The final set primarily included variables capturing real economic activity, such as the unemployment rate and industrial production growth, as well as inflation dynamics, represented by inflation indices. This process demonstrated that incorporating macroeconomic information of these two kinds significantly enhances the forecasting accuracy of NN models.

Beyond statistical accuracy, the practical economic value of forecasting models is a crucial consideration. \citet{dunis2007economic} addressed this by comparing Autoregressive Moving Average (ARMA), KF models, and NNs not only on statistical metrics like Root Mean Squared Error (RMSE) but also on their performance in a directional trading strategy. Their findings were significant: while the ARMA model achieved the lowest forecast error, the NN-based strategy generated the highest risk-adjusted returns, as measured by the Sharpe Ratio. This underscores that statistical superiority does not always translate to better economic outcomes, making application-specific evaluation essential.

Autoencoders (AEs) have been explored for their ability to perform non-linear dimensionality reduction on the zero-rate curve. \citet{suimon2020autoencoder} used an AE to decompose the Japanese zero-rate curve into interpretable factors analogous to the classic level, slope, and curvature of the DNS model. They then developed a trading strategy based on identifying mispricings, where the AE-reconstructed rate deviates from the actual rate. Although forecasting models like Long Short-Term Memory (LSTM) networks produced higher returns, the AE-based strategy yielded stable positive returns and offered greater interpretability, showcasing its utility in bond valuation and market analysis.

This study advances the literature on zero-rate curve forecasting by conducting a comprehensive and methodologically rigorous comparison of competing modeling approaches. Our key contributions are as follows:
\begin{enumerate}
    \item We demonstrate how European data can be adapted to provide the large volumes of data required for ML models, given the fact that European Central Bank (ECB) zero-rate data is only available from the early 2000s.
    \item We benchmark traditional econometric frameworks against modern NN architectures, exploring two distinct modeling strategies: (i) forecasting latent factors and mapping them into zero rates, and (ii) forecasting individual rates directly.
    \item The analysis is carried out for both the U.S. Treasury market and for bonds issued by the ECB and compares both results. Furthermore, macroeconomic variables are incorporated to enhance predictive performance.
    \item A robust evaluation framework is proposed, which combines traditional statistical accuracy metrics with an assessment of economic relevance through the performance of a bond trading strategy based on each model’s forecasts.
\end{enumerate}

This paper is structured as follows: Section 2 describes the data and the applied data augmentation for the European data, outlines the model architectures, and details the resources and methods employed. Section 3 focuses on the model performance and provides background on the underlying investment strategy. Section 4 presents and discusses the empirical results. Finally, Section 5 concludes the study by summarizing the key findings and their implications.

\section{Resources and Methods}

\subsection{Data}

\subsubsection{Zero rates}

Our analysis is built upon weekly zero-rate data for a consistent set of maturities. We formally define these two key components below.

The zero rate, denoted by $R(t, T)$, is the interest rate promised by the issuer (in our case, the government) to investors holding a bond with no intermediate coupon payments that matures at time $T$. Such a bond is called a zero-coupon bond. The zero rate typically varies across maturities $T$ --- a relationship known as the term structure of interest rates --- and evolves stochastically over time. In our dataset, the maturities considered (3-month, 6-month, 1-year, 2-year, 3-year, 5-year, and 10-year) represent specific points $T$ on this zero-rate curve. 

\begin{figure}[H]
\centering
\includegraphics[width=\linewidth]{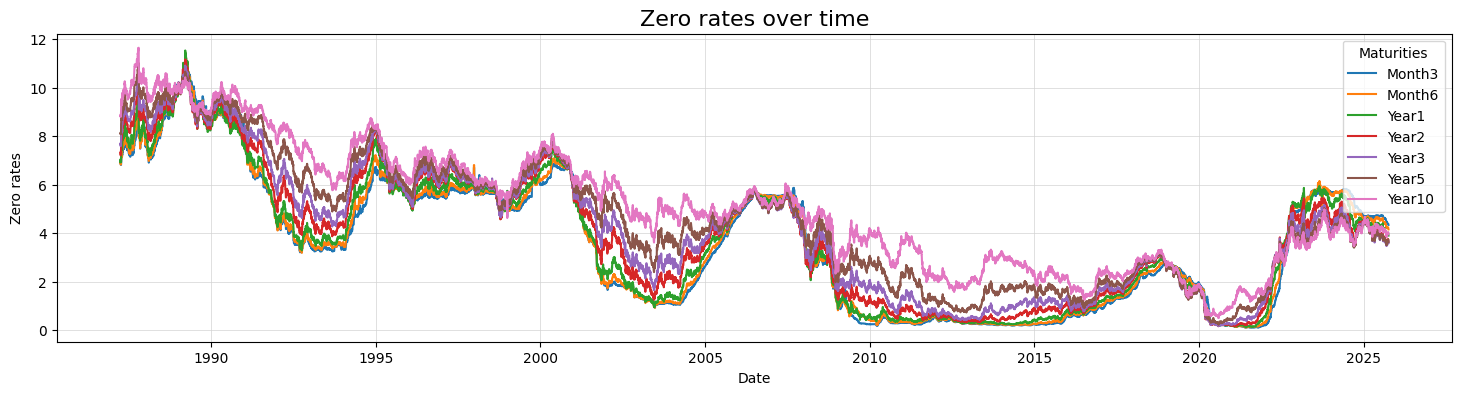}
\caption{U.S. zero-rate curve between April 1987 and February 2025.}
\label{fig:american_zero}
\end{figure}

The U.S. dataset consists of zero rates for these maturities, spanning the period from April 1987, to February 2025. This data was sourced from the FRED database and Thomson Reuters DataStream, with its evolution illustrated in Figure~\ref{fig:american_zero}.

Following the recommendation of \citet{engel2018forecasting}, we use the zero-rate curve of Euro area government bonds with a triple-A rating from the European Central Bank (ECB) as the European counterpart to U.S. zero rates. These bonds are selected because of their minimal default risk. However, the official ECB data series only begins in September 2004. For this analysis, we use data spanning from this start date until February 2025, as for the U.S. market. This selected time frame provides a limited history, which is insufficient for training data-intensive ML models, such as NNs.

To overcome this data limitation, we use a longer-running series of German government bonds as a proxy to backcast the European series, following the recommendation of \citet{bruggemann2008forecasting}. This strategy, however, introduces two distinct challenges. First, the German proxy data itself is partly incomplete, as described in Table \ref{tab:mats}, with some missing maturities that must be imputed. Second, a robust model must be established to link the German proxy to the official European series to construct the artificial pre-2004 data.

\begin{table}[H]
\caption{Availability of German zero-rate data from February 1992 to February 2025.}
\label{tab:mats}
\centering
\begin{tabularx}{\textwidth}{XX}
\toprule
\textbf{Maturity} & \textbf{Available since} \\
\midrule
3-month & October 2005 \\
6-month & November 1997 \\
1-year  & February 1994 \\
2-year  & Whole period \\
3-year  & Whole period \\
5-year  & Whole period \\
10-year & Whole period \\
\bottomrule
\end{tabularx}
\end{table}

To address the first challenge of incomplete German data, we imputed the missing maturities by fitting the Dynamic Nelson–Siegel (DNS) curve at each point in time (see Section~\ref{model_spec} for details on the DNS specification). To validate this imputation procedure, we conducted a robustness experiment by artificially removing selected maturities from a complete dataset and evaluating how accurately the DNS model could reconstruct them (see Table~\ref{tab:ns_missing_exp}). While the reconstruction performance was weaker for the 3-month and 6-month rates when three maturities were missing, this issue occurred continuously only between February 1992 and February 1994, representing less than 10\% of the total sample. For all other maturities, the DNS-imputed rates achieved an $R^2$ above $0.92$, confirming the robustness and reliability of the imputation approach across the majority of the data.

\begin{table}[H]
\caption{$R^2$ values for simulated missing-maturity experiments on the German zero-rate curve. Each scenario omits the indicated short-term maturities before re-estimating the DNS model.}
\label{tab:ns_missing_exp}
\centering
\begin{tabularx}{\textwidth}{X *{3}{>{\hsize=0.5\hsize}X}}
\toprule
\textbf{Scenario} & \textbf{3M} & \textbf{6M} & \textbf{1Y} \\
\midrule
(i) Missing 3M & 0.9840 & - & - \\
(ii) Missing 3M, 6M & 0.9213 & 0.9743 & - \\
(iii) Missing 3M, 6M, 1Y & 0.5781 & 0.7987 & 0.9612 \\
\bottomrule
\end{tabularx}
\end{table}

\begin{figure}[H]
\centering
\includegraphics[width=\linewidth]{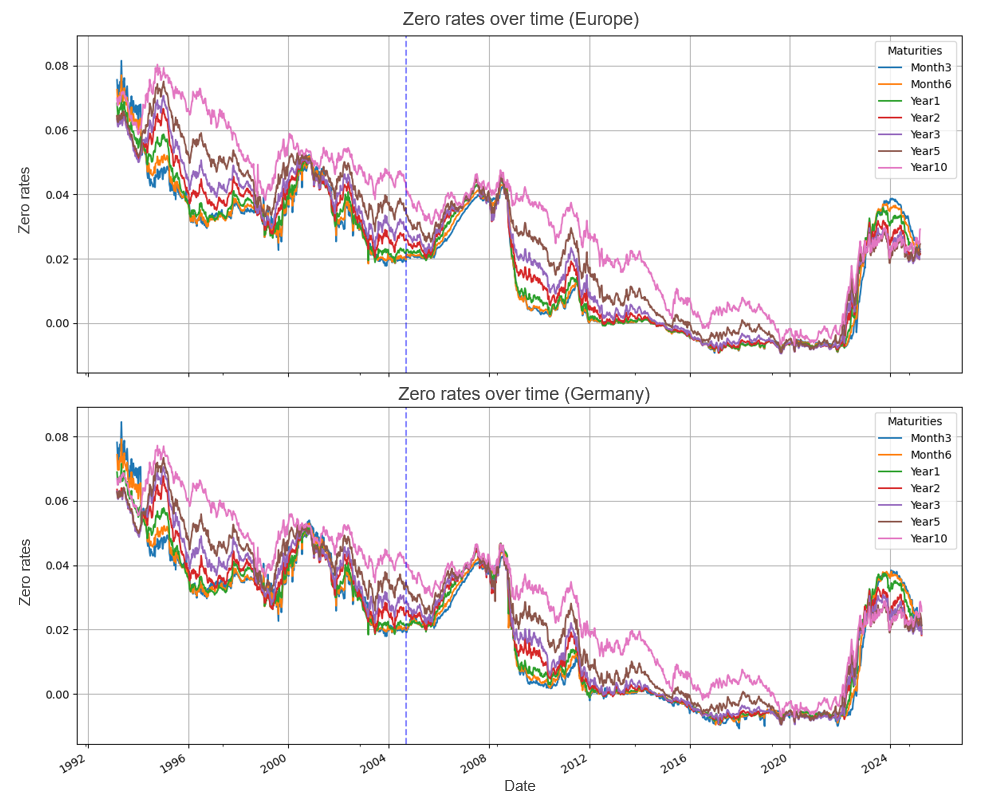}
\caption{Euro triple A zero-rate and German zero-rate curves between February 1992 and February 2025. The section to the left of the dashed line indicates the artificially extended dataset, generated via regression based on German zero-rate data.}
\label{fig:Euro_zero}
\end{figure}

With the complete German proxy series, we then addressed the initial problem of missing European data. We used an Ordinary Least Squares (OLS) model to establish the relationship between the German proxy data and the European zero rates during their overlapping period. Using this fitted model, we backcast the European data to February 1992, thereby extending the time series. The model demonstrated an excellent fit, with $R^2$ values above 0.99 for all maturities (Table~\ref{tab:r2}). The resulting coefficients were then used to predict the European zero rates for the period prior to 2004. The final, extended European zero-rate curve is depicted in Figure~\ref{fig:Euro_zero}.

\begin{table}[H]
\caption{$R^2$ from OLS regressions of European on German zero rates for each maturity.}
\label{tab:r2}
\centering
\begin{tabular}{lc}
\toprule
\textbf{Maturity} & \textbf{$R^2$} \\
\midrule
3-month & 0.9921 \\
6-month & 0.9964 \\
1-year & 0.9987 \\
2-year & 0.9985 \\
3-year & 0.9973 \\
5-year & 0.9946 \\
10-year & 0.9922 \\
\bottomrule
\end{tabular}
\end{table}

\subsubsection{Macroeconomic Data}

Following the approach of \citet{ang2003no}, we incorporate macroeconomic indicators of inflation and real economic activity. For the U.S. market, data is sourced from FRED, with the specific time series detailed in Table~\ref{tab:data_am}.

The corporate bond spread is calculated as the difference between Moody's seasoned BAA and AAA corporate bond yields:
\begin{align*}
\text{Spread}_t = \text{BAA}_t - \text{AAA}_t,
\end{align*}
where $\text{BAA}_t$ and $\text{AAA}_t$ are the yields in percentage points. Corporate bond spreads are often used as a proxy for credit market conditions and inflation expectations. Wider spreads may indicate increased uncertainty or risk premiums, often associated with higher future inflation expectations \citep{Kang2015Inflation}.

The other monthly macroeconomic series are transformed into yearly logarithmic growth rates using the formula:
\[
\log \left( \frac{I(t)}{I(t-12)} \right),
\]
where $I(t)$ is the value of the indicator at time $t$ in months. We rely on yearly growth because some macroeconomic indicators exhibit seasonal patterns. Comparing each month with the same month in the previous year naturally removes most of this seasonality, yielding a smoother and more informative signal for forecasting.

\begin{table}[H]
\caption{U.S. macroeconomic time series used in the analysis.}
\label{tab:data_am}
\centering
\begin{tabular}{p{3.5cm} p{4.5cm} p{2cm} p{2cm}} 
\toprule
\textbf{Indicator of} & \textbf{Time Series} & \textbf{Abbreviation} & \textbf{Frequency} \\
\midrule
Real economic activity & Total Employment in Private Industries & U.S.PRIV & Monthly \\ \hline
Real economic activity & Industrial Production Index & INDPRO & Monthly \\ \hline
Real economic activity & Non Farm Payroll & PAYEMS & Monthly \\ \hline
Inflation & CPI & CPIAUCSL & Monthly \\ \hline
Inflation & PPI & PPIACO & Monthly \\ \hline
Inflation & Corporate Bond Spreads & SPREAD & Daily \\ \hline
\end{tabular} 
\end{table}

For the European analysis, we utilize German macroeconomic data for the entire sample period, as it is more readily available and Germany plays a central role in the European Union's triple-A sovereign bond market. The indicators, shown in Table~\ref{tab:macro_de}, are chosen to align with the U.S. dataset and undergo the same yearly logarithmic growth rate transformation. The corporate bond spread is excluded from the German analysis. This omission is necessary because Germany lacks a corporate bond market with comparable depth, liquidity, and historical data to those of the U.S.. Therefore, a reliable, equivalent spread indicator is not available for our analysis.

\begin{table}[H]
\caption{German macroeconomic time series used in the analysis.}
\label{tab:macro_de}
\centering
\begin{tabular}{p{3.5cm} p{4.5cm} p{2cm} p{2cm}} 
\toprule
\textbf{Indicator of} & \textbf{Time Series} & \textbf{Abbreviation} & \textbf{Frequency} \\
\midrule
Real economic activity & Total Employment & EMPDE & Monthly \\ \hline
Real economic activity & Industrial Production Index & INDPRODE & Monthly \\ \hline
Real economic activity & Total Wages & PAYEMSDE & Monthly  \\ \hline
Inflation & CPI & CPIDE & Monthly \\ \hline
Inflation & PPI & PPIDE & Monthly \\ \hline
\end{tabular}
\end{table}

For the monthly series included in our weekly dataset, we apply a one-month lag to account for publication delays. Each lagged monthly value is then carried forward and repeated each week from its release date until the next monthly update. This procedure ensures the dataset remains continuous and aligned with the weekly frequency, while preventing the use of information that would not have been available at the time.

For the daily series, the data is aggregated to the weekly frequency by taking the last available observation within each week, ensuring consistency with the timing of other variables in the dataset.

\subsection{Model Architecture}\label{model_spec}

Our primary objective is to forecast the zero-rate curve. The methodologies we implement can be broadly understood by first examining traditional approaches and then exploring enhancements and alternative paradigms offered by Neural Networks (NNs). 

A traditional approach to forecasting the zero-rate curve involves a two-step process. First, a data shrinkage or dimensionality reduction technique is used to compress the information from the entire curve into a small set of latent factors. Then, to ensure stationarity, the log-differences of these factor time series are modeled using time series methods to produce forecasts. The forecasts are then reverted from log-differences and used to reconstruct the full forecasted zero-rate curve. In the traditional approaches, use the following forecasting models:

\begin{itemize}
    \item \textbf{Autoregressive (AR) model:} AR(1) models are fitted to each factor series independently.
    \item \textbf{Vector Autoregressive (VAR) model:} A joint VAR(1) model is fitted to all factors simultaneously. This allows for potential cross-correlations between the factors.
\end{itemize}

One of the most popular models in finance for the dimensionality reduction of zero-rate curve data is the Dynamic Nelson-Siegel (DNS) model \citep{diebold2006forecasting}. The zero rate $R(t,T)$ at time $t$ for a given maturity $T$ is described as a function of three time-varying factors:
\begin{equation}\label{eqns}
R(t,T) = X_1(t) + X_2(t) \left( \frac{1 - e^{-\lambda_t (T - t)}}{\lambda_t (T - t)} \right) + X_3(t) \left( \frac{1 - e^{-\lambda_t (T - t)}}{\lambda_t (T - t)} - e^{-\lambda_t (T - t)} \right),
\end{equation}
where $X_1(t), X_2(t), X_3(t) \in \mathbb{R}$ and $\lambda_t > 0$.

\begin{figure}[h!]
    \centering
    \includegraphics[width=\textwidth]{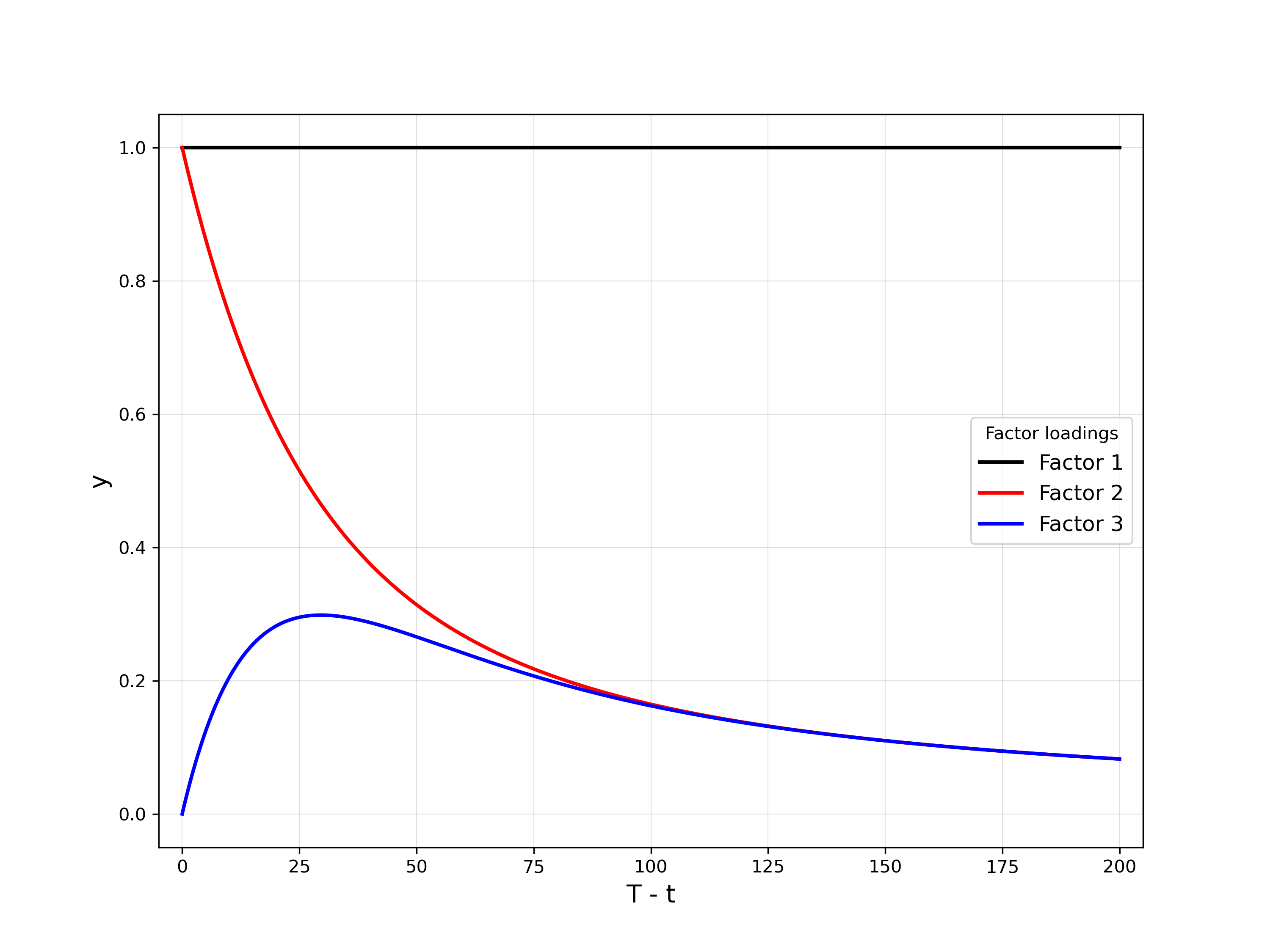}
    \caption{DNS factor loadings with $\lambda_t = 0.0606$ for different times to maturity, $T-t$.}
    \label{fig:fac_loadings}
\end{figure}

This decomposition provides an intuitive economic interpretation of the factors as level, slope, and curvature, respectively. These interpretations arise from how each factor influences rates of different maturities through their corresponding factor loadings—that is, the maturity-dependent weights that determine how sensitive the rate at each maturity is to changes in a given factor (as shown in Figure \ref{fig:fac_loadings}). The level factor, $X_1(t)$, has a constant loading of 1, affecting all maturities equally. The slope factor, $X_2(t)$, has a loading that decays to zero, thus primarily affecting short-term rates. The curvature factor, $X_3(t)$, is hump-shaped, primarily affecting medium-term rates. 

The non-linear decay parameter $\lambda_t$ is usually fixed to a constant value, $\lambda$. This simplifies Equation \eqref{eqns} into a model that is linear in its factors. Then, for each period $t$ in the dataset, the values of the three factors $X_1(t)$, $X_2(t)$, and $X_3(t)$ are estimated by running a single cross-sectional OLS regression of the observed rates for that period against the (now fixed) factor loadings. This process is repeated for every period, yielding the time series for each factor $X_i(t)$. While \citet{diebold2006forecasting} proposed $\lambda = 0.0609$, our preliminary tests indicated a better fit with $\lambda = 0.0606$, which was adopted for this study.

\begin{figure}[h!]
    \centering
    \includegraphics[width=\textwidth]{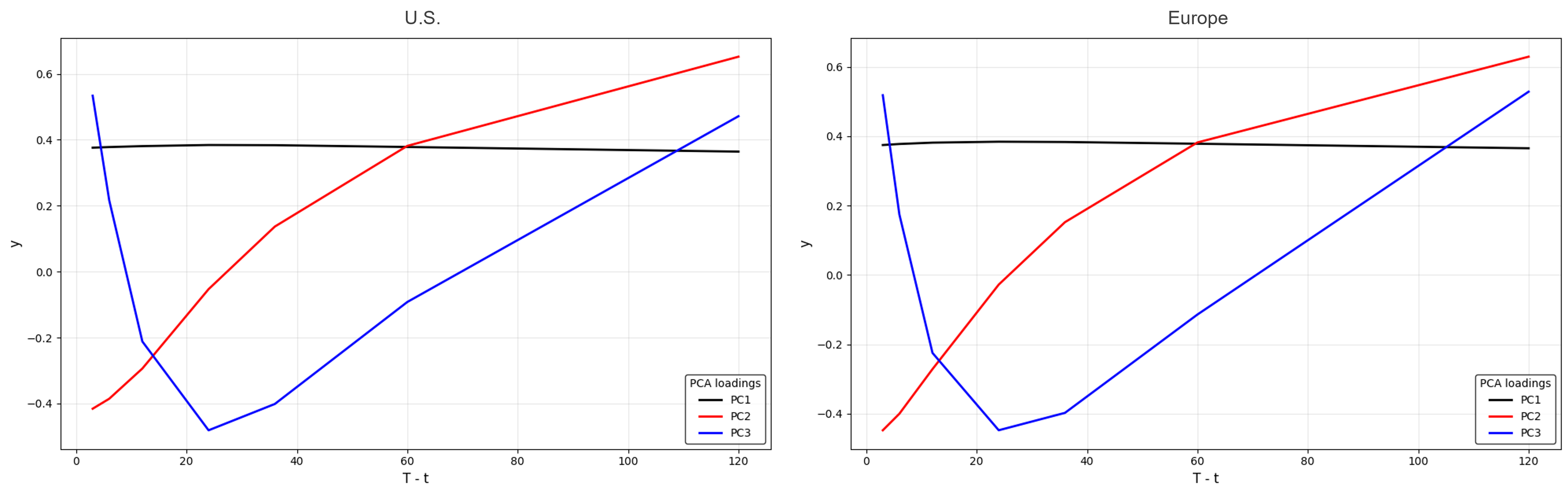}
    \caption{PCA factor loadings for different times to maturity, $T-t$, with U.S. data on the left and European data on the right.}
    \label{fig:fac_loadings_pca}
\end{figure}

Alternatively, we can use Principal Component Analysis (PCA) for dimensionality reduction. Unlike the DNS model, which imposes a specific functional form based on financial intuition, PCA is purely data-oriented. To maintain consistency, we retain the first three components for our forecasts, which can also be interpreted as level, slope, and curvature. Notice in Figure \ref{fig:fac_loadings_pca} that the first loading represents a constant, and if we invert the loadings of factors 2 and 3, they resemble the shapes of factors 2 and 3 from DNS shown in Figure \ref{fig:fac_loadings}.

As a more flexible, non-linear alternative, we also utilize Autoencoders (AEs), a type of NN specifically designed for dimensionality reduction \citep{suimon2020autoencoder}. An AE consists of two main components: an encoder and a decoder. The encoder network maps the input data (the full zero-rate curve) to a lower-dimensional latent representation $z$ (the factors). The decoder network then attempts to reconstruct the original zero-rate curve $\hat{x}$ from the compressed representation $z$ alone.
\begin{align*}
    z(x) &= \sigma^1(\omega^{1}x + b^{1}) && \text{(Encoding)} \\
    \hat{x} &= \sigma^2(\omega^{2}z(x) + b^{2}) && \text{(Decoding)}
\end{align*}
where $\omega^{1}$ and $\omega^{2}$ denote the weight matrices of the encoder and decoder, respectively, $b^{1}$ and $b^{2}$ are their corresponding bias vectors, and $\sigma^{1}(\cdot)$ and $\sigma^{2}(\cdot)$ represent the activation functions applied in each layer.

The network is trained by minimizing the Mean Squared Error (MSE) between the original input $x$ and the reconstructed output $\hat{x}$. The bottleneck created by the low-dimensional latent space forces the AE to learn the most salient features of the data. A linear AE trained with MSE loss is known to span the same subspace as PCA \citep{baldi1989neural}, but the use of non-linear activation functions ($\sigma^1, \sigma^2$) allows the AE to capture more complex, non-linear relationships in the zero-rate curve data. Also, to maintain consistency, we use AEs with three-dimensional bottlenecks. We also allow for architectures with additional layers in the encoder and the decoder to assess whether deeper representations further improve the model’s ability to learn zero-rate curve dynamics (See Section \ref{ae_fac}).

\begin{figure}[H]
    \centering
    \includegraphics[width=0.8\linewidth]{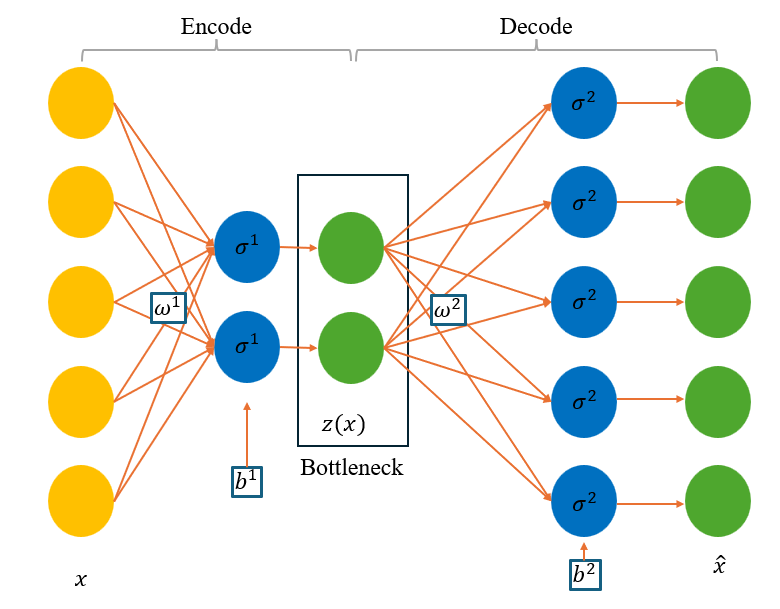}
    \caption{Single-layered Autoencoder.}
    \label{fig:ae_arch}
\end{figure}

While the DNS model provides an intuitive and flexible way to describe the shape and dynamics of the zero-rate curve, it does not prevent the possibility of arbitrage — that is, the existence of risk-free profit opportunities arising from inconsistencies in asset prices. In real financial markets, such situations should not persist: if they existed, traders would immediately exploit them, bringing prices back into balance. To address this limitation, \citet{christensen2011affine} reformulate the DNS model within a no-arbitrage framework, leading to the Arbitrage Free Nelson–Siegel (AFNS) model.

The AFNS model preserves the same three interpretable factors as the DNS --- level, slope, and curvature --- 
but imposes additional restrictions to ensure that rates of different maturities are internally consistent. 
This guarantees that the resulting zero-rate curve cannot imply unrealistic, risk-free profit opportunities. 
Formally, the factors are assumed to follow a mean-reverting process:
\begin{equation}
    dX_t = K (\theta - X_t) \, dt + \Sigma \, dW_t,
\end{equation}
where 
$X_t \in \mathbb{R}^3$ collects the three latent factors, 
$K \in \mathbb{R}^{3 \times 3}$ controls the speed of mean reversion toward the long-term mean 
$\theta \in \mathbb{R}^3$, 
$\Sigma \in \mathbb{R}^{3 \times 3}$ measures the instantaneous covariance of the factors, 
and $(W_{t})_{t\geq0}$ denotes a standard three-dimensional Brownian motion. This process ensures smooth and economically meaningful movements of the zero-rate curve.

To estimate the model empirically, the continuous-time process is expressed in discrete time, i.e. over a small time step $\Delta t$. The transition equation becomes
\[
X_{t+1} = \Phi X_t + c + \eta_t,
\]
where $\Phi = e^{-K\Delta t}$, $c = (I - \Phi)\theta$, and $\eta_t \sim \mathcal{N}(0, Q)$ with $Q$ denoting the covariance of the innovations implied by $\Sigma$. This discrete form describes how the latent factors evolve between observation dates.

The second equation in the state-space system links these latent factors to the observed rates:
\[
y_t(T) = f(T)^{\top} X_t + d(T) + \varepsilon_t,
\]
where $f(T)$ is the vector of Nelson–Siegel factor loadings for maturity $T$, $d(T)$ is a constant term (which includes an affine adjustment), and $\varepsilon_t$ captures measurement errors or small deviations between model-implied and actual zero rates.  

Thus, the AFNS model combines the transition equation, derived from the continuous-time mean-reverting dynamics, with the measurement equation that maps latent factors to observed rates. Together, these equations form the state-space representation estimated via the Kalman Filter, which jointly infers the unobserved factors and model parameters.

\begin{figure}[H]
    \centering
    \includegraphics[width=\linewidth]{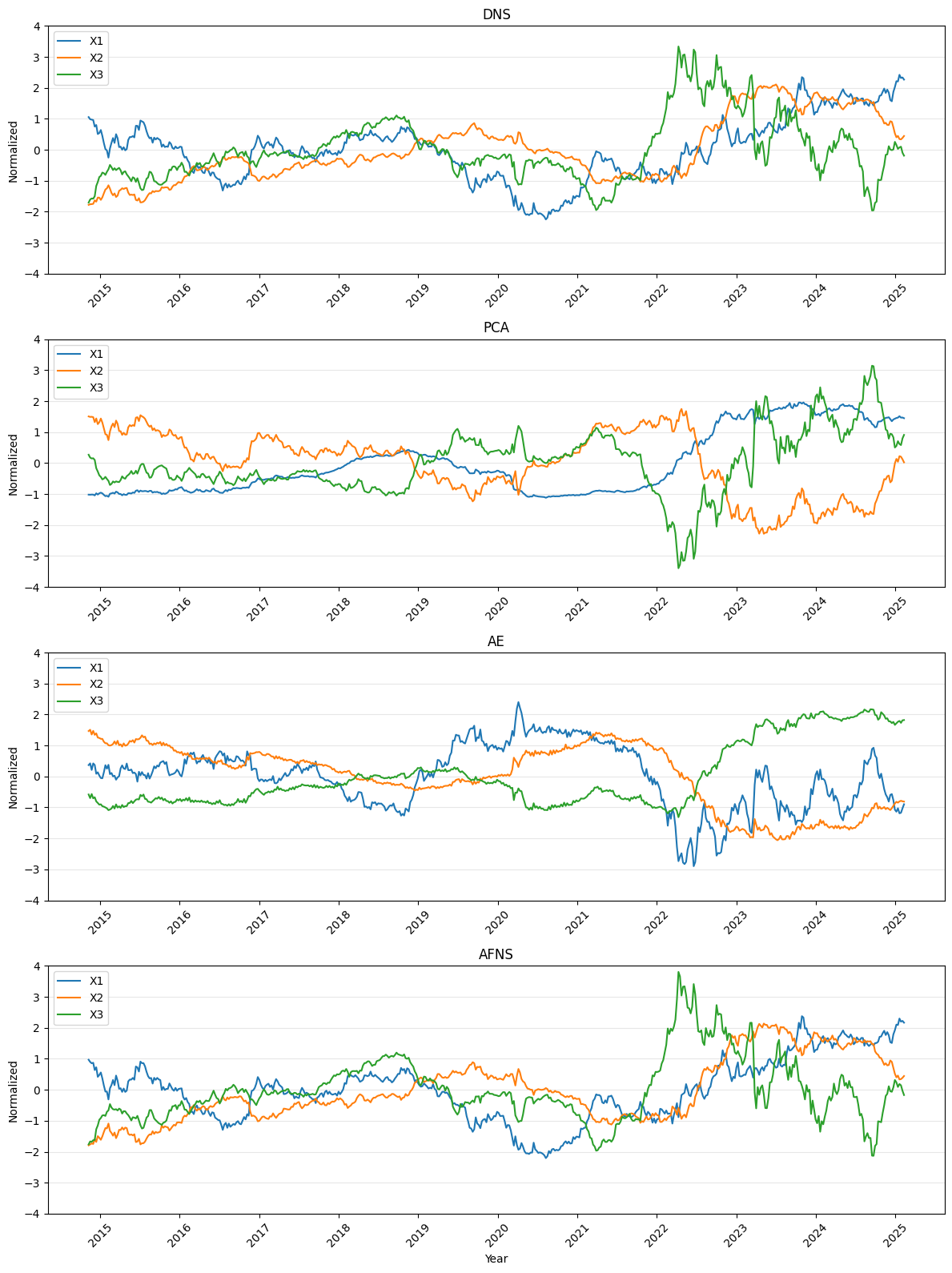}
    \caption{Normalized latent factors fitted for the U.S. data with DNS, PCA, AE, and AFNS between November 2014 and January 2025.}
    \label{fig:lat}
\end{figure}

Figure \ref{fig:lat} depicts the fitted latent factors for zero-rate U.S. data between December 2014 and February 2025.

To forecast the latent factors, traditional time series models like AR(1) and VAR(1) can be replaced with NNs that use lagged information as inputs. In this study, we employ a one-period lag. NNs offer greater flexibility, enabling an extension of the traditional forecasting framework in several ways. In this setup, we first perform the initial data compression step using PCA, DNS, AFNS, or AE, and subsequently train the NN to forecast the future values of these factors. 

Evidence supporting the feasibility of relaxing the stationarity assumption when using NNs for forecasting non-stationary time series has been provided by \citet{kim2004artificial} and \citet{bao2017deep}. Therefore, we do not apply differencing to the data when using NN in contrast to AR and VAR.

NNs can also be used to predict the zero-rate curve directly from compressed factors, modifying the traditional two-step factor-based approach. In this setup, the initial dimensionality reduction step is retained, but the NN is trained to map the current latent factors (e.g., PCA, DNS, AFNS, AE) directly to the future zero-rate curve. This eliminates the intermediate step of first forecasting future factors and then reconstructing the curve from those forecasted factors.

This approach also provides greater flexibility in input selection. Instead of latent factors, the NN can take selected raw zero rates as inputs. To maintain comparability with the three-factor models, we use three representative rates (e.g., 3-month, 60-month, and 120-month) as inputs in this setup.

Another key advantage of NNs is their flexibility to easily incorporate additional information. We leverage this by augmenting our models with macroeconomic data as supplementary predictive inputs, grouping these variables into the categories of real economic activity and inflation, as suggested in \citet{nunes2019comparison}.

These macroeconomic inputs can be fed into the NN in distinct ways:
\begin{itemize}
    \item \textbf{Raw Form:} The (normalized) macroeconomic variables can be used directly as inputs, with no prior dimensionality reduction.
    \item \textbf{Compressed Form:} Alternatively, we can first apply a dimensionality reduction technique (PCA or AE) to compress the information of each macroeconomic category (inflation or real economic activity) to one single factor. 
\end{itemize}

Finally, the resulting data is concatenated with the primary zero-rate-based inputs (i.e., the DNS/PCA/AE/AFNS factors, or the selected raw rates). This combined input vector is then used to train the NN, both for the models that forecast factors and for the models that perform direct prediction of the entire curve.

\begin{figure}[H]
    \centering
    \includegraphics[width=\linewidth]{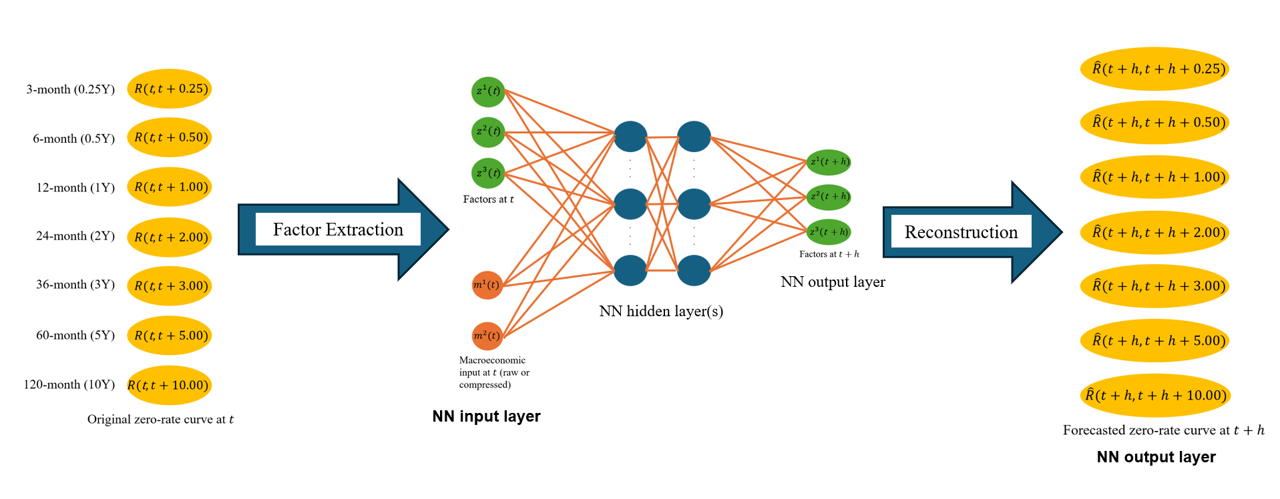}
    \caption{Architecture for the model employing NN for factor forecasting with a horizon $h$. The factors are first extracted from the zero-rate curve. In the illustration, the macroeconomic input is assumed to have already been extracted from the raw data (or provided directly in raw form, as described earlier). The NN then predicts the future values of the factors, and finally, the zero-rate curve is reconstructed from the predicted factors.}
    \label{fig:NNlat}
\end{figure}

\begin{figure}[H]
    \centering
    \includegraphics[width=\linewidth]{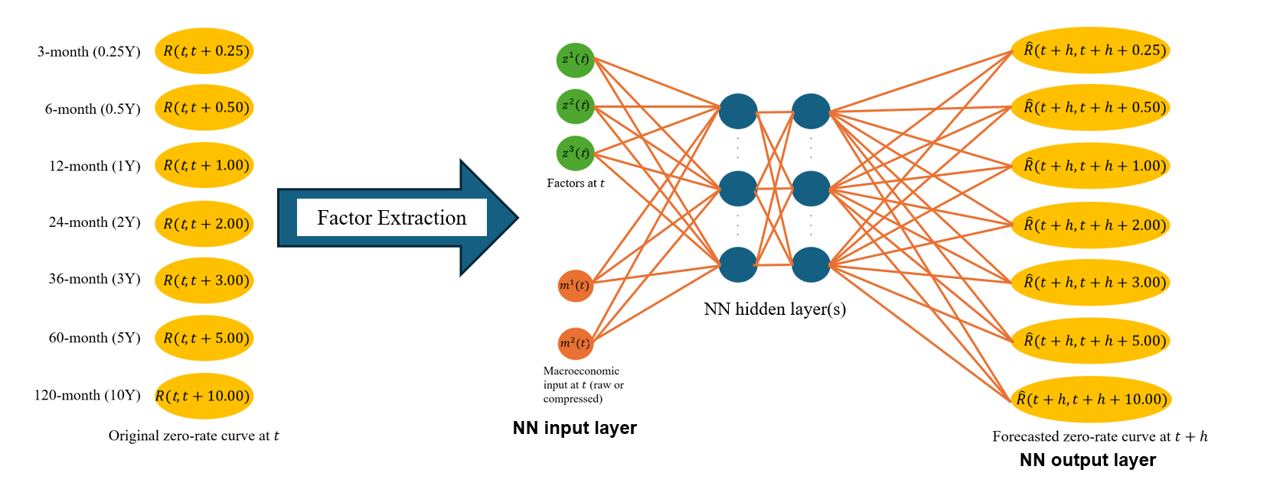}
    \caption{Architecture for the model employing NN for direct zero-rate forecasting for horizon $h$. In this illustration, macroeconomic inputs are assumed to be pre-extracted from the raw data (or provided directly in raw form, as described earlier). After extracting the latent factors, the NN directly predicts the future zero-rate curve, bypassing any intermediate reconstruction step.}
    \label{fig:NNdir}
\end{figure}

Figure \ref{fig:NNlat} illustrates the architecture of the model employing NN for factor forecasting, while Figure \ref{fig:NNdir} presents the corresponding architecture used for direct zero-rate forecasting.

Our training follows a two-tier strategy:
\begin{itemize}
    \item \textbf{Global Re-estimation:} Every two years (104 weeks), the model is retrained from scratch for multiple epochs with random weight initialization. This periodic retraining helps the model avoid local minima and adapt to long-term patterns \citep{tashman2000out}.
    \item \textbf{Local Updates:} On a weekly basis, the model is updated only with the new data by training for a single epoch.
\end{itemize}

For models that combine AFNS with a NN, the computationally intensive optimization of AFNS parameters is not repeated at every training epoch. Instead, previously estimated parameters are reused, and full recalibration occurs only during the global re-estimation step to keep the forecasting process computationally feasible.

Table \ref{tab:summ} summarizes all the 43 models that are tested. All inputs and outputs for NNs and AEs are normalized to have unit variance. The NNs and AEs are implemented in Python using the keras and tensorflow libraries. The RMSprop optimizer and Mean Squared Error (MSE) loss are used for training. 

\section{Model Performance}

In this section, we describe the procedure adopted for hyperparameter tuning of both the autoencoders AEs and the NNs used for factor-based and direct rate forecasting. Before detailing the model optimization process, we first outline the investment strategy that uses the zero-rate forecasts provided by the models, followed by the performance measures used to evaluate the strategy.

\subsection{Investment Strategy}\label{strat2}

To connect our zero-rate forecasts to investment decisions, we utilize duration as a primary measure of interest-rate risk. Duration quantifies a bond portfolio's price sensitivity to changes in interest rates.

\citep{zagst2002interest} For a coupon bond with coupon payments $C = (C(T_1, ..., C(T_n))$, $t_0 \leq T_1 < ... < T_n \leq T^*$, maturity $T_B = T_n$ and yield-to-maturity $y$, the Macaulay duration is given by
\[
D_{\text{Mac}}(y, t, T_B, C) = \frac{\sum_{i=1}^{n} (T_i - t)C(T_i)(1 + y)^{-(T_i - t)}}{\sum_{i=1}^{n} C(T_i)(1 + y)^{-(T_i - t)}}
\]
and the modified duration by
\[
D_{\text{mod}}(y, t, T_B, C) = \frac{D_{\text{Mac}}(y, t, T_B, C)}{1 + y}.
\]

Intuitively, duration is a first-order risk measure for the change in a bond's value. The approximate change in the bond's price ($\Delta \text{Bond}$) for a small change in yield ($\Delta y$) is given by:
\[
\Delta \text{Bond}(y, t, T_B, C) \approx -D_{\text{mod}}(y, t, T_B, C) \cdot \text{Bond}(y, t, T_B, C) \cdot \Delta y
\]
This relationship implies a straightforward core principle: an investor should increase the portfolio's duration when yields are expected to fall (to maximize gains from rising bond prices) and decrease the duration when yields are expected to rise (to mitigate losses from falling bond prices).

Our strategy's forecasts rely on par yields. We derive these par yields, $y(t)$ for {maturity $t$}, from our predicted zero-coupon rate curve $R(0,i)$ using the formula:
\[
y(t) = \frac{1 - \frac{1}{(1+R(0,t))^t}}{\sum_{i=1}^{t} \frac{1}{(1+R(0,i))^i}}.
\]
Since zero-coupon rates $R(0,i)$ are typically available only for a discrete set of maturities, we interpolate the missing rates for intermediate maturities. We apply simple linear interpolation between adjacent zero rates to obtain the curve, as the missing points primarily lie in the medium-to-long maturities, where zero rates change more gradually, making linear interpolation a suitable approach.

Based on our zero-rate forecasts, we employ a proprietary duration management strategy. Rather than making abrupt "all-or-nothing" changes, our approach progressively adjusts the portfolio's duration based on the strength of the forecast. A stronger signal of falling rates leads to a larger increase in duration, while a weaker signal results in a more modest adjustment, and vice versa for rising yields. To reflect industry standards, we employ lower and upper limits in these duration adjustments.

To balance signal stability with market responsiveness, we forecast one-month (4-week) changes in the zero-rate curve and update the signal on a weekly basis to adjust the portfolio’s duration. The longer forecast horizon reduces sensitivity to short-term noise, while weekly updates allow for timely reactions to new market information. This hybrid approach yields robust and consistent predictions. 

To implement the strategy in the U.S., we note that linear combinations of two assets are sufficient to achieve every target duration. We invest in the 1-month U.S. Treasury Bill, which serves as our risk-free asset, and the U.S. 10-year total return benchmark, an index that represents a portfolio of all securities issued with a maturity of 10 years. For Europe, we restrict to investments in the 1-month German Treasury Bill, which we use as a proxy for the risk-free rate, and the ICE BofA 10+ Year AAA Euro Government Index, a bond market index that tracks the performance of Euro-denominated government bonds rated AAA with a remaining maturity of at least 10 years.

For both strategies, we enforce a no short-selling constraint, meaning all portfolio weights must be non-negative, and the lower bound $2.5$ and upper bound $7.5$ for the duration. For performance comparison, we evaluate our active strategy against a passive benchmark portfolio that maintains a constant 5-year duration.

\subsection{Performance metrics}

To evaluate our investment strategy, several key performance measures are used. 

One of them is the Information Ratio (IR), which evaluates performance relative to a benchmark. It is defined as
\[
\text{IR} =
\frac{\mathbb{E}\left[R(\mathbf{x}) - R_b\right]}
{\text{STD}\left[R(\mathbf{x}) - R_b\right]},
\]
where $R(\mathbf{x})$ denotes the strategy's return, $R_b$ is the return of the benchmark, and $\text{STD}\left[R(\mathbf{x}) - R_b\right]$ is the standard deviation of the active return, also known as the tracking error (TE).

Moreover, we employ the Omega Ratio ($\Omega$). 
Following \citet{keating2002universal}, it is calculated as the ratio of the expected excess gains and expected excess losses against a benchmark. 
A value greater than 1 suggests that the strategy has a higher expectation of gains than losses relative to this benchmark. Its formula is:
\[
\Omega = \frac{\mathbb{E}[\max(R(x) - R_b, 0)]}{\mathbb{E}[\max(R_b - R(x), 0)]}.
\] 

Finally, to quantify downside risk, the (relative) Maximum Drawdown (MDD) is used. The MDD represents the largest relative loss from a historical peak to a subsequent trough in the value of an investment \citep{chekhlov2005drawdown}. The formula is given by: 
$$
\text{MDD} = 
\max_{t \in [0,T]} 
\left( 
1 - 
\frac{\text{V}_t}{
\max_{\tau \in [0,t]} \text{V}_\tau}
\right) \times 100.
$$
In this equation, $\text{V}_t$ represents the total value of the portfolio at time $t$.

\subsection{Hyperparameter Tuning}\label{hyp2}

To develop robust models that generalize effectively to new, unseen data, we designed a rigorous framework for hyperparameter optimization and validation. This process is critical, as hyperparameters chosen based on in-sample training data often lead to overfitting. Our strategy ensures that hyperparameters are selected based on their performance on a dedicated, out-of-training validation set, providing a more reliable estimate of how the model will perform in the future with unseen data.

While inspired by the principles of Blocked Cross-Validation recommended by \citet{bergmeir2018note} to preserve temporal causality, our methodology is a custom adaptation tailored to our research needs. Rather than creating multiple, independent validation blocks, we employ a sequential, expanding-window approach that more closely mimics a real-world forecasting workflow.

Our data is partitioned as follows:
\begin{enumerate}
    \item \textbf{Initial Training Set}: A fixed block of data from April 1987 to January 2005 (or February 1992 to January 2005 for European data), used to establish the initial model.
    \item \textbf{Validation Set}: The subsequent period from January 2005 to December 2014. The validation set is used exclusively for hyperparameter tuning. The tuning process involves retraining the model at the start of sequential two-year blocks within this period and expanding the training window weekly thereafter.
    \item \textbf{Test Set}: A final, completely untouched hold-out period from December 2014 to February 2025 used for the final out-of-sample performance evaluation. The chosen end date reflects the most recent complete observations accessible when the analysis was performed.
\end{enumerate}

The validation period was deliberately selected for its diverse market conditions—encompassing rising and falling interest rates, high volatility, and periods of stability—which mirror the diversity of dynamics of the test set (see Figures \ref{fig:american_zero}  and \ref{fig:Euro_zero}). This structural similarity increases the likelihood that hyperparameters identified as optimal during validation will maintain their performance on the final test data.

We employ Bayesian Optimization and Hyperband (BOHB) as the algorithm for efficient hyperparameter tuning. BOHB merges the strengths of two methods: the sample efficiency of Bayesian Optimization (BO) and the scalability of Hyperbands (HB). It uses BO to select promising hyperparameter configurations instead of random sampling and employs the HB successive halving strategy to allocate computational resources, quickly discarding poorly performing configurations and dedicating more resources to promising ones \citep{falkner2018bohb}.

We optimized the hyperparameters for three distinct model components: the AEs for factor extraction, the NNs for forecasting factors, and the NNs doing direct zero-rate prediction.

\subsubsection{Autoencoders for Factor Extraction}\label{ae_fac}
The primary optimization objective for the AEs was to find the set of hyperparameters that minimized the MSE of the input reconstruction on the validation set. For both the zero-rate curve and macroeconomic AEs, we tuned the following common parameters:

\begin{itemize}
    \item \textbf{Learning Rate}: A continuous value sampled on a logarithmic scale between $0.0001$ and $0.1$.
    \item \textbf{Activation Function}: A categorical hyperparameter defining the activation function between hidden layers. The choices are ReLu, sigmoid, and tanh .
    \item \textbf{Batch Size}: An integer value sampled uniformly between $25$ and $100$.
\end{itemize}

For the macroeconomic AE, hyperparameter tuning was conducted on a focused subset of indicators: USPRIV and INDPRO for U.S. economic activity (and EMPDE and INDPRODE for Europe), as well as CPIAUCSL and SPREAD for U.S. inflation (and CPIDE and PPIDE for Europe). We chose not to include some indicators during the hyperparameter optimization phase, in order to limit model complexity and ensure a more efficient tuning process. They were reintroduced for model estimation and evaluation on the test set (see Section~\ref{results}).

On the real activity side, PAYEMS was not included in the hyperparameter optimization, as it encompasses government employment, which can be significantly influenced by policy decisions and may not accurately capture the underlying economic fundamentals driving the term structure. On the inflation side, PPIACO was not included during this stage, since empirical evidence suggests that producer prices tend to lag consumer prices and do not contemporaneously reflect inflation dynamics \citep{clark1995producer}.

\begin{figure}[H]
    \centering
    \includegraphics[width=0.6\linewidth]{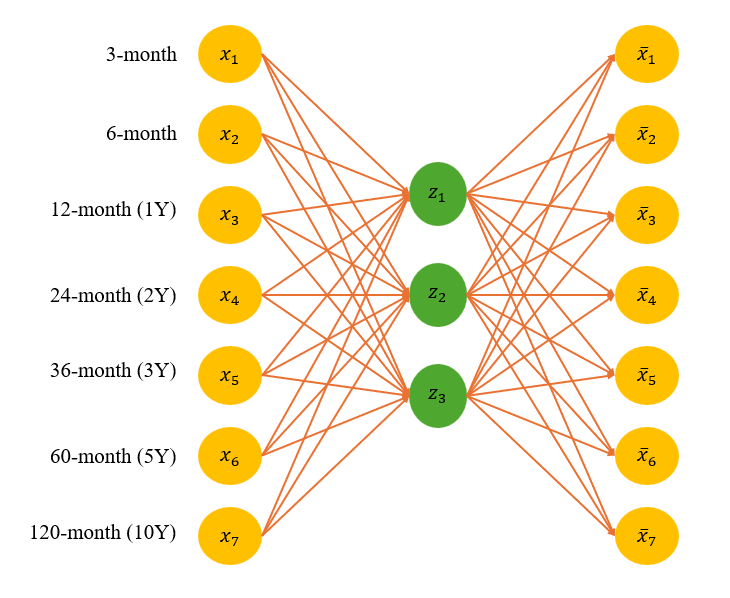}
    \caption{AE for factor extraction of the zero-rate curve with architecture 7-3-7.}
    \label{fig:ae1}
\end{figure}

\begin{figure}[H]
    \centering
    \includegraphics[width=0.8\linewidth]{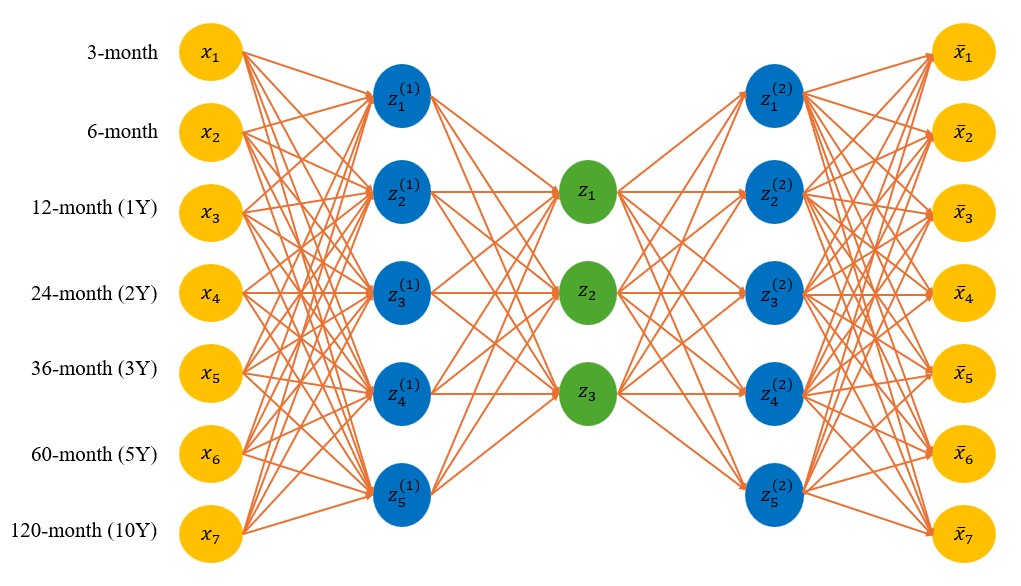}
    \caption{AE for factor extraction of the zero-rate curve with architecture 7-5-3-5-7.}
    \label{fig:ae2}
\end{figure}

For the zero-rate curve AE, we also optimized network depth, where a shallower 7-3-7 architecture (indicating 7 input neurons, 3 in the hidden layer, and 7 output neurons, see Figure \ref{fig:ae1}) proved superior to a deeper 7-5-3-5-7 structure (see Figure \ref{fig:ae2}).

\subsubsection{Neural Networks}

For the NNs, identifying the best hyperparameters required a multi-stage process to ensure the models were not only statistically accurate but also economically meaningful. In this context, economic meaning refers to the model’s ability to generate forecasts that capture relevant market dynamics and translate into actionable signals for portfolio management, rather than merely minimizing numerical prediction errors. A model with a low MSE may still produce forecasts that fail to accurately anticipate turning points or structural shifts in the zero-rate curve, thereby limiting its usefulness for investment or risk-management decisions.

The initial search focused on identifying configurations that minimized the average validation MSE. We then excluded models with clear issues in their directional accuracy or in the investment performance of the bond strategy based on their forecasts.

To further ensure that the selected models delivered stable and consistent performance across time, the final selection was guided by a custom stability-oriented metric, the Average RMSE Quantile Penalized (ARQP) score. The ARQP explicitly penalizes large, infrequent forecast errors—outliers that can severely affect investment outcomes or risk management metrics—thus favoring models that maintain reliability over different periods. Specifically, we split the validation data into five sequential folds of equal size and computed:
\begin{equation}
    \text{ARQP} = \overline{\text{RMSE}} + 0.02 \sum_{i=1}^{5} Q_{0.95}^{(i)}
\end{equation}
where $\overline{\text{RMSE}}$ is the average RMSE across all maturities in the validation set and $Q_{0.95}^{(i)}$ is the 95\% quantile of the squared residuals for fold $i$.

In addition to the learning rate, activation function, and batch size mentioned previously, we optimized the network architecture for each forecasting NN:
\begin{itemize}
    \item \textbf{Number of Hidden Layers}: A choice between 1 or 2 hidden layers.
    \item \textbf{Number of Neurons per Layer}: An integer value sampled uniformly between $3$ and $10$.
\end{itemize}

\section{Results}\label{results}

In this section, we apply, evaluate, and perform a deeper analysis of selected models from the duration management strategy described in Section \ref{strat2}, using one-month (four-week) ahead forecasts. The evaluation is conducted using both U.S. and European data under various macroeconomic settings. The specific macroeconomic indicators used for each setting are detailed in Table \ref{tab:macro_exp_combined}. We implemented a systematic selection procedure based on forecast accuracy and portfolio performance, ensuring the chosen models have strong predictive power and robust portfolio outcomes.

\begin{table}[H]
\caption{Macroeconomic settings for U.S. and European models. A checkmark \checkmark indicates that an indicator is included in a given setting. Setting N represents models with no macroeconomic inputs. Settings A (U.S.) and D (EU) were used for the hyperparameter tuning process detailed in Section \ref{hyp2}. Note that for setting C, only models with "Raw" macro inputs are applicable. For setting F, the inflation indicator (CPIDE) is included in its raw form even for models designed to use PCA or AE for macro inputs.\label{tab:macro_exp_combined}}
\begin{tabularx}{\textwidth}{lllccccccc}
\toprule
\textbf{Indicator of} & \textbf{Region} & \textbf{Abbreviation} & \textbf{N} & \textbf{A} & \textbf{B} & \textbf{C} & \textbf{D} & \textbf{E} & \textbf{F} \\
\midrule
Real Economic Activity & U.S. & USPRIV & - & \checkmark & \checkmark & - & - & - & - \\
Real Economic Activity & U.S. & INDPRO & - & \checkmark & \checkmark & - & - & - & - \\
Real Economic Activity & U.S. & PAYEMS & - & - & \checkmark & \checkmark & - & - & - \\
Inflation & U.S. & CPIAUCSL & - & \checkmark & \checkmark & - & - & - & - \\
Inflation & U.S. & PPIACO & - & - & \checkmark & - & - & - & - \\
Inflation & U.S. & SPREAD & - & \checkmark & \checkmark & \checkmark & - & - & - \\
\midrule
Real Economic Activity & Europe & EMPDE & - & - & - & - & \checkmark & \checkmark & \checkmark \\
Real Economic Activity & Europe & INDPRODE & - & - & - & - & \checkmark & \checkmark & \checkmark \\
Real Economic Activity & Europe & PAYEMSDE & - & - & - & - & - & \checkmark & - \\
Inflation & Europe & CPIDE & - & - & - & - & \checkmark & \checkmark & \checkmark \\
Inflation & Europe & PPIDE & - & - & - & - & \checkmark & \checkmark & - \\
\bottomrule
\end{tabularx}
\end{table}

This approach generates a large number of models to compare across different settings, which we will refer to as the setting name, along with the corresponding model number. Model A25, for instance, refers to Model 25 using macroeconomic setting A. Models using no macroeconomic data will have N as their prefix.

\begin{figure}[H]
    \centering
    \includegraphics[width=\linewidth]{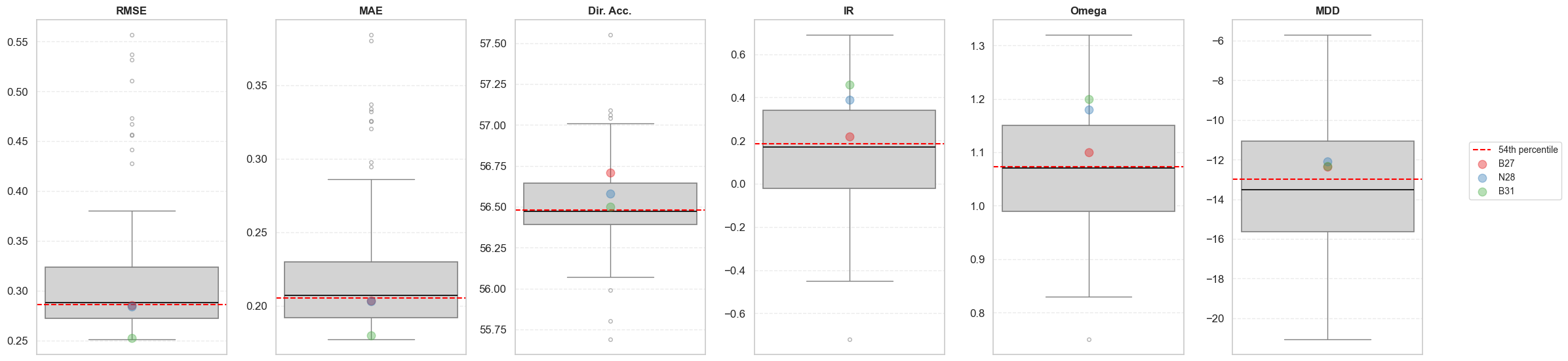}
    \caption{Box plots for the U.S. results depicting RMSE, MAE, directional accuracy across all maturities, IR, $\Omega$, and MDD. The 54th percentile was used here for filtering out models.}
    \label{fig:us_box}
\end{figure}

\begin{figure}[H]
    \centering
    \includegraphics[width=\linewidth]{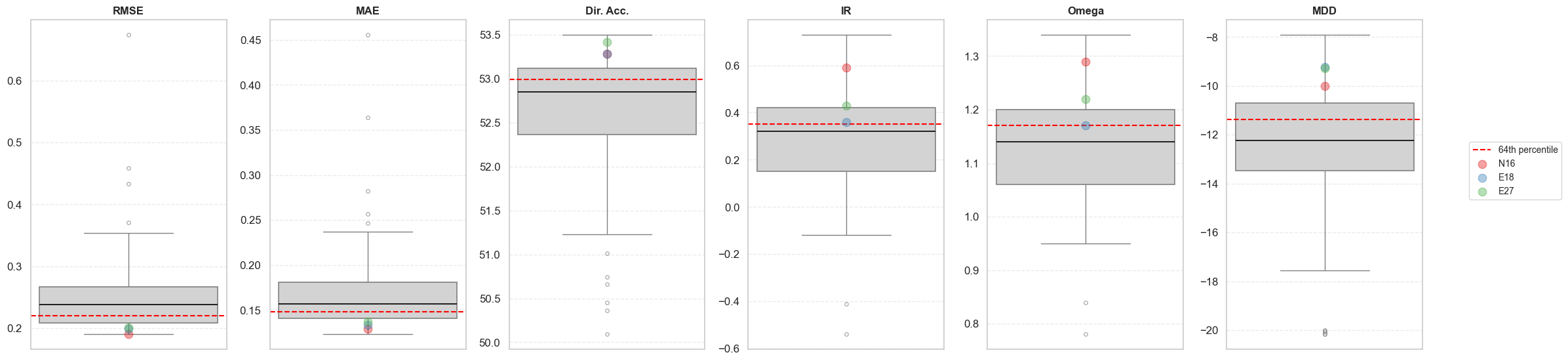}
    \caption{Box plots for the Europe results depicting RMSE, MAE, directional accuracy across all maturities, IR, $\Omega$, and MDD. The 64th percentile was used here for filtering out models.}
    \label{fig:eu_box}
\end{figure}

Since a detailed analysis of every model was not feasible, we applied a systematic filtering procedure to identify the most promising candidates. At each iteration, models falling outside a specified percentile threshold were removed—namely, those exhibiting the highest average RMSE or MAE values across maturities, or the lowest performance in directional accuracy, IR, $\Omega$, or MDD. Only models that consistently remained within the threshold for all performance metrics were retained. The percentile cutoff was then progressively tightened until exactly three models met all the selection criteria. If a stricter threshold excluded too many models, the previous level was restored to ensure a sufficient number of top-performing candidates for final analysis. Tables 2.1 and 2.2 in the Supplementary Material show the analysis in detail, while Table \ref{tab:selected_models_combined} lists the three selected models for the U.S. and Europe. Figures \ref{fig:us_box} and \ref{fig:eu_box} illustrate their comparative performance relative to the full model set. The higher percentile threshold required for Europe (64th versus 54th for the U.S.) indicates that model performance was more heterogeneous across metrics in the U.S. dataset. This suggests that predictive consistency was higher for the European market, whereas U.S. models exhibited greater variability depending on the evaluation criterion.

\begin{table}[H]
\centering
\caption{Selected models for the U.S. and Europe that satisfy all performance criteria.}
\label{tab:selected_models_combined}
\begin{tabular}{p{1.5cm} p{1cm} p{3.0cm} p{2.5cm} p{4cm}}
\toprule
\textbf{Region} & \textbf{Model} & \textbf{Input Zero Rates} & \textbf{Input Macro} & \textbf{Prediction Method} \\
\midrule
\multirow{3}{*}{\textbf{U.S.}} & B27 & DNS & AE & NN for direct prediction \\
& N28 & AFNS & - & NN for direct prediction \\
& B31 & AFNS & AE & NN for direct prediction \\
\midrule
\multirow{3}{*}{\textbf{Europe}} & N16 & PCA & -- & NN for factor prediction \\
& E18 & PCA & PCA & NN for factor prediction \\
& E27 & DNS & AE & NN for direct prediction \\
\bottomrule
\end{tabular}
\end{table}

While in a practical investment context, one might focus solely on return-based metrics, we deliberately included forecast accuracy measures (RMSE, MAE, and directional accuracy) in our selection criteria. This dual requirement ensures that the selected models exhibit not only strong portfolio performance but also reliable predictive power. By prioritizing models whose strong performance is underpinned by reliable zero-rate forecasts, we aim to enhance the robustness and interpretability of our findings, thereby mitigating the risk of selecting models that perform well due to chance or overfitting.

\subsection{Analysis}

We conduct a two-stage analysis to evaluate the filtered models. In the first stage, we construct a base score to assess overall performance across the full test period, combining three accuracy-oriented metrics (RMSE, MAE, and directional accuracy) with three performance-oriented metrics (IR, $\Omega$, and MDD). Models are ranked for each metric, and the ranks are summed to obtain the base score. Table \ref{tab:perf_res_combined} reports the metric values, while Table \ref{tab:perf_rank_combined} summarizes the resulting rankings.

One important observation is that, even though most traditional models were not filtered out by forecast accuracy measures RMSE and MAE, they do not appear in the final selection because of their inferior directional accuracy or investment performance. This highlights the importance of assessing models within their specific context of use; metrics that measure statistical error alone may not capture a model's practical value for financial decision-making \citep{dunis2007economic}.

\begin{table}[H]
\centering
\caption{Comparison of model performance metrics for the U.S. and Europe. Return metrics are annualized.}
\begin{tabular}{l l c c c c c c}
\toprule
\textbf{Region} & \textbf{Model} & \textbf{RMSE} & \textbf{MAE} & \textbf{Dir.} & \textbf{IR} & \textbf{$\Omega$} & \textbf{MDD} \\
\midrule
\multirow{3}{*}{\textbf{U.S.}} 
 & B27 & 0.2855 & 0.2034 & 56.71 &  0.22 & 1.10 & -12.36\\
 & N28 & 0.2841 & 0.2031 & 56.58 & 0.39 & 1.18 & -12.10\\
 & B31 & 0.2527 & 0.1798 & 56.50 & 0.46 & 1.20 & -12.34\\
\midrule
\multirow{3}{*}{\textbf{Europe}} 
 & N16 & 0.1901 & 0.1288 & 53.28 & 0.59 & 1.29 & -10.02 \\ 
 & E18 & 0.1989 & 0.1336 & 53.28 & 0.36 & 1.17 & -9.23 \\
 & E27 & 0.2013 & 0.1368 & 53.42 & 0.43 & 1.22 & -9.27\\
\bottomrule
\end{tabular}

\label{tab:perf_res_combined}
\end{table}

\begin{table}[H]
\centering
\caption{Consolidated performance rankings for the selected U.S. and European models. The ranking is based on Table \ref{tab:perf_res_combined}.}
\label{tab:perf_rank_combined}
\begin{tabular}{l l c c c c c c c c}
\toprule
\textbf{Region} & \textbf{Model} & \textbf{RMSE} & \textbf{MAE} & \textbf{Dir.} & \textbf{IR} & \textbf{$\Omega$} & \textbf{MDD} & \textbf{Base Score}  \\
\midrule
\multirow{3}{*}{\textbf{U.S.}} & B27 & 3 & 3 & 1 & 3 & 3 & 3 & 16 \\
& N28 & 2 & 2 & 2 & 2 & 2 & 1 & 11  \\
& B31 & 1 & 1 & 3 & 1 & 1 & 2 & 9  \\
\midrule
\multirow{3}{*}{\textbf{Europe}} & N16 & 1 & 1 & 2 & 1 & 1 & 3 & 9  \\
& E18 & 2 & 2 & 2 & 3 & 3 & 1 & 13  \\
& E27 & 3 & 3 & 1 & 2 & 2 & 2 & 13  \\
\bottomrule
\end{tabular}
\end{table}

Next, we conduct a period-based analysis to evaluate the consistency of model performance across four distinct market phases, defined separately for the U.S. and Europe. This approach enables us to assess how models behave under varying zero-rate curve dynamics and monetary policy environments.

For the U.S., the first phase ($\text{P}_1$, December 2014 – February 2018) is characterized by a contracting zero-rate curve. The second phase ($\text{P}_2$, February 2018 – July 2021) marks a period of falling zero rates, encompassing the market disruptions during the Covid-19 crisis. The third phase ($\text{P}_3$, July 2021 – November 2022) reflects sharply rising zero rates as the Federal Reserve implemented anti-inflationary policies. Finally, the fourth phase ($\text{P}_4$, November 2022 – February 2025) was marked by heightened volatility, particularly in short-term maturities.

For Europe, the first phase ($\text{P}_1$, December 2014 – September 2019) corresponds to gradually decreasing rates. The second phase ($\text{P}_2$, September 2019 – November 2021) captures stable but negative rates, again overlapping with the Covid-19 crisis. The third phase ($\text{P}_3$, December 2021 – October 2023) represents a regime of sharply rising rates as the ECB adopted anti-inflationary measures. The fourth phase ($\text{P}_4$, October 2023 – February 2025) is defined by broad-based volatility across maturities. The separation of U.S. and European phases reflects the lag in monetary policy adjustments between the two regions.

For each period, models are ranked according to $\Omega$ and MDD. We do not use the IR here because, in shorter subperiods, model returns may fall below the benchmark, resulting in negative IRs. In such cases, ranking becomes misleading, as higher TE would mechanically lead to less negative IR values, which does not correspond to better performance. The ranks are summed to produce a score specific to that period. These scores are then aggregated across all phases and combined with the base score to yield a comprehensive assessment of performance consistency across market conditions. This ensures that the final ranking reflects not only the average level of performance, but also the stability of that performance across distinct environments. We emphasize that this period-wise evaluation relies exclusively on performance metrics and therefore introduces a tilt toward performance measures. Therefore, in the event of a tie in the final ranking, preference is given to the model with the lower base score, as it incorporates a broader set of metrics.

Table \ref{tab:period_rankings_combined} depicts the results after the period analysis. Integrating this into the Table \ref{tab:perf_rank_combined}, we obtain the results depicted in Table \ref{tab:final_rankings_combined}.

\begin{table}[H]
\centering
\caption{Period-based performance rankings for selected models in the U.S. and Europe.}
\label{tab:period_rankings_combined}
\begin{tabular}{ll cccccc}
\toprule
\textbf{Region} & \textbf{Model} & \textbf{$\text{P}_1$} & \textbf{$\text{P}_2$} & \textbf{$\text{P}_3$} & \textbf{$\text{P}_4$} & \textbf{Sum} & \textbf{Period Score} \\
\midrule
\multirow{3}{*}{\textbf{U.S.}} & B27 & 1 & 1 & 3 & 2 & 7 & 2 \\
& N28 & 1 & 3 & 1 & 1 & 6 & 1 \\
& B31 & 1 & 2 & 2 & 2 & 7 & 2 \\
\midrule
\multirow{3}{*}{\textbf{Europe}} & N16 & 2 & 3 & 1 & 1 & 7 & 2 \\
& E18 & 3 & 1 & 1 & 2 & 7 & 2 \\
& E27 & 1 & 2 & 1 & 2 & 6 & 1 \\
\bottomrule
\end{tabular}
\end{table}

\begin{table}[H]
\centering
\caption{Consolidated final rankings for selected models in the U.S. and Europe. The final selected models are marked in \textbf{bold}.}
\label{tab:final_rankings_combined}
\begin{tabular}{ll cccc}
\toprule
\textbf{Region} & \textbf{Model} & \textbf{Base Score} & \textbf{Period Score} & \textbf{Total Score} & \textbf{Final Rank} \\
\midrule
\multirow{3}{*}{\textbf{U.S.}} & B27 & 16 & 2 & 18 & 3 \\
& N28 & 11 & 1 & 12 & 2 \\
& \textbf{B31} & 9 & 2 & 11 & 1 \\
\midrule
\multirow{3}{*}{\textbf{Europe}} & \textbf{N16} & 9 & 2 & 11 & 1 \\
& E18 & 13 & 2 & 15 & 3 \\
& E27 & 13 & 1 & 14 & 2 \\
\bottomrule
\end{tabular}
\end{table}

For the U.S. market, B31 is the final recommendation. This model uses NN for direct prediction with AFNS (Figure \ref{fig:NNdir}) to compress information about the zero rates along with AE for macroeconomic setting B, which includes all macroeconomic indicators (Table \ref{tab:macro_exp_combined}). Its performance profile shows a favorable asymmetry: in the specific periods and metrics where it underperforms relative to others, the underperformance is only minor, whereas its outperformance is substantially stronger (see Table \ref{tab:perf_res_combined} or Section 4 in the Supplementary Document). Figure \ref{fig:us_gen} illustrates that Model B31 consistently made appropriate duration adjustments across different environments. During periods of declining yields—such as between 2015 and mid-2016, and again between 2019 and mid-2020—the model maintained a high duration. Conversely, during phases of rising yields—such as from mid-2017 to mid-2018 and between 2022 and 2024—it maintained a low duration. Overall, Model B31 made a better duration decision compared to the benchmark (i.e., duration above 5 during falling yields or below 5 during rising yields) in 52.6\% of the observed periods, as illustrated in Figure \ref{fig:us_gen}.

\begin{figure}[H]
    \centering
    \includegraphics[width=\linewidth]{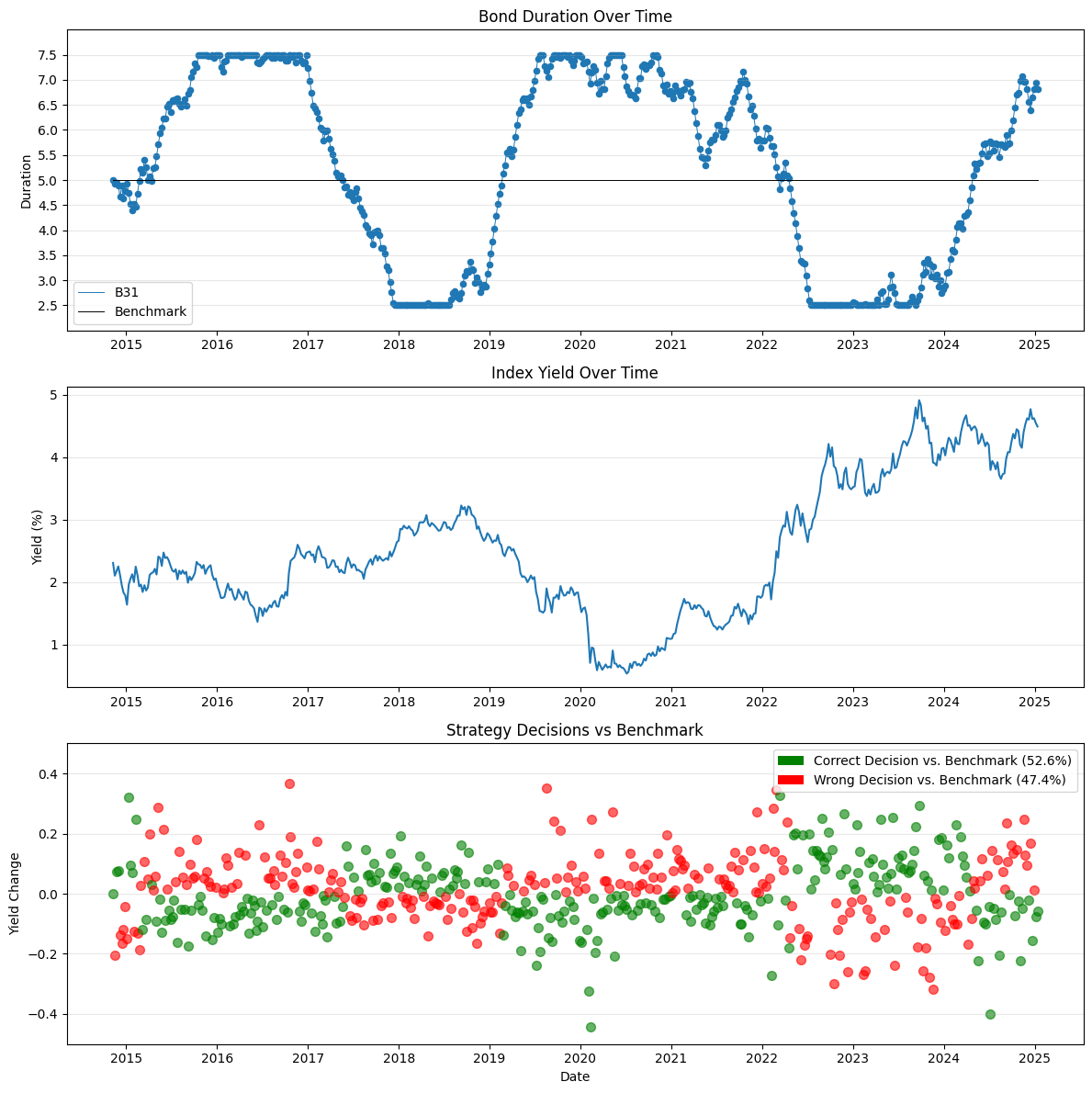}
    \caption{The first panel illustrates the portfolio duration implied by Model B31 based on U.S. data. The second panel displays the yield of the traded index, specifically the U.S. 10-year total return benchmark. The third panel illustrates yield changes over time. Green points highlight periods when the portfolio allocation recommended by Model B31 outperforms the 5-year constant-duration benchmark—specifically, when duration exceeds 5 during falling yields (leading to higher gains) or falls below 5 during rising yields (limiting losses). Red points, in contrast, indicate periods when the strategy is less favorably positioned relative to the benchmark. }
    \label{fig:us_gen}
\end{figure}

\begin{figure}[H]
    \centering
    \includegraphics[width=\linewidth]{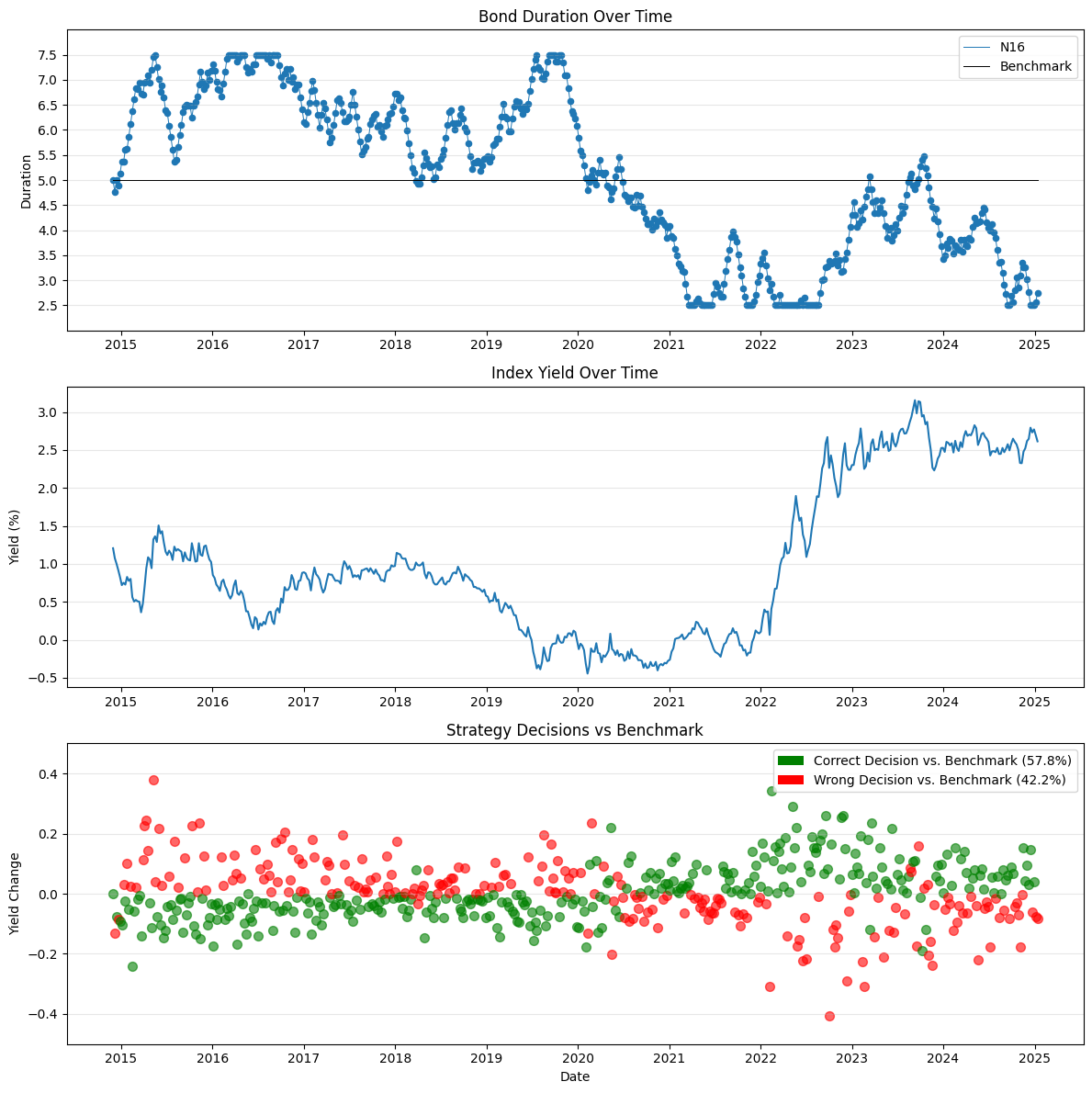}
    \caption{The first panel illustrates the portfolio duration implied by Model N16 based on Europe data. The second panel displays the yield of the traded index, specifically the ICE BofA 10+ Year AAA Euro Government Index. The third panel illustrates yield changes over time. Green points highlight periods when the portfolio allocation recommended by Model N16 outperforms the 5-year constant-duration benchmark—specifically, when duration exceeds 5 during falling yields (leading to higher gains) or falls below 5 during rising yields (limiting losses). Red points, in contrast, indicate periods when the strategy is less favorably positioned relative to the benchmark.}
    \label{fig:eu_gen}
\end{figure}

For Europe, N16 is the final recommendation. This model uses NN for factor prediction with zero rates being compressed with PCA (Figure \ref{fig:NNlat}) and without macroeconomic input. As for B31 in the U.S., its periods of underperformance relative to the other models are relatively modest, however, whenever N16 does outperform the competing specifications, it does so by a substantial margin. This is particularly clear in $P_1$ and $P_4$ (see Section 5 in the Supplementary Document), where N16 achieves significantly higher returns than the other models. Figure \ref{fig:eu_gen} illustrates that Model N16 also consistently made appropriate duration adjustments across different environments. Markedly, in the period of declining yields between 2018 and mid-2019, it kept its duration high, while during the period of increasing yields between 2022 and 2023, it kept its duration low, making adjustments when necessary. Model N16 made a better duration decision compared to the benchmark (i.e., duration above 5 during falling yields or below 5 during rising yields) in 57.8\% of the observed period.

Interestingly, comparing the second and third panels of Figures~\ref{fig:us_gen} and~\ref{fig:eu_gen} highlights the higher volatility of the U.S. market relative to the European market. For the test set, the U.S. exhibits a standard deviation of zero-rate changes of $\sigma = 0.12$, while Europe shows $\sigma = 0.10$. European financial and macroeconomic time series tend to exhibit smoother dynamics and lower volatility. In contrast, U.S. markets tend to react more sharply to high-frequency data releases and unexpected policy signals from the Federal Reserve, introducing greater unpredictability that increases forecast errors. Structural differences in central bank communication and market behavior across regions may further contribute to these disparities in model performance.

Consequently, the median forecast errors (RMSE and MAE) are consistently lower in Europe than in the U.S., while the median investment performance measures are generally higher, as shown in Figure~\ref{fig:us_box} and Figure~\ref{fig:eu_box}. Overall, the European models exhibit better median performance across metrics, except for directional accuracy. Notably, the final European model does not include macroeconomic variables, which may reflect the fact that European markets are less sensitive to macroeconomic fluctuations and that including such information provides limited predictive benefit. Moreover, the higher volatility in the U.S. often causes the bond strategy to frequently hit its upper or lower bounds. By contrast, in Europe, this effect is less pronounced, as illustrated by the first panels of Figure~\ref{fig:us_gen} and Figure~\ref{fig:eu_gen}.

\subsection{Observations}

First and foremost, it is essential to note that all six models selected for in-depth analysis consistently demonstrated strong and meaningful performance. Each outperformed the benchmark and achieved a positive IR during the evaluation period, with certain models standing out in specific time frames or according to particular performance metrics. The selection process was guided by the goal of identifying the most comprehensive and robust models across diverse market conditions and evaluation criteria. Nonetheless, we recognize that alternative models may also be suitable, depending on individual investor objectives and strategic preferences.

Additionally, it is interesting to note the diversity of hyperparameters among the analyzed models. For the six models, two different activation functions were used, with learning rates ranging from $10^{-3}$ to $10^{-2}$, both with and without deep architectures. This highlights the importance of proper specific hyperparameter optimization when using NNs.

We do not observe any clear advantage or disadvantage between using NN for direct prediction of zero rates versus predicting latent factors. Both methodologies appear viable, as they consistently generate accurate forecasts. However, factor-based approaches offer the added benefit of interpretability, as they enable understanding of the underlying dynamics that drive zero-rate curve changes. In contrast, direct zero-rate prediction may capture non-linearities more flexibly but at the cost of reduced transparency.

None of the selected models employs AEs for zero-rate compression; however, three of the four using macroeconomic input utilize AEs to process it. Remarkably, the versions that were ultimately selected were always those using the full macroeconomic input set (setting B for the U.S. and setting E for Europe). We interpret this as follows: the AEs are capable of handling a broad set of macroeconomic features, effectively filtering out redundant or noisy variables and extracting meaningful signals. This pre-processing step provides the main NNs with a cleaner, condensed dataset, effectively stripping away random fluctuations and combining repetitive information. Additionally, this capacity was achieved despite hyperparameter tuning being conducted solely on setting A for the U.S. and setting D for Europe.

That said, we do not dismiss the potential value of AEs for interest rates. We believe AEs for zero-rate compression can have even greater value in higher-dimensional settings, such as when dealing with a larger number of maturities, where non-linear structures are more likely to emerge. AEs for interest rates may also prove useful in contexts beyond forecasting, such as data reconstruction, scenario generation, risk management, or stress testing applications \citep{flaig2023validation}.

While \citet{nunes2019comparison} documented improvements from incorporating macroeconomic information in the European market, our results suggest that, within our evaluation framework, these benefits are more pronounced in the U.S. case. This does not imply that macroeconomic variables are uninformative for European rates; indeed, two of the three top-performing models in our final selection included macroeconomic factors. Overall, this highlights that the value of macroeconomic information may vary across regions and that model performance should be interpreted in the context of the specific application, as emphasized by \citet{dunis2007economic}.

\begin{figure}
    \centering
    \includegraphics[width=\textwidth]{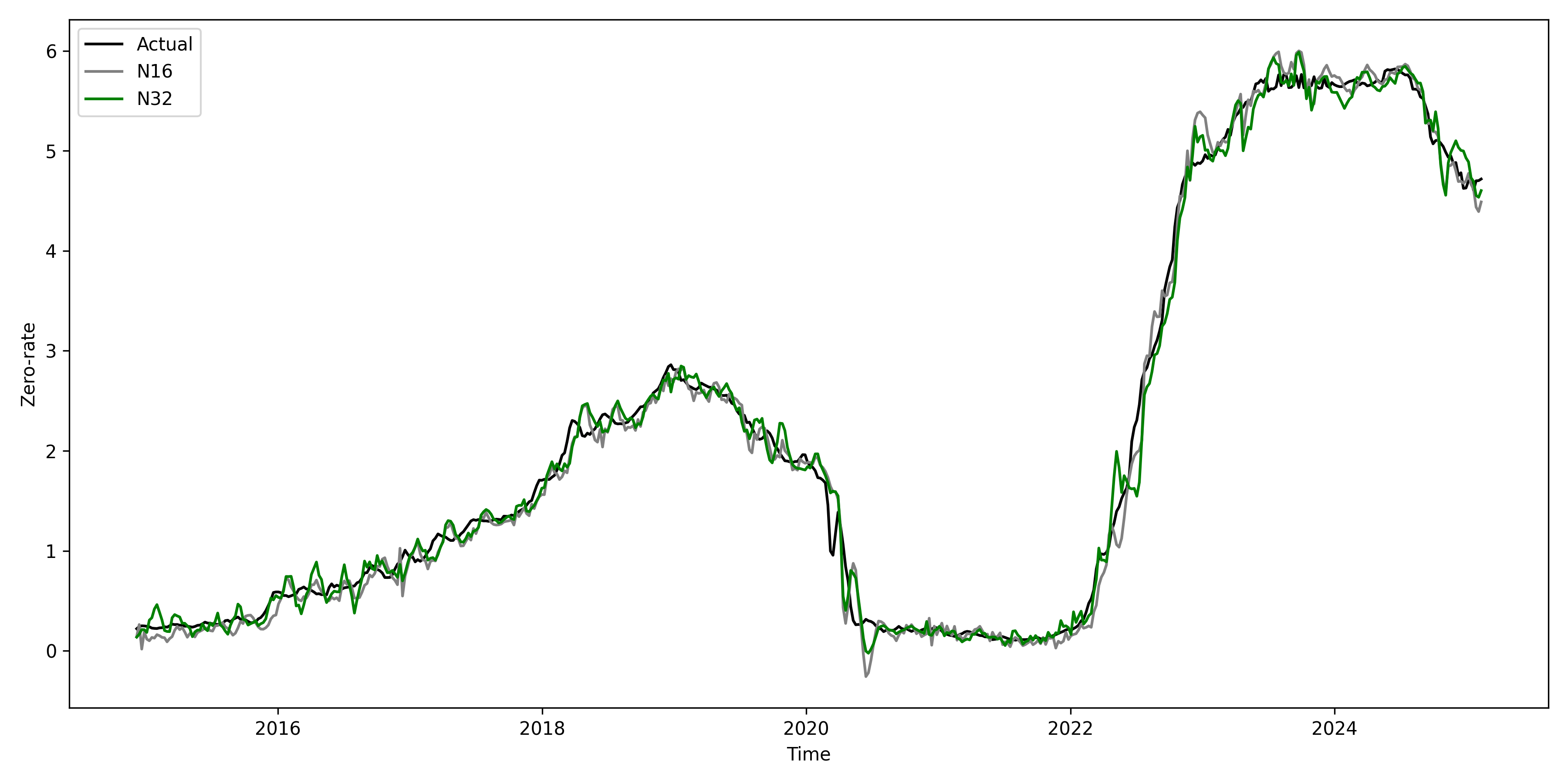}
    \caption{Comparison between the one-month forecasts of Models N16 and N32 and the actual U.S. three-month zero rate.}
    \label{fig:sec3}
\end{figure}

Finally, it is important to consider why no traditional model was selected for the final comparison, despite most of them showing satisfactory forecasting accuracy in terms of RMSE and MAE (Tables 2.1 and 2.2 in the Supplementary Document). For example, Figure \ref{fig:sec2} illustrates the U.S. forecasts generated by models N2 and N5. Both models employ PCA for zero-rate curve decomposition, while N2 uses an AR model and N5 a VAR model to predict the future factors, respectively. Although their average forecast errors are reasonable, the predicted zero-rate curves largely resemble a rightward shift of the current curve, suggesting a persistent lag in capturing market dynamics. This behavior fails to provide actionable investment signals, revealing a key limitation of traditional models that is not reflected in standard error metrics alone. In contrast, the NN counterparts, also using PCA without macroeconomic inputs, shown in Figure \ref{fig:sec3}, produce richer forecasts. Notably, during the rates increases of 2022, these models accurately anticipated the sustained upward trend.

\begin{figure}
    \centering
    \includegraphics[width=\textwidth]{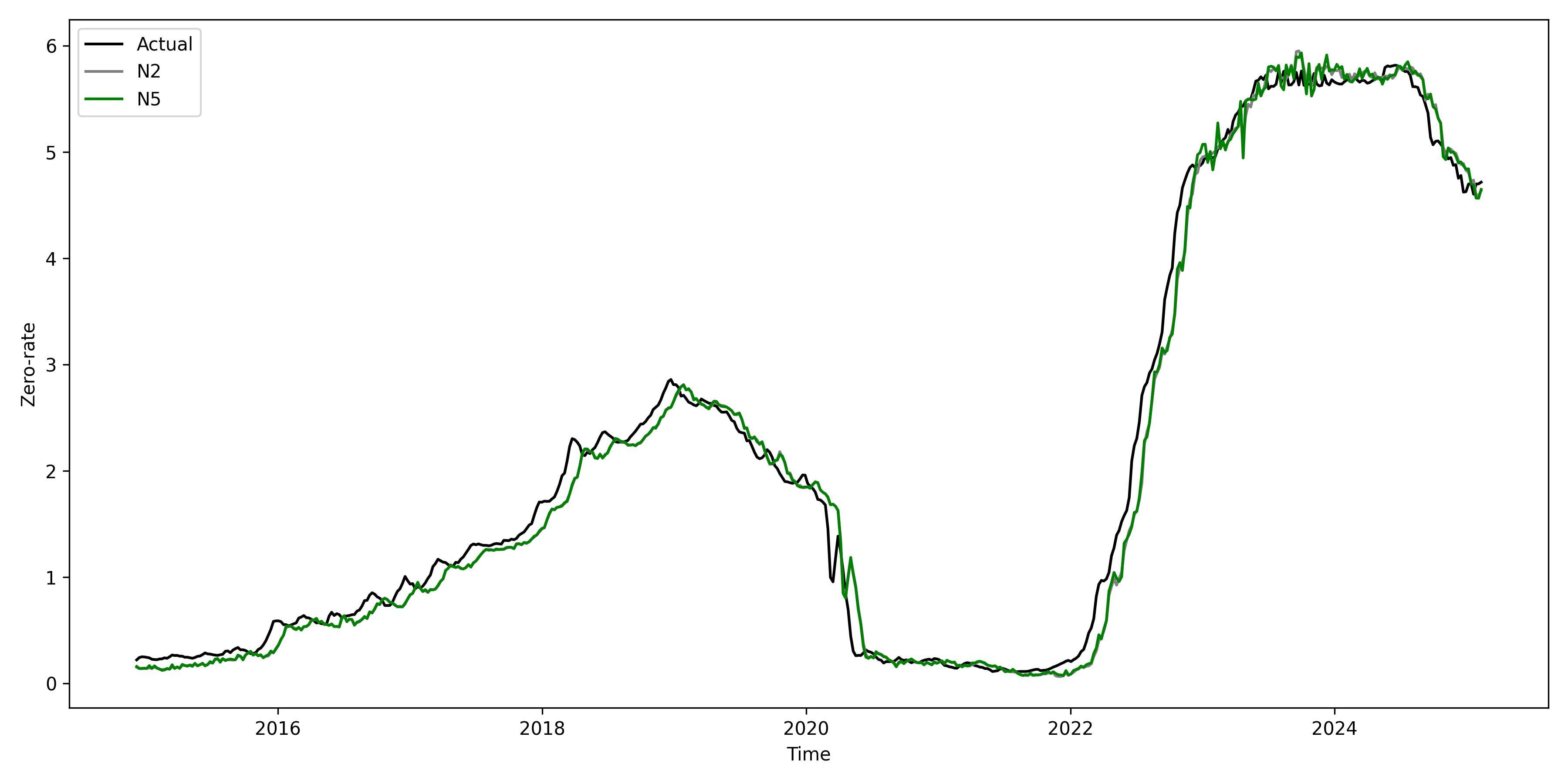}
    \caption{Comparison between the one-month forecasts of Models N2, N5, and the actual U.S. three-month zero rate. The curves from N2 and N5 are indeed very similar, which is to be expected, as the models use, respectively, PCA for zero-rate compression with AR and VAR for factor forecasting. Since PCA produces orthogonal factor loadings, the extracted components are already statistically uncorrelated, meaning that the subsequent AR and VAR dynamics operate on largely the same information. Some subtle differences can be noticed, however, especially on the right part of the graph.}
    \label{fig:sec2}
\end{figure}

Overall, NN models are better equipped to capture nonlinear patterns and regime shifts in the data, allowing them to respond more effectively to changing market conditions. Their flexibility enables them to generate more dynamic and informative zero-rate curve forecasts, which can translate into stronger investment signals, ultimately making them more valuable in practical portfolio applications.

\section{Conclusion}

This paper compares traditional and Machine Learning (ML) approaches for forecasting the term structure of U.S. and European zero rates.

Overall, this study makes several important contributions. First, it contributes methodologically by addressing the limited historical coverage of the European zero-rate dataset, reconstructing a pre-2004 extension through the integration of German government bond data with the ECB’s AAA-rated zero-rate curve. The suggested approach delivers an excellent in-sample fit ($R^2 > 0.99$ across all maturities) and provides a longer time series suitable for ML applications. Second, the study finds that well-tuned Neural Networks (NNs) achieve robust, accurate zero-rate curve forecasts, significantly outperforming traditional models. Although traditional methods have reasonable accuracy, they fail to capture market dynamics, making their forecasts less actionable. Notably, different NN architectures (factor-based vs. direct prediction) and dimensionality reduction techniques perform comparably well. Third, while Autoencoder (AE)-based models for dimensionality reduction of rates were not among the top performers, AEs proved useful for compressing macroeconomic information, especially in the U.S. market. Finally, the study proposes a unified evaluation framework that combines statistical and investment-based metrics, including a bond-trading strategy to evaluate the economic significance of the forecasts.

Some possible extensions remain. The analysis focuses on seven maturities from U.S. and European government bonds, with European data restricted to AAA-rated issuers. Expanding the maturity range or incorporating more macroeconomic indicators could yield further insights. Additionally, while Bayesian Optimization and Hyperband (BOHB) hyperparameter tuning produced solid results, further iterations or a different search space might enhance performance. The retraining schedule was fixed at two-year intervals. Future research could examine dynamic retraining frequencies based on the current level of market volatility. Other potential avenues include adding further lags or employing ensemble methods. Performance results are also tied to a specific trading strategy and may differ under alternative setups.

These findings offer practical implications for researchers and practitioners. The results underscore the benefits of integrating nonlinear learning methods with domain knowledge to produce more robust and flexible term structure forecasts. For practitioners, particularly fixed-income investors, ML-based term-structure models can generate adaptive and actionable forecasts that support better duration and allocation decisions under changing market conditions.

\section*{Funding}

The authors have nothing to report.

\section*{Conflicts of Interest}

The authors declare no conflicts of interest.

\section*{Data Availability Statement}

The data that support the findings of this study are openly available in the Federal Reserve Bank (FRED) at \texttt{https://fred.stlouisfed.org/}, the European Central Bank (ECB) at \texttt{https://data.ecb.europa.eu/data/datasets} and from LSEG Datastream. Restrictions apply to the availability of the last data source, which was used under license for this study.

\newpage
\appendix
\section[\appendixname~\thesection]{Appendix}
\setcounter{table}{0}
\renewcommand{\thetable}{A\arabic{table}}

\begin{longtable}{llll}
\caption{Overview of zero-rate curve forecasting models.}
\label{tab:summ}\\
\toprule
\textbf{Model} & \textbf{Input Zero Rate} & \textbf{Input Macro} & \textbf{Prediction Method} \\
\midrule
\endfirsthead

\toprule
\textbf{Model} & \textbf{Input Zero Rate} & \textbf{Input Macro} & \textbf{Prediction Method} \\
\midrule
\endhead

\bottomrule
\endfoot

\textbf{1} & DNS & - & AR for factors forecasting \\
\textbf{2} & PCA & - & AR for factors forecasting \\
\textbf{3} & AE & - & AR for factors forecasting \\
\textbf{4} & DNS & - & VAR for factors forecasting \\
\textbf{5} & PCA & - & VAR for factors forecasting \\
\textbf{6} & AE & - & VAR for factors forecasting \\
\textbf{7} & AFNS & - & KF for factors forecasting \\
\textbf{8} & DNS & - & NN for factors forecasting \\
\textbf{9} & DNS & Raw & NN for factors forecasting \\
\textbf{10} & DNS & PCA & NN for factors forecasting \\
\textbf{11} & DNS & AE & NN for factors forecasting \\
\textbf{12} & AFNS & - & NN for factors forecasting \\
\textbf{13} & AFNS & Raw & NN for factors forecasting \\
\textbf{14} & AFNS & PCA & NN for factors forecasting \\
\textbf{15} & AFNS & AE & NN for factors forecasting \\
\textbf{16} & PCA & - & NN for factors forecasting \\
\textbf{17} & PCA & Raw & NN for factors forecasting \\
\textbf{18} & PCA & PCA & NN for factors forecasting \\
\textbf{19} & PCA & AE & NN for factors forecasting \\
\textbf{20} & AE & - & NN for factors forecasting \\
\textbf{21} & AE & Raw & NN for factors forecasting \\
\textbf{22} & AE & PCA & NN for factors forecasting \\
\textbf{23} & AE & AE & NN for factors forecasting \\
\textbf{24} & DNS & - & NN for direct prediction \\
\textbf{25} & DNS & Raw & NN for direct prediction \\
\textbf{26} & DNS & PCA & NN for direct prediction \\
\textbf{27} & DNS & AE & NN for direct prediction \\
\textbf{28} & AFNS & - & NN for direct prediction \\
\textbf{29} & AFNS & Raw & NN for direct prediction \\
\textbf{30} & AFNS & PCA & NN for direct prediction \\
\textbf{31} & AFNS & AE & NN for direct prediction \\
\textbf{32} & PCA & - & NN for direct prediction \\
\textbf{33} & PCA & Raw & NN for direct prediction \\
\textbf{34} & PCA & PCA & NN for direct prediction \\
\textbf{35} & PCA & AE & NN for direct prediction \\
\textbf{36} & AE & - & NN for direct prediction \\
\textbf{37} & AE & Raw & NN for direct prediction \\
\textbf{38} & AE & PCA & NN for direct prediction \\
\textbf{39} & AE & AE & NN for direct prediction \\
\textbf{40} & Rate & - & NN for direct prediction \\
\textbf{41} & Rate & Raw & NN for direct prediction \\
\textbf{42} & Rate & PCA & NN for direct prediction \\
\textbf{43} & Rate & AE & NN for direct prediction \\
\end{longtable}

\begin{table}[H]
\caption{Optimized hyperparameters for the AE models used for factor extraction from zero-rate curve and macroeconomic data for both U.S. and European datasets.}
\label{tab:hyp_aes}
\begin{tabularx}{\textwidth}{lXXXXX}
\toprule
\textbf{Model Component} & \textbf{Data} & \textbf{Learning Rate} & \textbf{Activation Function} & \textbf{Batch Size} & \textbf{Epochs} \\
\midrule
AE for Zero-Rate Factors & U.S. & 0.001711 & sigmoid & 30 & 2000 \\
AE for Zero-Rate Factors & Europe & 0.000277 & tanh & 27 & 2000 \\
\addlinespace
AE for Macro Factors & U.S. & 0.035207 & ReLU & 30 & 1000 \\
AE for Macro Factors & Europe & 0.058653 & ReLU & 43 & 1000 \\
\bottomrule
\end{tabularx}
\end{table}

\newpage

\listoffigures

\newpage
\bibliographystyle{plainnat}
\bibliography{bibliography}

\end{document}


\pagenumbering{roman}\setcounter{page}{0}
\pagestyle{headings}
\pagenumbering{arabic}

\section*{Supplementary Document: Data-Driven Duration Management}
\setcounter{section}{0}
\renewcommand{\thesection}{\arabic{section}}

\section{Optimized Hyperparameters for the NNs}
\setcounter{page}{1}

\begin{longtable}{|p{1.2cm}|p{2cm}|p{2cm}|p{1.2cm}|p{1.5cm}|p{1.5cm}|p{1.5cm}|p{1.5cm}|}
\caption{Optimized hyperparameters for the models presented in Table A1 of the paper with U.S. data. Here we test different values for the number of training epochs. In some cases, model performance improved when the number of training epochs was reduced from 2000 to 1000. Further reductions led to additional gains for specific models—most notably, model 30 benefitted from decreasing epochs to 500, though performance declined when reduced further to 250.}\label{tab:best_ref} \\
\hline
\textbf{Model} & \textbf{Learning Rate} & \textbf{Activation Function} & \textbf{Batch Size} & \textbf{Number of Layers} & \textbf{Neurons in Layer 1} & \textbf{Neurons in Layer 2} & \textbf{Epochs} \\ \hline
\endfirsthead

\hline
\textbf{Model} & \textbf{Learning Rate} & \textbf{Activation Function} & \textbf{Batch Size} & \textbf{Number of Layers} & \textbf{Neurons in Layer 1} & \textbf{Neurons in Layer 2} & \textbf{Epochs} \\ \hline
\endhead
\textbf{8} & 0.001530 & tanh & 54 & 1 & 6 & 0 & 2000 \\ \hline
\textbf{9} & 0.005330 & ReLu & 30 & 1 & 4 & 0 & 2000 \\ \hline
\textbf{10} & 0.004675 & ReLu & 26 & 1 & 5 & 0 & 2000 \\ \hline
\textbf{11} & 0.002496 & tanh & 27 & 2 & 3 & 4 & 2000 \\ \hline
\textbf{12} & 0.001760 & tanh & 47 & 2 & 3 & 4 & 2000 \\ \hline
\textbf{13} & 0.001020 & ReLu & 59 & 1 & 8 & 0 & 2000 \\ \hline
\textbf{14} & 0.036207 & ReLu & 98 & 1 & 3 & 0 & 2000 \\ \hline
\textbf{15} & 0.000555 & ReLu & 63 & 1 & 5 & 0 & 1000 \\ \hline
\textbf{16} & 0.004539 & ReLu & 59 & 1 & 5 & 0 & 2000 \\ \hline
\textbf{17} & 0.000876 & ReLu & 62 & 1 & 3 & 0 & 2000 \\ \hline
\textbf{18} & 0.000140 & ReLu & 100 & 1 & 6 & 0 & 2000 \\ \hline
\textbf{19} & 0.008846 & sigmoid & 47 & 1 & 4 & 0 & 2000 \\ \hline
\textbf{20} & 0.000166 & sigmoid & 50 & 2 & 3 & 4 & 2000 \\ \hline
\textbf{21} & 0.001838 & sigmoid & 42 & 2 & 3 & 6 & 2000 \\ \hline
\textbf{22} & 0.006962 & sigmoid & 47 & 1 & 4 & 0 & 1000 \\ \hline
\textbf{23} & 0.000283 & ReLu & 25 & 1 & 4 & 0 & 1000 \\ \hline
\textbf{24} & 0.000215 & tanh & 46 & 1 & 3 & 0 & 2000 \\ \hline
\textbf{25} & 0.000262 & ReLu & 44 & 1 & 3 & 0 & 2000 \\ \hline
\textbf{26} & 0.001404 & tanh & 47 & 1 & 3 & 0 & 2000 \\ \hline
\textbf{27} & 0.003753 & ReLu & 93 & 1 & 3 & 0 & 2000 \\ \hline
\textbf{28} & 0.000256 & tanh & 91 & 1 & 3 & 0 & 2000 \\ \hline
\textbf{29} & 0.000216 & ReLu & 37 & 1 & 5 & 0 & 2000 \\ \hline
\textbf{30} & 0.000512 & ReLu & 54 & 1 & 4 & 0 & 500 \\ \hline
\textbf{31} & 0.000214 & tanh & 35 & 2 & 3 & 10 & 2000 \\ \hline
\textbf{32} & 0.001999 & sigmoid & 35 & 1 & 4 & 0 & 2000 \\ \hline
\textbf{33} & 0.003195 & ReLu & 27 & 1 & 6 & 0 & 1000 \\ \hline
\textbf{34} & 0.007464 & ReLu & 41 & 1 & 3 & 0 & 2000 \\ \hline
\textbf{35} & 0.001374 & ReLu & 71 & 1 & 5 & 0 & 2000 \\ \hline
\textbf{36} & 0.004535 & ReLu & 41 & 1 & 4 & 0 & 2000 \\ \hline
\textbf{37} & 0.009403 & ReLu & 37 & 1 & 5 & 0 & 1000 \\ \hline
\textbf{38} & 0.000287 & ReLu & 33 & 2 & 4 & 6 & 1000 \\ \hline
\textbf{39} & 0.004397 & sigmoid & 52 & 2 & 3 & 4 & 2000 \\ \hline
\textbf{40} & 0.017537 & ReLu & 40 & 1 & 6 & 0 & 2000 \\ \hline
\textbf{41} & 0.032508 & ReLu & 54 & 1 & 4 & 0 & 2000 \\ \hline
\textbf{42} & 0.002808 & ReLu & 87 & 1 & 4 & 0 & 2000 \\ \hline
\textbf{43} & 0.005177 & tanh & 57 & 1 & 3 & 0 & 2000 \\ \hline
\end{longtable}

\begin{longtable}{|p{1.2cm}|p{2cm}|p{2cm}|p{1.2cm}|p{1.5cm}|p{1.5cm}|p{1.5cm}|p{1.5cm}|}

\caption{Optimized hyperparameters for the models presented in Table A1 of the paper with European data. In some cases, model performance improved when the number of training epochs was reduced from 2000 to 1000. Further reductions did not lead to additional gains.}\label{tab:best_ref_eu}\\
\hline
\textbf{Model} & \textbf{Learning Rate} & \textbf{Activation Function} & \textbf{Batch Size} & \textbf{Number of Layers} & \textbf{Neurons in Layer 1} & \textbf{Neurons in Layer 2} & \textbf{Epochs} \\ \hline
\endfirsthead

\hline
\textbf{Model} & \textbf{Learning Rate} & \textbf{Activation Function} & \textbf{Batch Size} & \textbf{Number of Layers} & \textbf{Neurons in Layer 1} & \textbf{Neurons in Layer 2} & \textbf{Epochs} \\ \hline
\endhead
\textbf{8} & 0.001074 & tanh & 44 & 1 & 6 & 0 & 2000 \\ \hline
\textbf{9} & 0.001450 & tanh & 90 & 1 & 4 & 0 & 2000 \\ \hline
\textbf{10} & 0.000234 & tanh & 26 & 2 & 5 & 5 & 2000 \\ \hline
\textbf{11} & 0.000204 & tanh & 38 & 2 & 3 & 9 & 2000 \\ \hline
\textbf{12} & 0.059644 & ReLu & 66 & 2 & 7 & 9 & 2000 \\ \hline
\textbf{13} & 0.010095 & ReLu & 38 & 1 & 6 & 0 & 2000 \\ \hline
\textbf{14} & 0.001049 & tanh & 60 & 1 & 4 & 0 & 2000 \\ \hline
\textbf{15} & 0.000671 & sigmoid & 42 & 2 & 5 & 6 & 2000 \\ \hline
\textbf{16} & 0.001672 & ReLu & 55 & 1 & 5 & 0 & 1000 \\ \hline
\textbf{17} & 0.006547 & sigmoid & 30 & 2 & 4 & 9 & 2000 \\ \hline
\textbf{18} & 0.000604 & tanh & 72 & 1 & 5 & 0 & 2000 \\ \hline
\textbf{19} & 0.008259 & sigmoid & 57 & 2 & 10 & 10 & 2000 \\ \hline
\textbf{20} & 0.005348 & sigmoid & 46 & 2 & 8 & 8 & 2000 \\ \hline
\textbf{21} & 0.000512 & tanh & 35 & 1 & 3 & 0 & 2000 \\ \hline
\textbf{22} & 0.001354 & tanh & 36 & 2 & 3 & 5 & 2000 \\ \hline
\textbf{23} & 0.002556 & tanh & 45 & 1 & 6 & 0 & 2000 \\ \hline
\textbf{24} & 0.001124 & tanh & 45 & 2 & 9 & 9 & 1000 \\ \hline
\textbf{25} & 0.000231 & tanh & 28 & 1 & 3 & 0 & 2000 \\ \hline
\textbf{26} & 0.001057 & tanh & 29 & 2 & 3 & 3 & 2000 \\ \hline
\textbf{27} & 0.000306 & ReLu & 46 & 1 & 7 & 0 & 2000 \\ \hline
\textbf{28} & 0.000268 & sigmoid & 63 & 2 & 8 & 10 & 2000 \\ \hline
\textbf{29} & 0.000706 & tanh & 39 & 2 & 3 & 4 & 2000 \\ \hline
\textbf{30} & 0.085164 & tanh & 86 & 2 & 4 & 8 & 2000 \\ \hline
\textbf{31} & 0.009967 & ReLu & 26 & 2 & 3 & 5 & 2000 \\ \hline
\textbf{32} & 0.000921 & sigmoid & 63 & 2 & 3 & 10 & 2000 \\ \hline
\textbf{33} & 0.030772 & sigmoid & 49 & 2 & 3 & 5 & 2000 \\ \hline
\textbf{34} & 0.002494 & tanh & 96 & 2 & 4 & 9 & 2000 \\ \hline
\textbf{35} & 0.002274 & tanh & 37 & 2 & 6 & 4 & 2000 \\ \hline
\textbf{36} & 0.001750 & ReLu & 55 & 1 & 5 & 0 & 2000 \\ \hline
\textbf{37} & 0.000915 & tanh & 49 & 2 & 3 & 5 & 1000 \\ \hline
\textbf{38} & 0.001505 & ReLu & 59 & 2 & 3 & 10 & 2000 \\ \hline
\textbf{39} & 0.001638 & ReLu & 30 & 1 & 6 & 0 & 2000 \\ \hline
\textbf{40} & 0.001884 & sigmoid & 32 & 2 & 4 & 7 & 2000 \\ \hline
\textbf{41} & 0.000491 & tanh & 34 & 2 & 3 & 6 & 2000 \\ \hline
\textbf{42} & 0.012733 & tanh & 67 & 1 & 6 & 0 & 2000 \\ \hline
\textbf{43} & 0.002390 & tanh & 65 & 2 & 8 & 7 & 2000 \\ \hline
\end{longtable}

\section{Tables with the overall results from Section 4}

\begin{longtable}{|p{1.5cm}|p{1.5cm}|p{1.5cm}|p{1.5cm}|p{1.5cm}|p{1.cm}|p{1.5cm}|}
\caption{Model selection for in-depth analysis in the U.S. market. We remove models that fall within the bottom 54th percentile based on average RMSE, average MAE, average directional accuracy, IR, $\Omega$ and MDD. For each filtering step, the models that remain are highlighted in \textbf{bold}.} 
\label{tab:us_model_selection} \\
\hline
\textbf{Model} & \textbf{RMSE} & \textbf{MAE} & \textbf{Dir} & \textbf{IR} & \textbf{$\Omega$} & \textbf{MDD} \\ 
\hline
\endfirsthead

\hline
\textbf{Model} & \textbf{RMSE} & \textbf{MAE} & \textbf{Dir} & \textbf{IR} & \textbf{$\Omega$} & \textbf{MDD} \\
\endhead
N1 & \textbf{0.2668} & \textbf{0.1828} & 56.47 & -0.22 & 0.92 & -21.02 \\ \hline
N2 & \textbf{0.2627} & \textbf{0.1799} & 56.47 & -0.38 & 0.85 & -15.69 \\ \hline
N3 & \textbf{0.2650} & \textbf{0.1853} & 56.44 & -0.23 & 0.92 & -19.57 \\ \hline
N4 & \textbf{0.2658} & \textbf{0.1827} & 56.47 & -0.22 & 0.92 & -21.06 \\ \hline
N5 & \textbf{0.2634} & \textbf{0.1806} & 56.47 & -0.42 & 0.83 & -16.11 \\ \hline
N6 & \textbf{0.2660} & \textbf{0.1863} & 56.47 & -0.21 & 0.92 & -19.20 \\ \hline
N7 & \textbf{0.2671} & \textbf{0.1820} & 56.47 & -0.21 & 0.92 & -19.66 \\ \hline
N8 & \textbf{0.2693} & \textbf{0.1916} & 56.42 & \textbf{0.36} & \textbf{1.16} & \textbf{-11.98} \\ \hline
A9 & 0.3551 & 0.2579 & 56.10 & -0.26 & 0.90 & -19.20 \\ \hline
B9 & 0.3628 & 0.2543 & \textbf{56.55} & \textbf{0.21} & \textbf{1.09} & -15.07 \\ \hline
C9 & 0.3259 & 0.2437 & \textbf{57.55} & \textbf{0.33} & \textbf{1.15} & -15.24 \\ \hline
A10 & 0.3526 & 0.2428 & \textbf{57.09} & -0.20 & 0.92 & -16.76 \\ \hline
B10 & 0.3212 & 0.2354 & 56.07 & 0.15 & 1.06 & -16.33 \\ \hline
A11 & 0.2997 & 0.2231 & \textbf{56.71} & \textbf{0.24} & \textbf{1.10} & -13.53 \\ \hline
B11 & 0.3239 & 0.2336 & \textbf{56.85} & \textbf{0.25} & \textbf{1.11} & -14.08 \\ \hline
N12 & 0.2878 & \textbf{0.2038} & 56.28 & \textbf{0.24} & \textbf{1.10} & -14.56 \\ \hline
A13 & 0.5370 & 0.2945 & \textbf{57.06} & \textbf{0.44} & \textbf{1.20} & \textbf{-12.16} \\ \hline
B13 & 0.4558 & 0.2977 & \textbf{56.77} & -0.01 & 1.00 & -19.95 \\ \hline
C13 & 0.2863 & 0.2054 & \textbf{57.01} & -0.45 & 0.83 & -14.44 \\ \hline
A14 & 0.5567 & 0.3843 & 55.80 & -0.01 & 0.99 & -19.88 \\ \hline
B14 & 0.5317 & 0.3805 & 55.69 & -0.21 & 0.92 & -19.27 \\ \hline
A15 & 0.3086 & 0.2128 & \textbf{56.53} & -0.45 & 0.83 & \textbf{-12.19} \\ \hline
B15 & 0.3234 & 0.2194 & \textbf{57.04} & -0.08 & 0.96 & \textbf{-9.71} \\ \hline
N16 & \textbf{0.2733} & \textbf{0.1942} & \textbf{56.69} & 0.07 & 1.03 & \textbf{-11.87} \\ \hline
A17 & \textbf{0.2526} & \textbf{0.1797} & 56.36 & 0.17 & 1.07 & \textbf{-11.42} \\ \hline
B17 & \textbf{0.2513} & \textbf{0.1778} & 56.36 & \textbf{0.41} & \textbf{1.18} & \textbf{-8.62} \\ \hline
C17 & \textbf{0.2811} & \textbf{0.1939} & 56.44 & 0.11 & 1.05 & \textbf{-8.23} \\ \hline
A18 & \textbf{0.2758} & \textbf{0.1995} & \textbf{56.61} & -0.13 & 0.95 & \textbf{-10.38} \\ \hline
B18 & \textbf{0.2734} & \textbf{0.1980} & \textbf{56.93} & -0.72 & 0.75 & -20.12 \\ \hline
A19 & 0.2859 & 0.2069 & \textbf{56.53} & \textbf{0.34} & \textbf{1.16} & \textbf{-12.51} \\ \hline
B19 & 0.2902 & 0.2116 & \textbf{56.50} & \textbf{0.28} & \textbf{1.13} & \textbf{-9.78} \\ \hline
N20 & 0.3085 & 0.2250 & 56.42 & 0.00 & 1.00 & \textbf{-9.20} \\ \hline
A21 & 0.3052 & 0.2189 & 56.44 & 0.03 & 1.01 & -16.88 \\ \hline
B21 & 0.3092 & 0.2250 & 56.36 & 0.09 & 1.04 & -16.38 \\ \hline
C21 & 0.3048 & 0.2206 & \textbf{56.61} & \textbf{0.31} & \textbf{1.13} & -13.58 \\ \hline
A22 & 0.3604 & 0.2765 & 56.26 & \textbf{0.50} & \textbf{1.24} & \textbf{-12.20} \\ \hline
B22 & 0.3801 & 0.2859 & 56.18 & \textbf{0.27} & \textbf{1.11} & -15.56 \\ \hline
A23 & 0.3272 & 0.2127 & \textbf{56.74} & 0.12 & 1.05 & \textbf{-8.57} \\ \hline
B23 & 0.5104 & 0.3256 & 56.39 & \textbf{0.62} & \textbf{1.27} & \textbf{-7.01} \\ \hline
N24 & \textbf{0.2549} & \textbf{0.1771} & 56.47 & \textbf{0.20} & \textbf{1.08} & \textbf{-8.14} \\ \hline
A25 & \textbf{0.2716} & \textbf{0.1921} & 56.44 & 0.07 & 1.03 & \textbf{-5.71} \\ \hline
B25 & \textbf{0.2622} & \textbf{0.1866} & 56.47 & -0.24 & 0.91 & \textbf{-12.03} \\ \hline
C25 & \textbf{0.2723} & \textbf{0.1990} & 56.47 & 0.05 & 1.02 & \textbf{-7.42} \\ \hline
A26 & \textbf{0.2760} & \textbf{0.1960} & 56.31 & \textbf{0.37} & \textbf{1.17} & \textbf{-10.15} \\ \hline
B26 & \textbf{0.2774} & \textbf{0.1970} & 56.36 & \textbf{0.22} & \textbf{1.10} & \textbf{-12.01} \\ \hline
A27 & \textbf{0.2852} & 0.2050 & \textbf{56.79} & \textbf{0.29} & \textbf{1.13} & \textbf{-11.63} \\ \hline
\textbf{B27} & \textbf{0.2855} & \textbf{0.2034} & \textbf{56.71} & \textbf{0.22} & \textbf{1.10} & \textbf{-12.36} \\ \hline
\textbf{N28} & \textbf{0.2841} & \textbf{0.2031} & \textbf{56.58} & \textbf{0.39} & \textbf{1.18} & \textbf{-12.10} \\ \hline
A29 & 0.2983 & 0.2091 & \textbf{56.58} & 0.18 & 1.07 & \textbf{-7.18} \\ \hline
B29 & 0.2865 & \textbf{0.1998} & \textbf{56.74} & -0.03 & 0.99 & -14.24 \\ \hline
C29 & \textbf{0.2823} & \textbf{0.1973} & 56.23 & 0.04 & 1.02 & -14.63 \\ \hline
A30 & 0.4416 & 0.3323 & 56.34 & \textbf{0.45} & \textbf{1.19} & \textbf{-10.08} \\ \hline
B30 & 0.4729 & 0.3341 & 56.39 & -0.24 & 0.90 & -19.42 \\ \hline
A31 & \textbf{0.2540} & \textbf{0.1786} & \textbf{56.63} & 0.03 & 1.01 & \textbf{-10.36} \\ \hline
\textbf{B31} & \textbf{0.2527} & \textbf{0.1798} & \textbf{56.50} & \textbf{0.46} & \textbf{1.20} & \textbf{-12.34} \\ \hline
N32 & \textbf{0.2589} & \textbf{0.1842} & 56.26 & \textbf{0.49} & \textbf{1.22} & \textbf{-11.62} \\ \hline
A33 & \textbf{0.2802} & \textbf{0.1990} & \textbf{56.66} & -0.03 & 0.99 & -17.04 \\ \hline
B33 & \textbf{0.2835} & \textbf{0.2026} & \textbf{56.77} & 0.10 & 1.04 & -14.66 \\ \hline
C33 & 0.3000 & 0.2068 & 56.44 & \textbf{0.45} & \textbf{1.21} & \textbf{-10.76} \\ \hline
A34 & 0.2889 & 0.2106 & \textbf{56.69} & \textbf{0.37} & \textbf{1.16} & -13.40 \\ \hline
B34 & \textbf{0.2840} & 0.2101 & \textbf{56.63} & \textbf{0.46} & \textbf{1.21} & -12.65 \\ \hline
A35 & \textbf{0.2584} & \textbf{0.1816} & 56.31 & \textbf{0.35} & \textbf{1.15} & \textbf{-12.01} \\ \hline
B35 & \textbf{0.2738} & \textbf{0.1911} & 56.39 & \textbf{0.65} & \textbf{1.30} & \textbf{-7.63} \\ \hline
N36 & \textbf{0.2656} & \textbf{0.1902} & \textbf{56.53} & 0.13 & 1.05 & -13.63 \\ \hline
A37 & 0.3236 & 0.2273 & \textbf{56.61} & \textbf{0.20} & \textbf{1.09} & -13.14 \\ \hline
B37 & 0.3278 & 0.2325 & \textbf{56.69} & 0.07 & 1.03 & -14.52 \\ \hline
C37 & 0.3155 & 0.2235 & 56.47 & 0.13 & 1.06 & -15.39 \\ \hline
A38 & \textbf{0.2546} & \textbf{0.1790} & 56.47 & \textbf{0.36} & \textbf{1.15} & \textbf{-6.17} \\ \hline
B38 & \textbf{0.2568} & \textbf{0.1805} & 56.47 & \textbf{0.34} & \textbf{1.14} & \textbf{-10.15} \\ \hline
A39 & 0.2979 & 0.2178 & 56.47 & \textbf{0.28} & \textbf{1.12} & -13.90 \\ \hline
B39 & 0.2933 & 0.2161 & \textbf{56.71} & \textbf{0.20} & \textbf{1.08} & -14.50 \\ \hline
N40 & 0.3264 & 0.2363 & \textbf{56.55} & \textbf{0.39} & \textbf{1.17} & \textbf{-11.41} \\ \hline
A41 & 0.4274 & 0.3203 & \textbf{56.63} & -0.28 & 0.89 & -16.27 \\ \hline
B41 & 0.4675 & 0.3369 & 56.36 & 0.01 & 1.00 & -13.57 \\ \hline
C41 & 0.4564 & 0.3260 & 55.99 & 0.04 & 1.02 & -16.11 \\ \hline
A42 & \textbf{0.2811} & \textbf{0.2045} & \textbf{56.50} & \textbf{0.41} & \textbf{1.18} & -14.07 \\ \hline
B42 & 0.3157 & 0.2258 & 56.42 & \textbf{0.69} & \textbf{1.32} & \textbf{-8.20} \\ \hline
A43 & 0.2882 & 0.2105 & \textbf{56.82} & \textbf{0.20} & \textbf{1.08} & -14.53 \\ \hline
B43 & 0.2928 & 0.2135 & \textbf{56.58} & 0.17 & 1.07 & -14.32 \\ \hline
\end{longtable}

\begin{longtable}{|p{1.5cm}|p{1.5cm}|p{1.5cm}|p{1.5cm}|p{1.5cm}|p{1.cm}|p{1.5cm}|}
\caption{Model selection for in-depth analysis in the European market. We remove models that fall within the bottom 64th percentile based on average RMSE, average MAE, average directional accuracy, IR, $\Omega$ and MDD. For each filtering step, the models that remain are highlighted in \textbf{bold}.}
\label{tab:eu_model_selection} \\
\hline
\textbf{Model} & \textbf{RMSE} & \textbf{MAE} & \textbf{Dir} & \textbf{IR} & \textbf{$\Omega$} & \textbf{MDD} \\ 
\hline
\endfirsthead

\hline
\textbf{Model} & \textbf{RMSE} & \textbf{MAE} & \textbf{Dir} & \textbf{IR} & \textbf{$\Omega$} & \textbf{MDD} \\
\endhead
N1 & \textbf{0.2016} & \textbf{0.1324} & 52.50 & -0.11 & 0.96 & -20.17 \\ \hline
N2 & \textbf{0.1945} & \textbf{0.1229} & \textbf{53.28} & 0.22 & 1.09 & \textbf{-10.77} \\ \hline
N3 & \textbf{0.2032} & \textbf{0.1309} & \textbf{53.31} & -0.04 & 0.99 & -20.00 \\ \hline
N4 & \textbf{0.2015} & \textbf{0.1322} & 52.47 & -0.11 & 0.96 & -20.16 \\ \hline
N5 & \textbf{0.1947} & \textbf{0.1230} & \textbf{53.28} & 0.22 & 1.09 & \textbf{-10.84} \\ \hline
N6 & \textbf{0.2036} & \textbf{0.1305} & \textbf{53.23} & -0.04 & 0.99 & -20.02 \\ \hline
N7 & \textbf{0.2018} & \textbf{0.1350} & 52.36 & -0.09 & 0.97 & -12.21 \\ \hline
N8 & \textbf{0.2192} & \textbf{0.1473} & 52.42 & 0.21 & 1.08 & -15.14 \\ \hline
D9 & 0.2883 & 0.1979 & 51.47 & 0.13 & 1.05 & -16.21 \\ \hline
E9 & 0.2895 & 0.2097 & 52.69 & 0.33 & 1.13 & -13.48 \\ \hline
F9 & 0.2569 & 0.1862 & 52.34 & 0.18 & 1.08 & -15.28 \\ \hline
D10 & 0.3536 & 0.2564 & 50.45 & 0.00 & 1.00 & -14.97 \\ \hline
E10 & 0.2745 & 0.1987 & 51.85 & -0.41 & 0.84 & -20.09 \\ \hline
F10 & 0.3488 & 0.2369 & 50.09 & -0.12 & 0.95 & -17.12 \\ \hline
D11 & \textbf{0.1990} & \textbf{0.1395} & 52.39 & \textbf{0.43} & \textbf{1.17} & -13.13 \\ \hline
E11 & \textbf{0.2053} & \textbf{0.1434} & 51.44 & \textbf{0.44} & \textbf{1.18} & -12.96 \\ \hline
F11 & 0.2231 & 0.1511 & 51.74 & \textbf{0.63} & \textbf{1.28} & \textbf{-10.36} \\ \hline
N12 & 0.6740 & 0.4558 & 51.63 & 0.26 & 1.12 & -14.69 \\ \hline
D13 & 0.3079 & 0.2175 & 52.55 & 0.32 & 1.13 & -13.17 \\ \hline
E13 & 0.3705 & 0.2464 & 52.04 & 0.34 & 1.15 & -13.19 \\ \hline
F13 & 0.3040 & 0.2130 & 52.42 & 0.28 & 1.12 & -13.09 \\ \hline
D14 & 0.3022 & 0.2071 & 50.66 & \textbf{0.47} & \textbf{1.22} & \textbf{-10.13} \\ \hline
E14 & 0.2560 & 0.1843 & 52.71 & \textbf{0.53} & \textbf{1.25} & \textbf{-9.56} \\ \hline
F14 & 0.3426 & 0.2176 & 50.36 & -0.04 & 0.98 & -15.82 \\ \hline
D15 & 0.4337 & 0.2824 & 51.01 & \textbf{0.56} & \textbf{1.24} & -12.78 \\ \hline
E15 & 0.2557 & 0.1706 & 50.74 & \textbf{0.73} & \textbf{1.34} & -12.15 \\ \hline
F15 & 0.2710 & 0.1831 & 51.23 & \textbf{0.38} & \textbf{1.17} & \textbf{-11.37} \\ \hline
\textbf{N16} & \textbf{0.1901} & \textbf{0.1288} & \textbf{53.28} & \textbf{0.59} & \textbf{1.29} & \textbf{-10.02} \\ \hline
D17 & 0.2541 & 0.1760 & 51.77 & \textbf{0.45} & \textbf{1.21} & \textbf{-10.10} \\ \hline
E17 & 0.2598 & 0.1724 & 52.96 & 0.29 & 1.15 & \textbf{-11.05} \\ \hline
F17 & 0.2260 & 0.1547 & 52.90 & \textbf{0.40} & \textbf{1.19} & \textbf{-10.23} \\ \hline
D18 & \textbf{0.2162} & \textbf{0.1456} & \textbf{53.04} & 0.35 & 1.16 & \textbf{-10.34} \\ \hline
\textbf{E18} & \textbf{0.1989} & \textbf{0.1336} & \textbf{53.28} & \textbf{0.36} & \textbf{1.17} & \textbf{-9.23} \\ \hline
F18 & 0.2211 & 0.1483 & 52.55 & 0.31 & 1.14 & \textbf{-10.15} \\ \hline
D19 & 0.2739 & 0.1728 & \textbf{53.07} & 0.10 & 1.05 & -11.39 \\ \hline
E19 & 0.2767 & 0.1807 & 52.63 & 0.17 & 1.08 & \textbf{-9.98} \\ \hline
F19 & 0.2777 & 0.1778 & \textbf{53.07} & 0.16 & 1.07 & \textbf{-10.70} \\ \hline
N20 & 0.2388 & 0.1696 & 52.90 & 0.26 & 1.12 & -12.66 \\ \hline
D21 & \textbf{0.1994} & \textbf{0.1307} & 52.77 & 0.35 & \textbf{1.17} & -12.42 \\ \hline
E21 & \textbf{0.1999} & \textbf{0.1321} & \textbf{53.39} & 0.35 & 1.16 & -12.84 \\ \hline
F21 & \textbf{0.1950} & \textbf{0.1289} & \textbf{53.39} & \textbf{0.47} & \textbf{1.22} & -12.26 \\ \hline
D22 & 0.2378 & 0.1567 & \textbf{53.20} & \textbf{0.47} & \textbf{1.23} & \textbf{-11.21} \\ \hline
E22 & 0.2382 & 0.1551 & 52.71 & 0.27 & 1.13 & \textbf{-10.65} \\ \hline
F22 & 0.2363 & 0.1592 & \textbf{53.12} & \textbf{0.38} & \textbf{1.17} & -12.16 \\ \hline
D23 & 0.2821 & 0.1846 & 52.23 & 0.25 & 1.14 & -11.39 \\ \hline
E23 & 0.2468 & 0.1718 & \textbf{53.09} & 0.33 & 1.16 & -11.87 \\ \hline
F23 & 0.2598 & 0.1843 & 52.20 & 0.03 & 1.01 & -12.05 \\ \hline
N24 & 0.2202 & 0.1490 & \textbf{53.09} & \textbf{0.42} & \textbf{1.22} & -11.43 \\ \hline
D25 & \textbf{0.2161} & 0.1510 & 51.80 & 0.10 & 1.04 & \textbf{-8.10} \\ \hline
E25 & \textbf{0.2003} & \textbf{0.1357} & 52.12 & 0.10 & 1.04 & \textbf{-8.26} \\ \hline
F25 & \textbf{0.2119} & \textbf{0.1470} & 51.66 & 0.11 & 1.04 & \textbf{-7.91} \\ \hline
D26 & 0.2386 & 0.1526 & \textbf{53.23} & \textbf{0.42} & \textbf{1.18} & -12.02 \\ \hline
E26 & 0.2376 & 0.1523 & \textbf{53.07} & \textbf{0.46} & \textbf{1.22} & \textbf{-11.38} \\ \hline
F26 & 0.2452 & 0.1558 & \textbf{53.09} & \textbf{0.48} & \textbf{1.21} & -11.49 \\ \hline
D27 & 0.2215 & \textbf{0.1468} & 52.74 & 0.21 & 1.09 & \textbf{-9.58} \\ \hline
\textbf{E27} & \textbf{0.2013} & \textbf{0.1368} & \textbf{53.42} & \textbf{0.43} & \textbf{1.22} & \textbf{-9.27} \\ \hline
F27 & 0.2594 & 0.1649 & \textbf{53.20} & -0.07 & 0.97 & -15.91 \\ \hline
N28 & 0.4587 & 0.3641 & 51.96 & \textbf{0.48} & \textbf{1.29} & \textbf{-8.39} \\ \hline
D29 & 0.2672 & 0.1879 & 52.36 & \textbf{0.42} & \textbf{1.18} & -11.84 \\ \hline
E29 & 0.2768 & 0.1956 & 52.96 & 0.20 & 1.09 & -12.54 \\ \hline
F29 & 0.2530 & 0.1786 & 52.71 & \textbf{0.51} & \textbf{1.23} & \textbf{-10.23} \\ \hline
D30 & 0.2297 & 0.1478 & \textbf{53.44} & 0.14 & 1.06 & -13.43 \\ \hline
E30 & \textbf{0.2195} & \textbf{0.1452} & \textbf{53.50} & 0.17 & 1.07 & -15.10 \\ \hline
F30 & \textbf{0.2105} & \textbf{0.1372} & \textbf{53.39} & 0.07 & 1.03 & -13.64 \\ \hline
D31 & \textbf{0.1991} & \textbf{0.1326} & 52.98 & -0.11 & 0.96 & -14.47 \\ \hline
E31 & \textbf{0.2054} & \textbf{0.1380} & 52.17 & \textbf{0.48} & \textbf{1.23} & -13.00 \\ \hline
F31 & \textbf{0.2070} & \textbf{0.1346} & 52.80 & 0.34 & 1.16 & \textbf{-10.79} \\ \hline
N32 & \textbf{0.2003} & \textbf{0.1326} & 52.93 & \textbf{0.48} & \textbf{1.21} & -12.42 \\ \hline
D33 & 0.2767 & 0.1897 & 52.36 & 0.15 & 1.08 & -13.69 \\ \hline
E33 & 0.2895 & 0.1943 & 52.82 & -0.01 & 1.00 & -13.93 \\ \hline
F33 & 0.2444 & 0.1716 & 52.98 & -0.05 & 0.98 & -15.64 \\ \hline
D34 & 0.2416 & 0.1677 & \textbf{53.25} & 0.32 & 1.15 & -12.62 \\ \hline
E34 & 0.2439 & 0.1670 & 52.88 & 0.15 & 1.06 & -15.20 \\ \hline
F34 & 0.2330 & 0.1600 & 53.01 & 0.14 & 1.06 & -16.25 \\ \hline
D35 & 0.2918 & 0.1836 & \textbf{53.15} & 0.21 & 1.11 & \textbf{-10.50} \\ \hline
E35 & 0.3011 & 0.1929 & 52.88 & \textbf{0.47} & \textbf{1.22} & \textbf{-10.54} \\ \hline
F35 & 0.2557 & 0.1745 & \textbf{53.12} & 0.28 & 1.14 & \textbf{-10.35} \\ \hline
N36 & \textbf{0.1935} & \textbf{0.1311} & \textbf{53.28} & \textbf{0.36} & \textbf{1.17} & -11.81 \\ \hline
D37 & 0.2272 & 0.1536 & 52.71 & \textbf{0.37} & \textbf{1.20} & -11.44 \\ \hline
E37 & 0.2290 & 0.1535 & \textbf{53.09} & \textbf{0.50} & \textbf{1.23} & \textbf{-11.31} \\ \hline
F37 & 0.2224 & 0.1481 & \textbf{53.07} & \textbf{0.47} & \textbf{1.22} & \textbf{-11.19} \\ \hline
D38 & 0.2763 & 0.1710 & 52.39 & \textbf{0.53} & \textbf{1.24} & -13.26 \\ \hline
E38 & \textbf{0.2166} & \textbf{0.1454} & 52.90 & 0.25 & 1.11 & -13.79 \\ \hline
F38 & 0.2627 & 0.1657 & 52.98 & \textbf{0.59} & \textbf{1.27} & -12.67 \\ \hline
D39 & \textbf{0.2013} & \textbf{0.1340} & \textbf{53.17} & \textbf{0.38} & \textbf{1.17} & -12.56 \\ \hline
E39 & \textbf{0.2053} & \textbf{0.1379} & \textbf{53.20} & 0.33 & 1.14 & -13.19 \\ \hline
F39 & \textbf{0.2087} & \textbf{0.1420} & 52.85 & 0.33 & 1.15 & -12.24 \\ \hline
N40 & \textbf{0.2100} & \textbf{0.1438} & 52.88 & 0.28 & 1.13 & -12.96 \\ \hline
D41 & \textbf{0.2077} & \textbf{0.1386} & \textbf{53.12} & 0.27 & 1.13 & -12.68 \\ \hline
E41 & \textbf{0.2098} & \textbf{0.1413} & \textbf{53.12} & 0.15 & 1.06 & \textbf{-9.22} \\ \hline
F41 & \textbf{0.2131} & \textbf{0.1410} & \textbf{53.25} & 0.32 & 1.15 & -14.73 \\ \hline
D42 & 0.2518 & 0.1790 & 52.66 & 0.33 & \textbf{1.17} & \textbf{-10.60} \\ \hline
E42 & 0.2420 & 0.1727 & 52.71 & 0.35 & \textbf{1.17} & -12.30 \\ \hline
F42 & 0.2493 & 0.1779 & 52.52 & 0.33 & 1.14 & -12.32 \\ \hline
D43 & 0.2447 & 0.1649 & \textbf{53.07} & -0.54 & 0.78 & -17.55 \\ \hline
E43 & 0.2381 & 0.1650 & 52.58 & \textbf{0.50} & \textbf{1.24} & \textbf{-10.60} \\ \hline
F43 & 0.2581 & 0.1737 & 52.52 & \textbf{0.42} & \textbf{1.21} & \textbf{-11.03} \\ \hline
\end{longtable}

\renewcommand{\thesection}{\arabic{section}}
\section{Whole Period Evaluations}

\subsection{U.S.}

\begin{center}
\includegraphics[width=\textwidth]{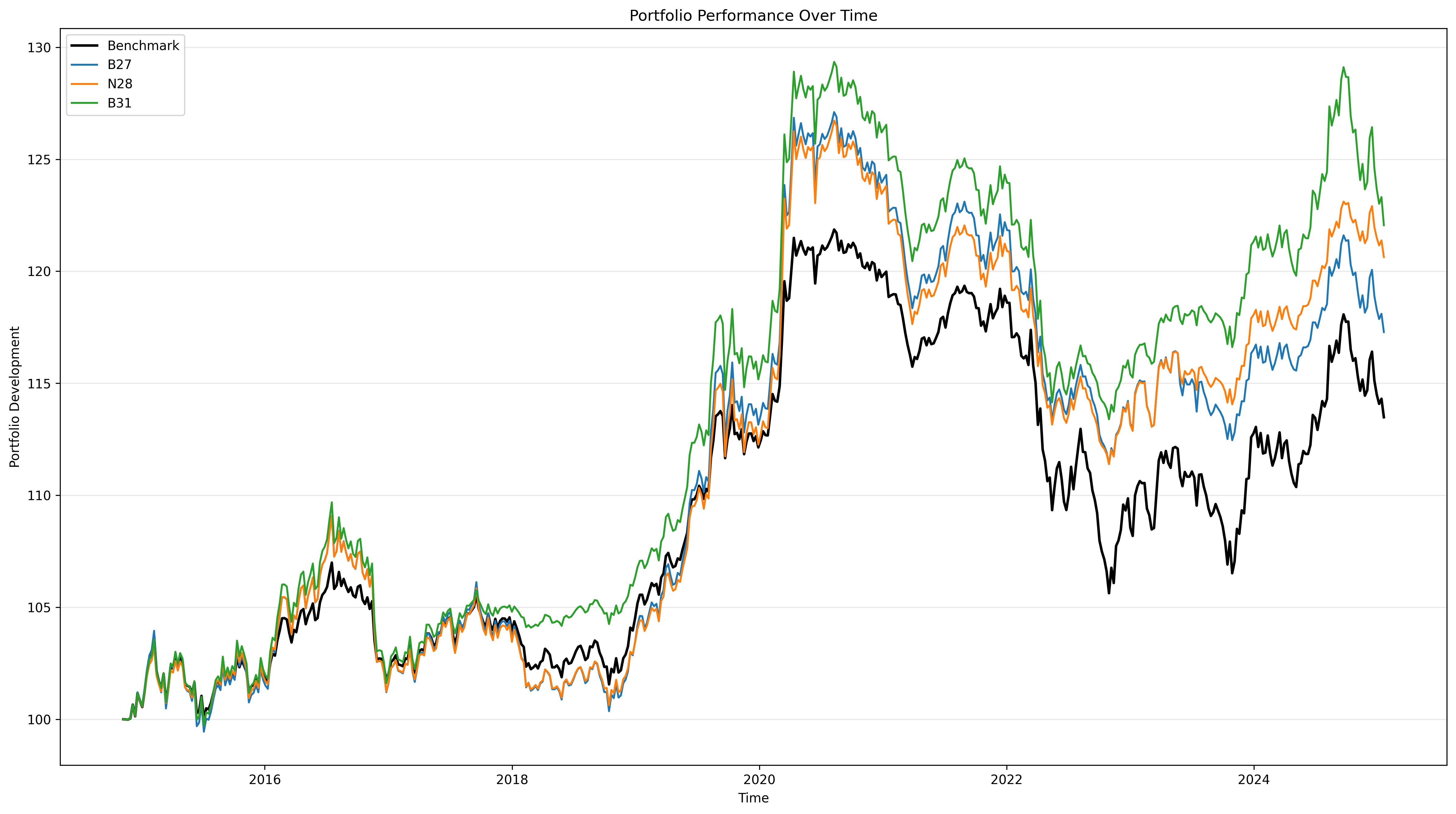}
\captionof{figure}{Performance from the different models and the benchmark between July 2014 and February 2025.} \label{fig:perf_gen}
\end{center}

\begin{center}
\includegraphics[width=\textwidth]{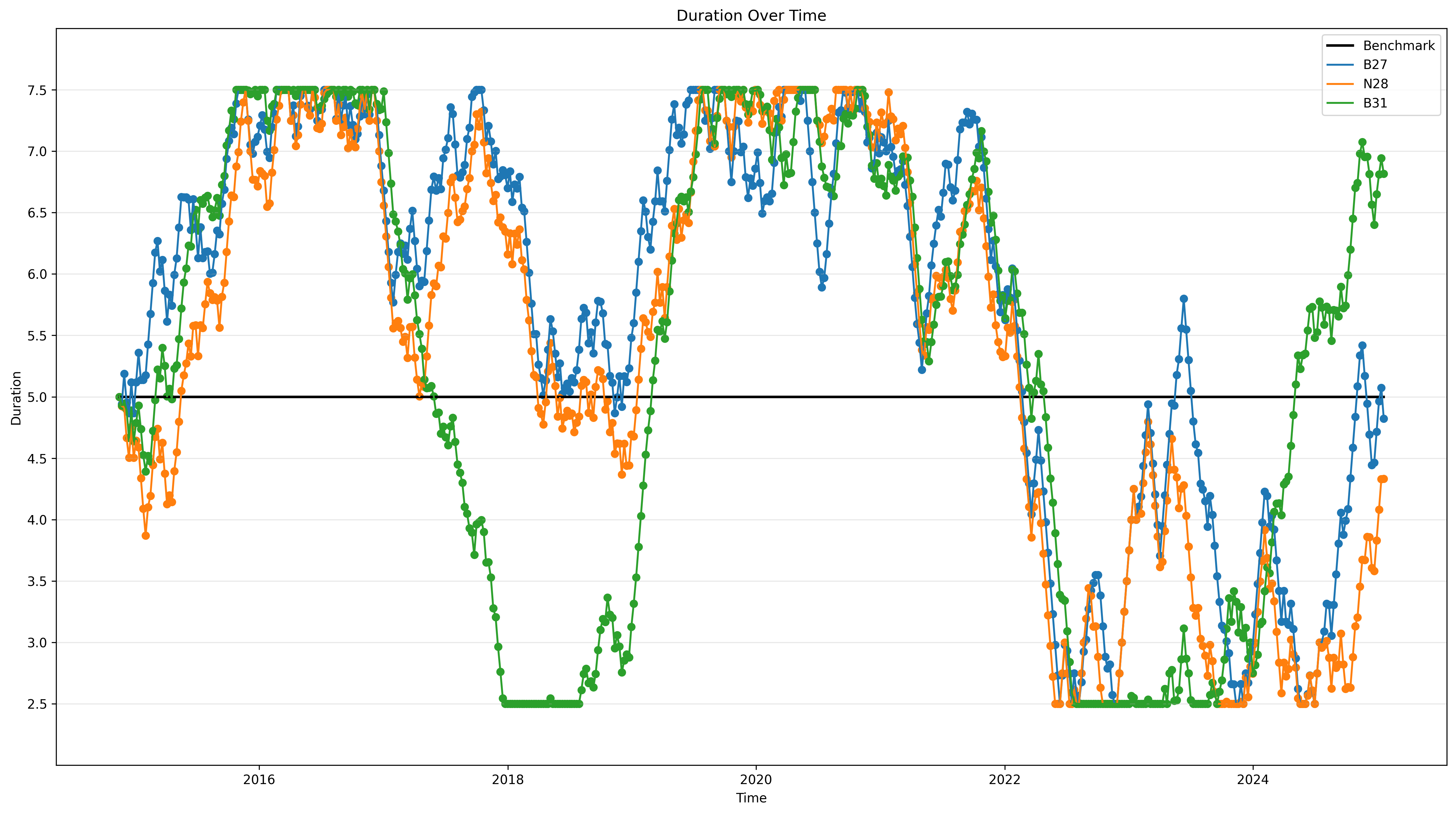}
\captionof{figure}{Duration from the different portfolios between July 2014 and February 2025.} \label{fig:duration_total}
\end{center}

\subsection{Europe}

\begin{center}
\includegraphics[width=\textwidth]{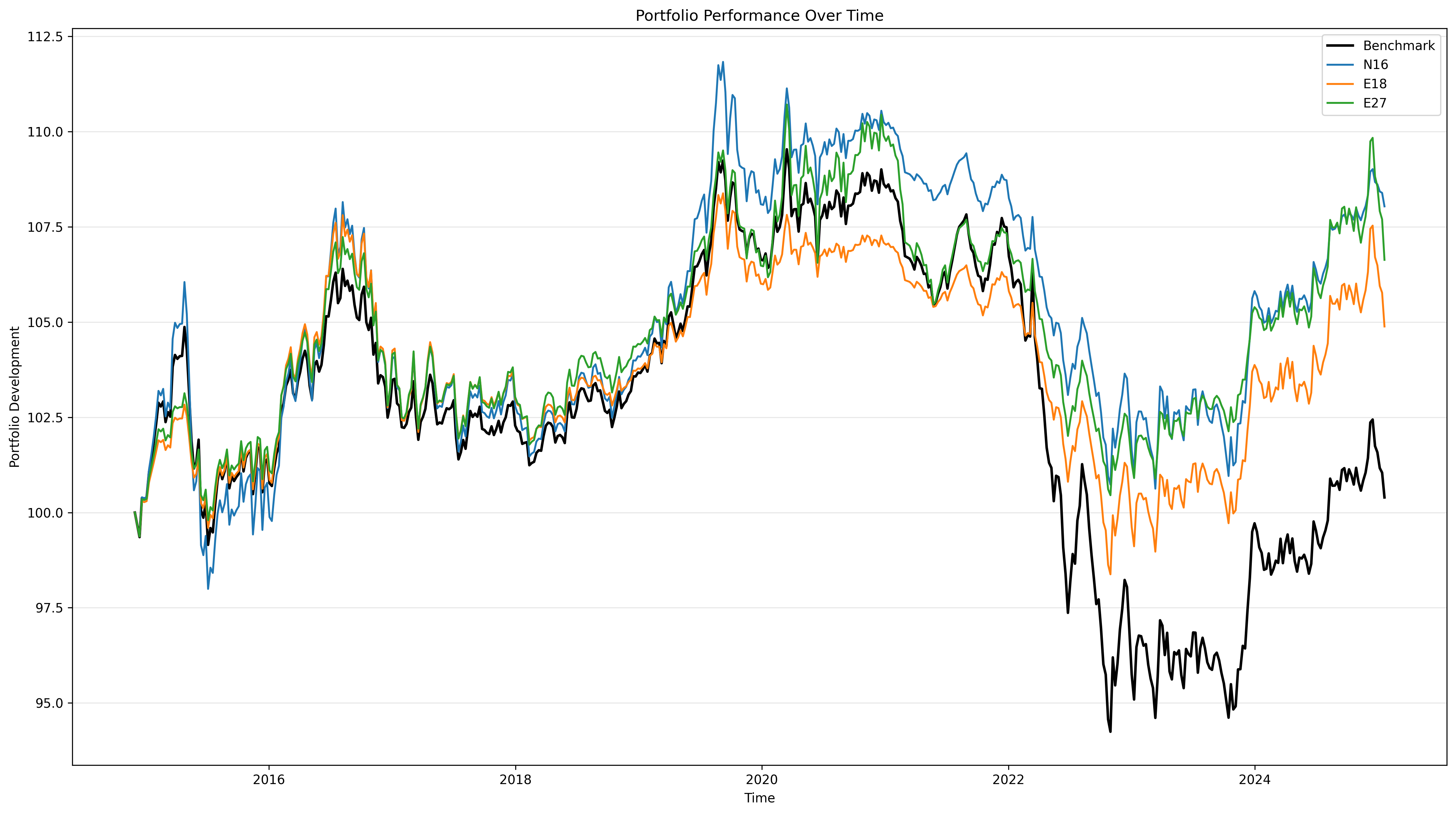}
\captionof{figure}{Performance from the different models and the benchmark between December 2014 and February 2025.} \label{fig:perf_gen_eu}
\end{center}

\begin{center}
\includegraphics[width=\textwidth]{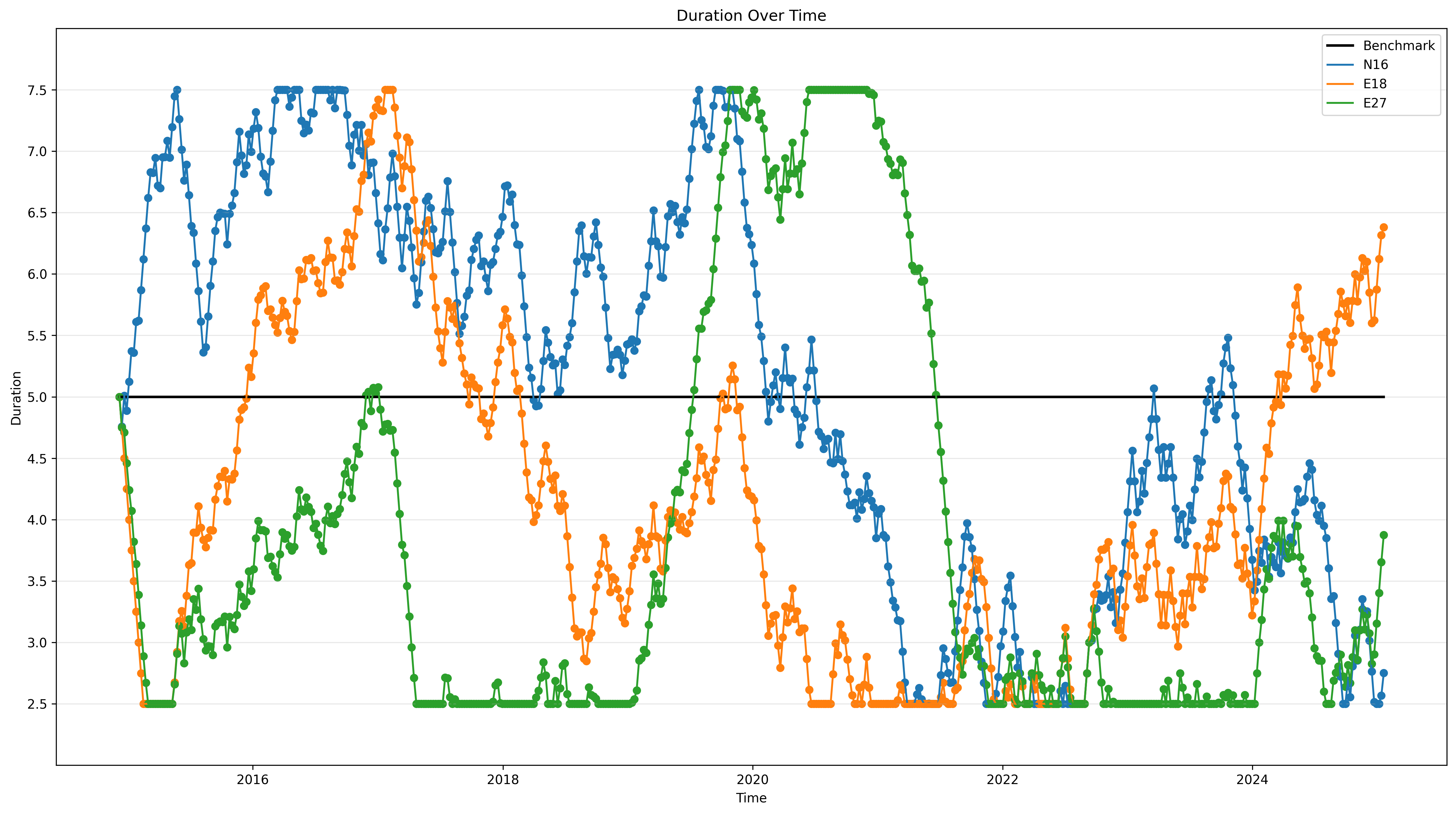}
\captionof{figure}{Duration from the different portfolios in Europe between December 2014 and February 2025.} \label{fig:duration_total_eu}
\end{center}

\section{Periodic Evaluation - U.S.}

\subsection{Period 1 ($\text{P}_1$)}

\begin{table}[ht]
\caption{Results for the different models in the period between July 2014 and February 2018. }\label{tab:result1}
\centering
\begin{tabular}{|p{1.5cm}|p{1.4cm}|p{1.4cm}|p{1.4cm}|p{1.4cm}|p{1.4cm}|p{1.4cm}|p{1.4cm}|}
\hline
\textbf{Model} &\textbf{Ret.} & \textbf{$\Omega$} & \textbf{MDD}  & \textbf{Rank. $\Omega$} & \textbf{Rank. MDD} & \textbf{Sum Rank.} & \textbf{Rank. $\text{P}_1$} \\
\hline
\textbf{B27} & 0.78  & 0.95 & -7.21  & 2 & 2 & 4 & 1 \\ \hline
\textbf{N28} & 0.77 & 0.93 & -7.17  & 3 & 1 & 4 & 1\\ \hline
\textbf{B31} & 1.36 & 1.06 & -7.37  & 1 & 3 & 4 & 1\\ \hline
\textbf{Bench.} & 0.98 & - &  -4.92 & - & - & - & - \\ \hline
\end{tabular}
\end{table}

\begin{center}
\includegraphics[width=\textwidth]{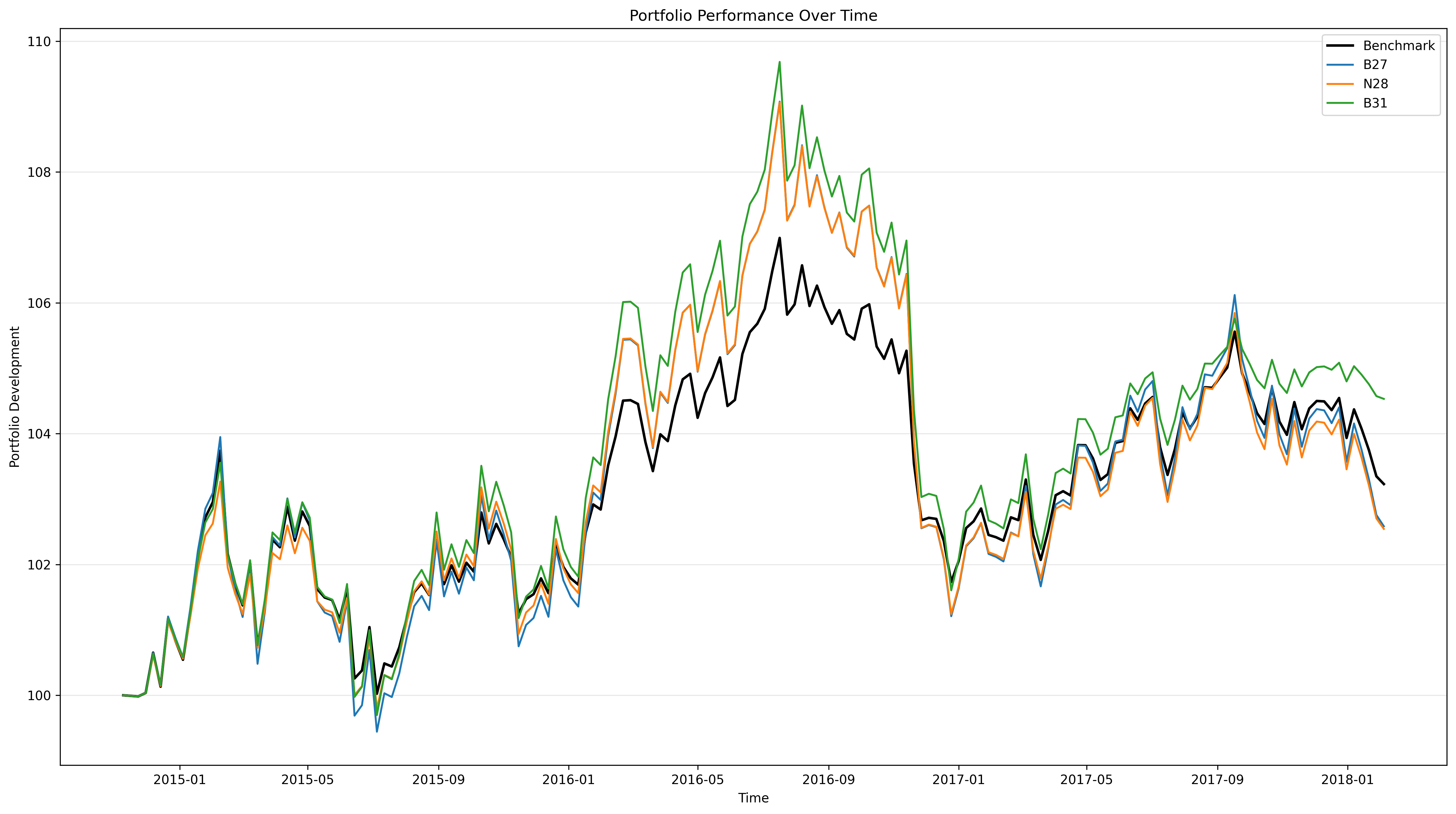}
\captionof{figure}{Performance of the different portfolios between July 2014 and February 2018. } \label{fig:result1}
\end{center}

\begin{center}
\includegraphics[width=\textwidth]{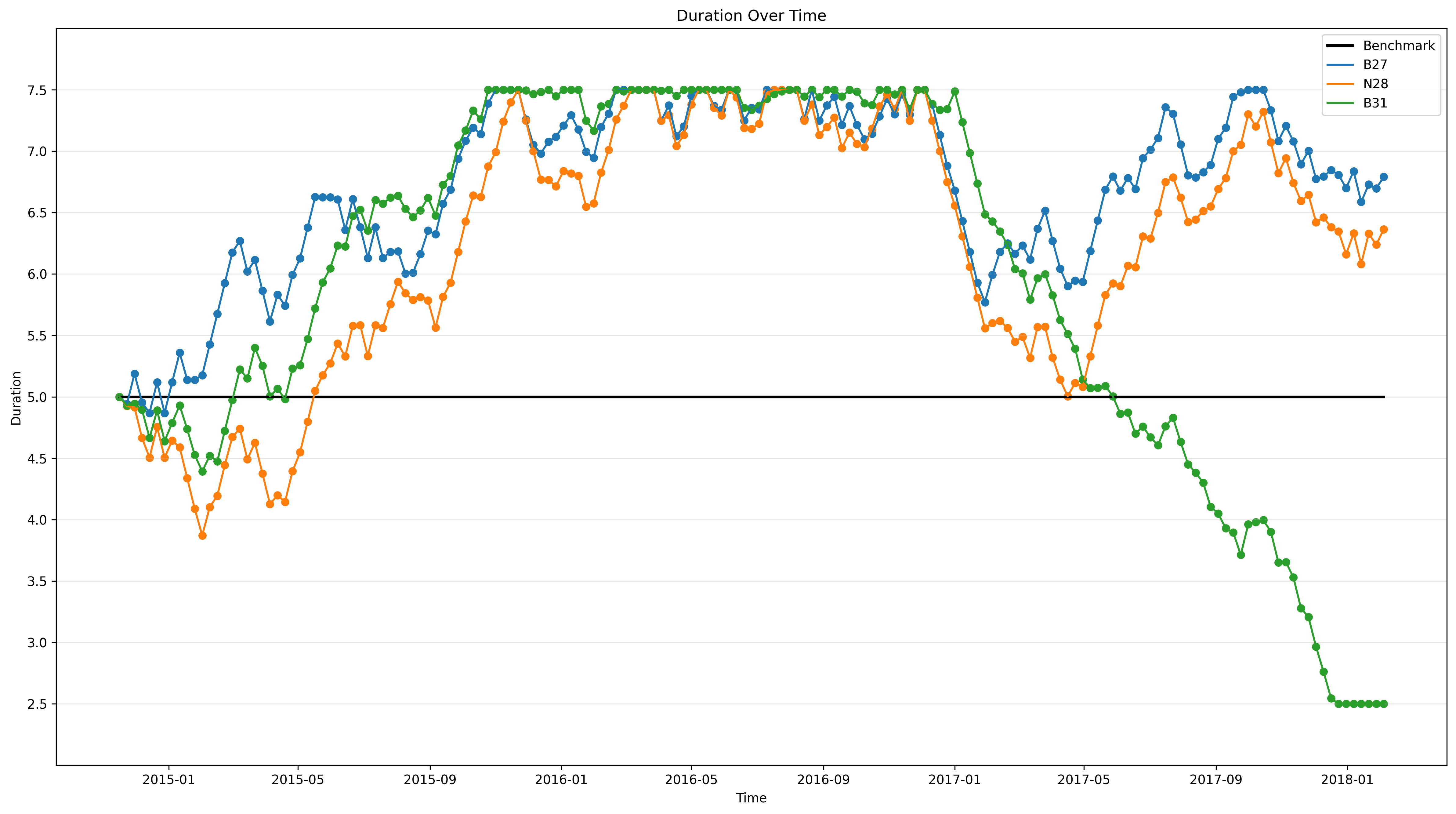}
\captionof{figure}{Duration of the different portfolios between July 2014 and February 2018. } \label{fig:duration1}
\end{center}

\subsection{Period 2 ($\text{P}_2$)}

\begin{table}[ht]
\caption{Results for the different models in the period between February 2018 and July 2021. }\label{tab:result2}
\centering
\begin{tabular}{|p{1.5cm}|p{1.4cm}|p{1.4cm}|p{1.4cm}|p{1.4cm}|p{1.4cm}|p{1.4cm}|p{1.4cm}|}
\hline
\textbf{Model} &\textbf{Ret.} & \textbf{$\Omega$} & \textbf{MDD}  & \textbf{Rank. $\Omega$} & \textbf{Rank. MDD} & \textbf{Sum Rank.} & \textbf{Rank. $\text{P}_2$} \\
\hline

\textbf{B27} & 5.24 & 1.41 & -6.89 & 1 & 2 & 3 & 1 \\ \hline
\textbf{N28} & 5.03 & 1.32 & -7.17 & 2 & 3 & 5 & 3\\ \hline
\textbf{B31} & 4.98 & 1.24 & -6.88 & 3 & 1 & 4 & 2\\ \hline
\textbf{Bench.} & 4.18 & - & -5.02 & - & - & - & - \\ \hline
\end{tabular}
\end{table}

\begin{center}
\includegraphics[width=\textwidth]{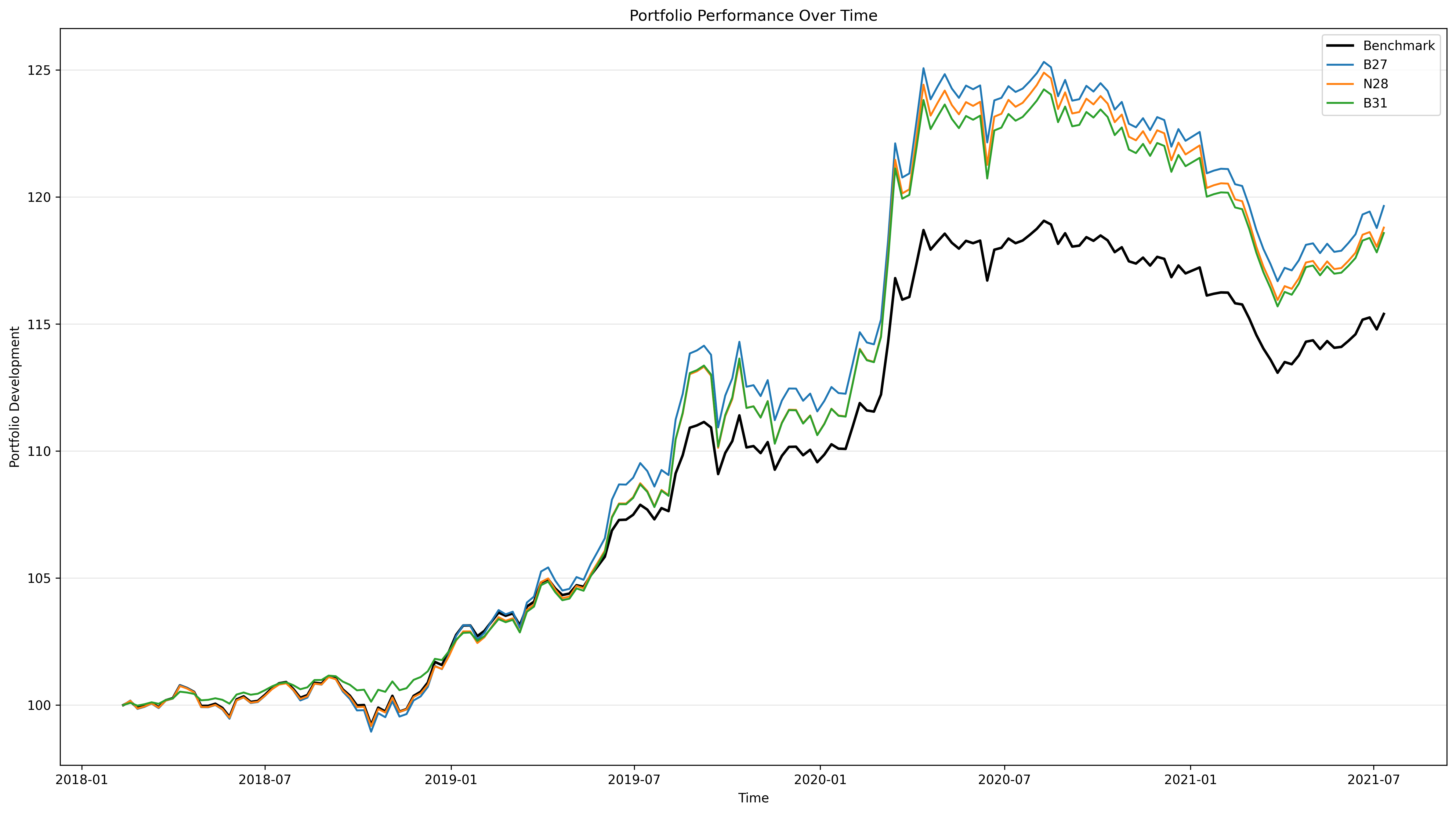}
\captionof{figure}{Performance of the different portfolios between February 2018 and July 2021. } \label{fig:result2}
\end{center}

\begin{center}
\includegraphics[width=\textwidth]{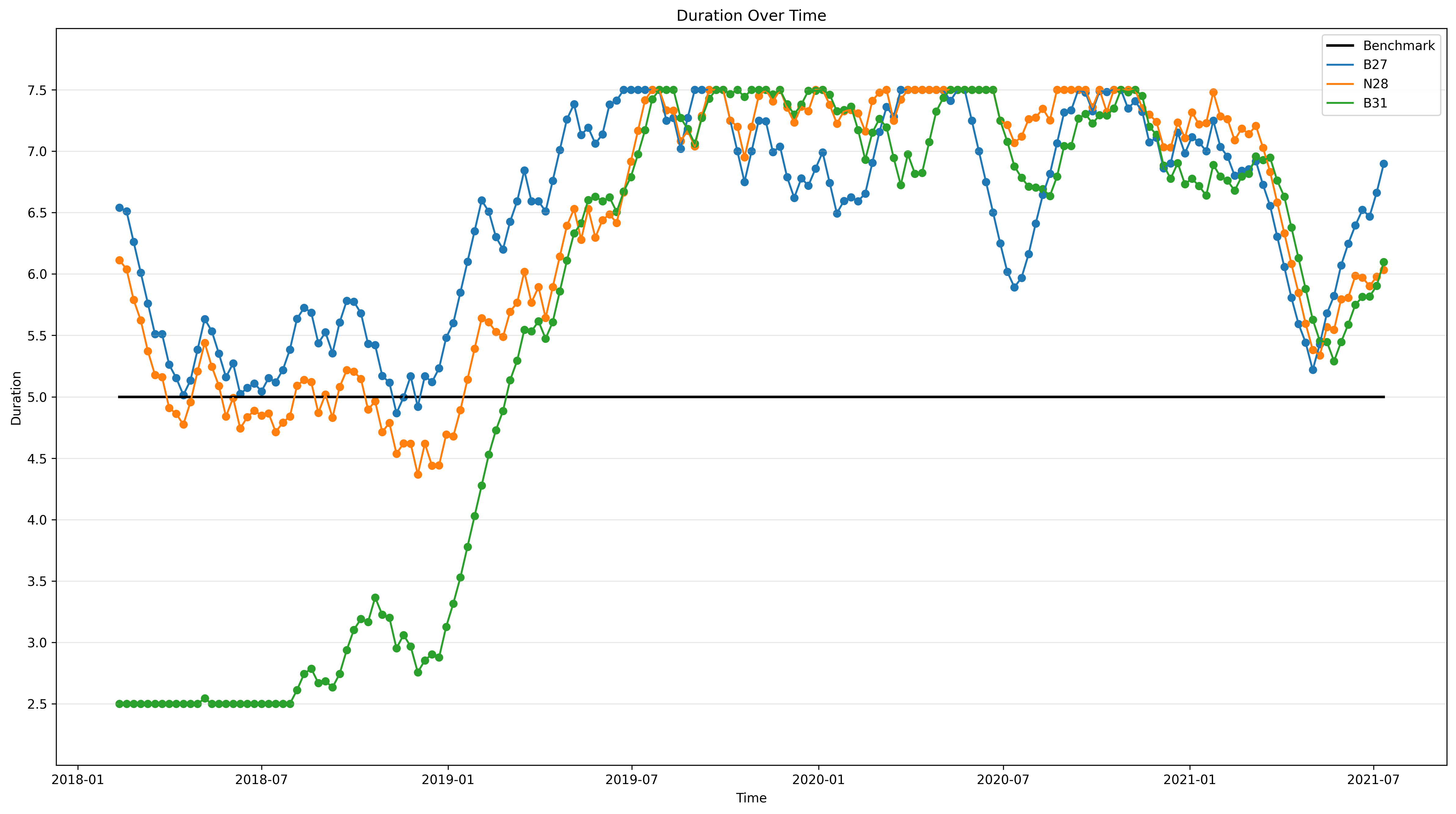}
\captionof{figure}{Duration of the different portfolios between February 2018 and July 2021. } \label{fig:duration2}
\end{center}

\subsection{Period 3 ($\text{P}_3$)}

\begin{table}[ht]
\caption{Results for the different models in the period between July 2021 and November 2022. }\label{tab:result3}
\centering
\begin{tabular}{|p{1.5cm}|p{1.4cm}|p{1.4cm}|p{1.4cm}|p{1.4cm}|p{1.4cm}|p{1.4cm}|p{1.4cm}|}
\hline
\textbf{Model} &\textbf{Ret.} & \textbf{$\Omega$} & \textbf{MDD}  & \textbf{Rank. $\Omega$} & \textbf{Rank. MDD} & \textbf{Sum Rank.} & \textbf{Rank. $\text{P}_3$} \\
\hline
\textbf{B27} & -6.48  & 1.43 & -9.51 & 3 & 3 & 6 & 3 \\ \hline
\textbf{N28} & -5.96 & 1.58 & -8.73 & 1 & 1 & 2 & 1\\ \hline
\textbf{B31} & -6.43  & 1.49 & -9.32 & 2 & 2 & 4 & 2\\ \hline
\textbf{Bench.} & -8.02 & - & -11.50 & - & - & - & - \\ \hline
\end{tabular}
\end{table}

\begin{center}
\includegraphics[width=\textwidth]
{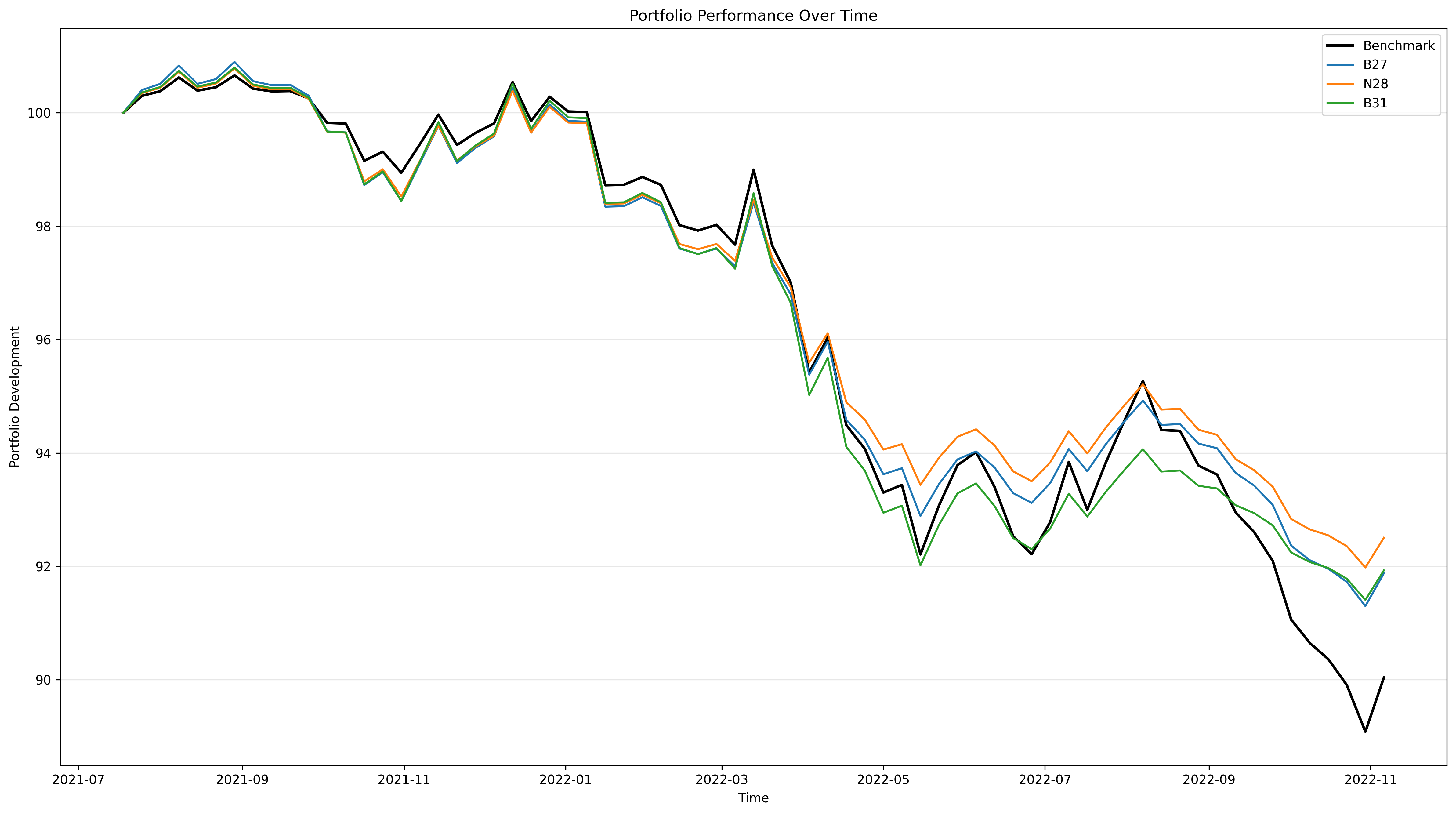}
\captionof{figure}{Performance of the different portfolios between July 2021 and November 2022. } \label{fig:result3}
\end{center}

\begin{center}
\includegraphics[width=\textwidth]{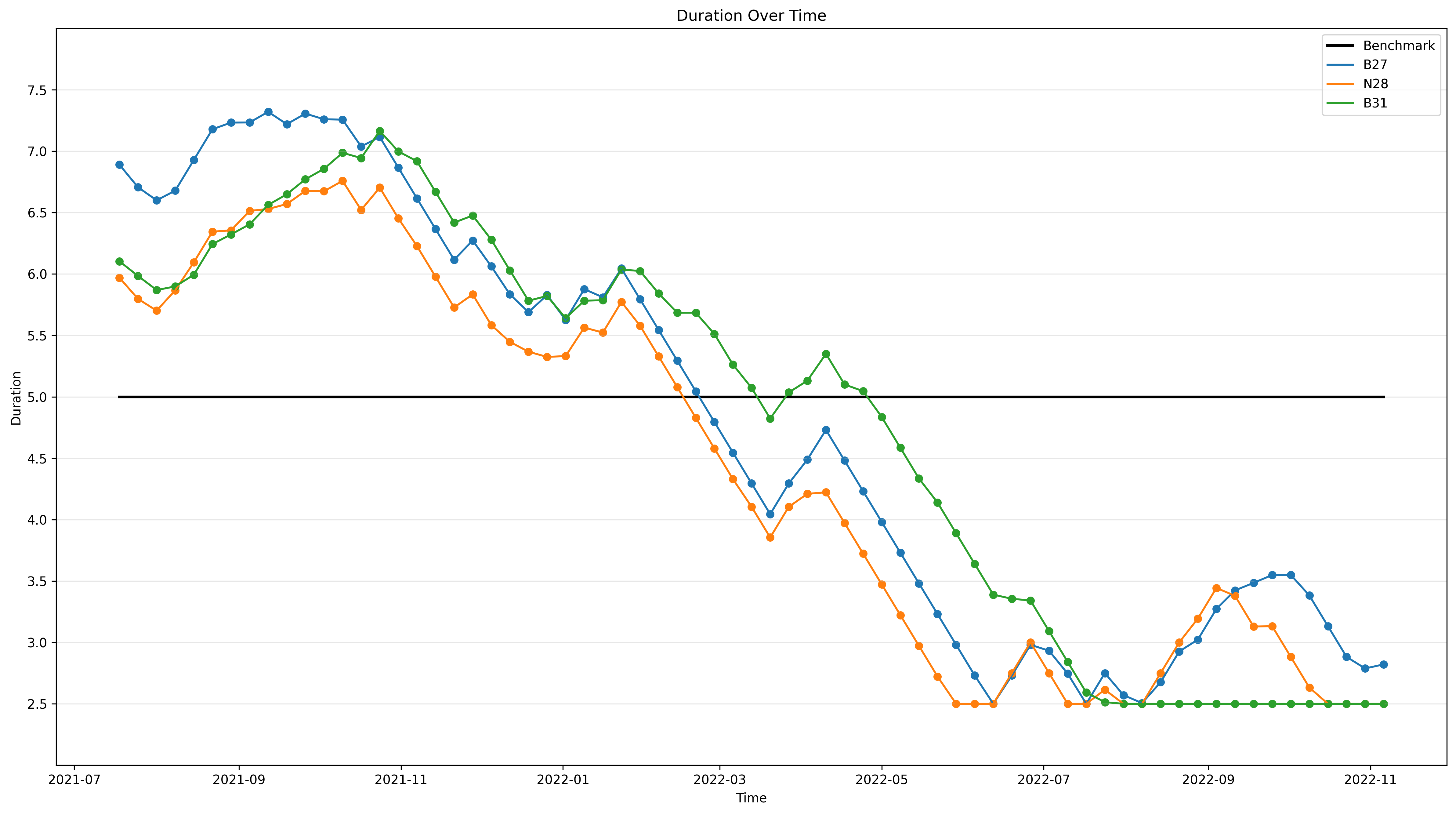}
\captionof{figure}{Duration of the different portfolios between July 2021 and November 2022. } \label{fig:duration3}
\end{center}

\subsection{Period 4 ($\text{P}_4$)}

\begin{table}[ht]
\caption{Results for the different models in the period between November 2022 and February 2025. }\label{tab:result4}
\centering
\begin{tabular}{|p{1.5cm}|p{1.4cm}|p{1.4cm}|p{1.4cm}|p{1.4cm}|p{1.4cm}|p{1.4cm}|p{1.4cm}|}
\hline
\textbf{Model} &\textbf{Ret.} & \textbf{$\Omega$} & \textbf{MDD}  & \textbf{Rank. $\Omega$} & \textbf{Rank. MDD} & \textbf{Sum Rank.} & \textbf{Rank. $\text{P}_4$} \\
\hline
\textbf{B27} & 2.19 & 0.79 & -3.55 & 3 & 2 & 5 & 2 \\ \hline
\textbf{N28} & 3.50 & 1.09 & -2.02 & 1 & 1 & 2 & 1\\ \hline
\textbf{B31} & 3.22 & 1.03 & -5.46 & 2 & 3 & 5 & 2\\ \hline
\textbf{Bench.} & 3.08 & - & -5.93 & - & - & - & - \\ \hline
\end{tabular}
\end{table}
\begin{center}
\includegraphics[width=\textwidth]{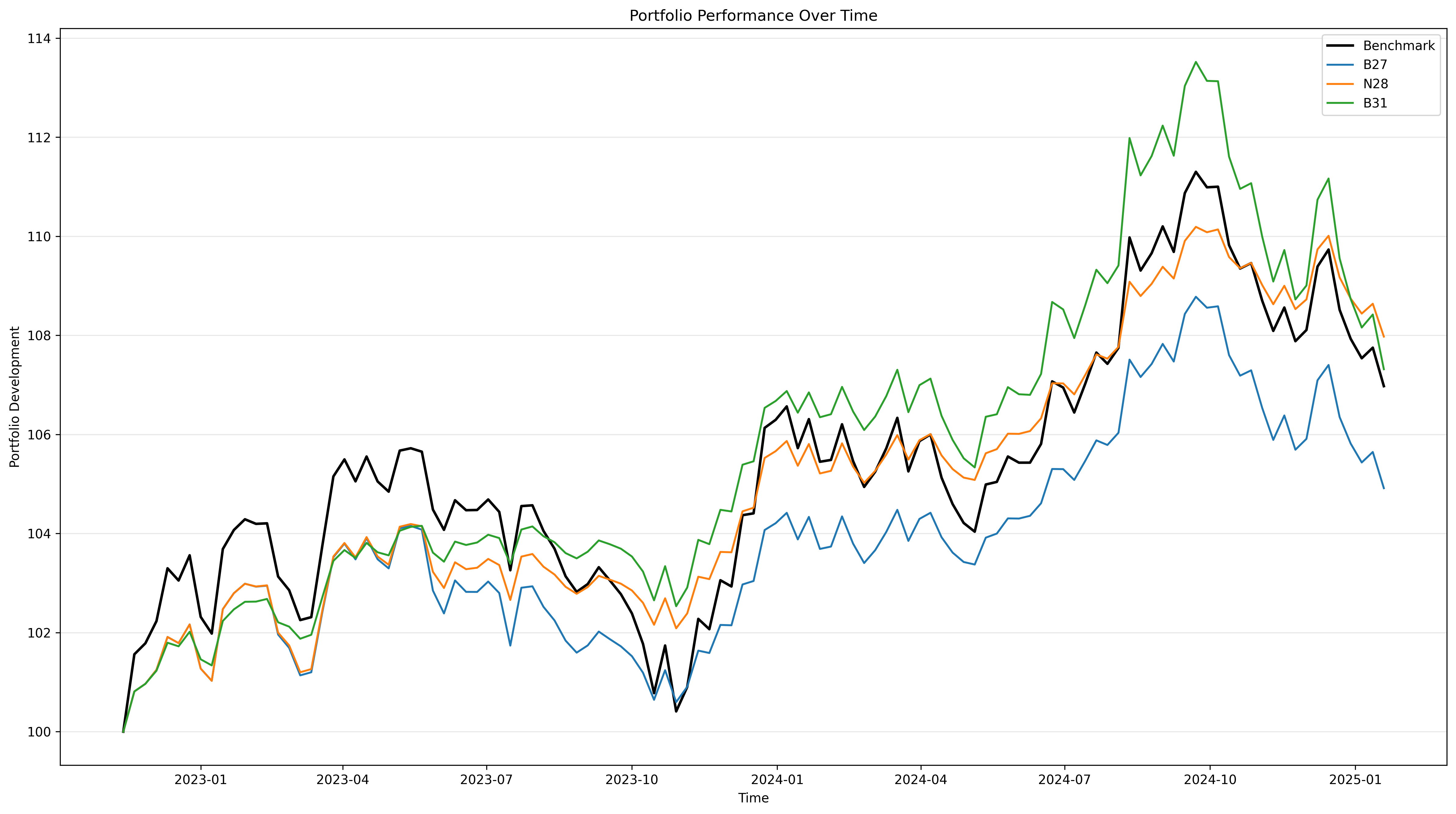}
\captionof{figure}{Performance of the different portfolios between November 2022 and February 2025. } \label{fig:result4}
\end{center}

\begin{center}
\includegraphics[width=\textwidth]{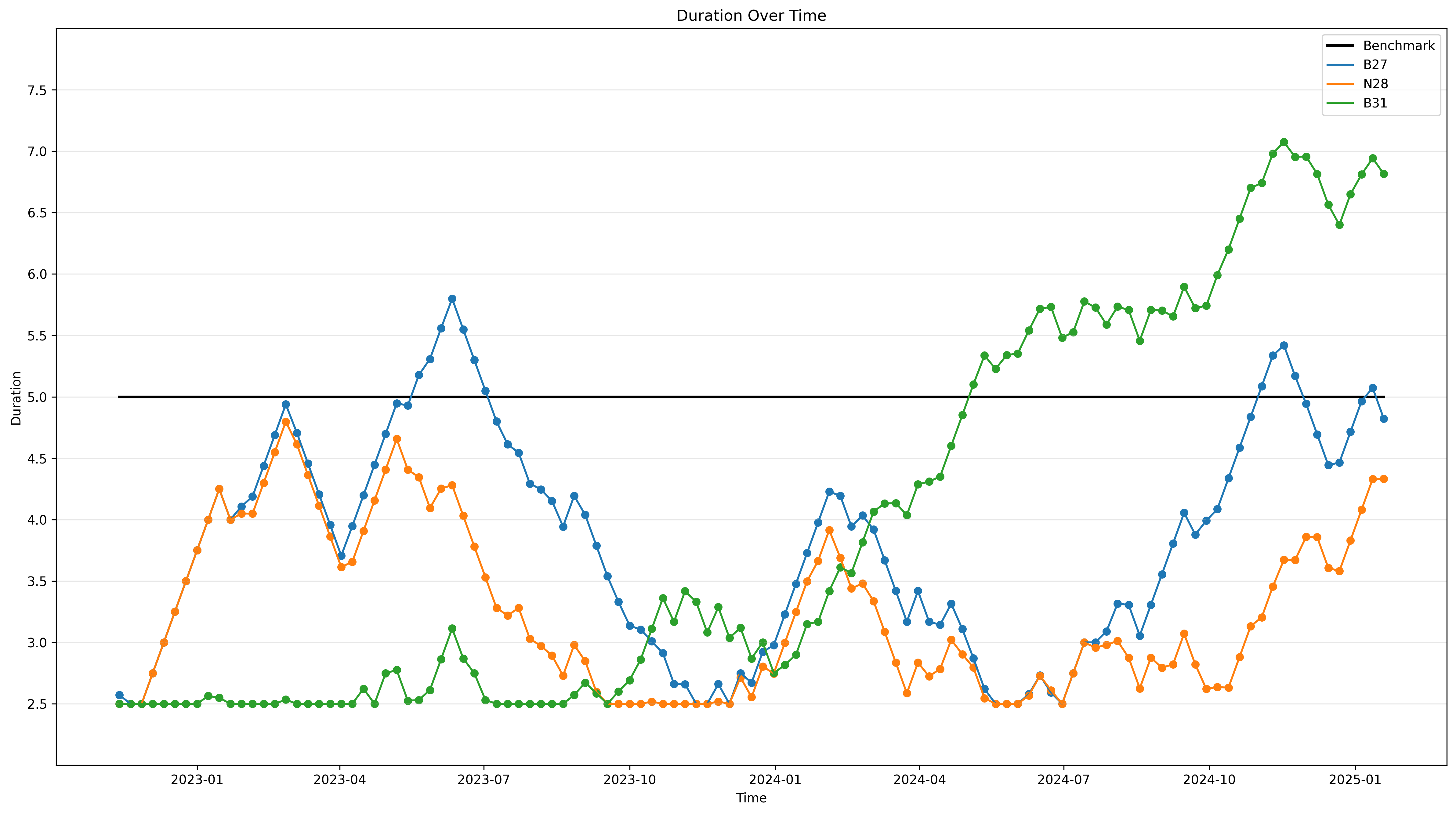}
\captionof{figure}{Duration of the different portfolios between November 2022 and February 2025.} \label{fig:duration4}
\end{center}

\renewcommand{\thesection}{\arabic{section}}
\section{Periodic Evaluation - Europe}

\subsection{Period 1 ($\text{P}_1$)}

\begin{table}[ht]
\caption{Results for the different models in the period between December 2014 and September 2019. }\label{tab:result1_eu}
\centering
\begin{tabular}{|p{1.5cm}|p{1.4cm}|p{1.4cm}|p{1.4cm}|p{1.4cm}|p{1.4cm}|p{1.4cm}|p{1.4cm}|}
\hline
\textbf{Model} &\textbf{Ret.} & \textbf{$\Omega$} & \textbf{MDD}  & \textbf{Rank. $\Omega$} & \textbf{Rank. MDD} & \textbf{Sum Rank.} & \textbf{Rank. $\text{P}_1$} \\
\hline

\textbf{N16} & 2.26  & 1.19 & -7.60 & 1 & 3 & 4 & 2\\ \hline
\textbf{E18} & 1.64 & 0.93 & -5.49 & 3 & 2 & 5 & 3\\ \hline
\textbf{E27} &  1.85  & 1.04 & -5.09 &  2 & 1 & 3 & 1\\ \hline
\textbf{Bench.} & 1.79 & - & -5.45 &  - & - & - & - \\ \hline
\end{tabular}
\end{table}

\begin{center}
\includegraphics[width=\textwidth]{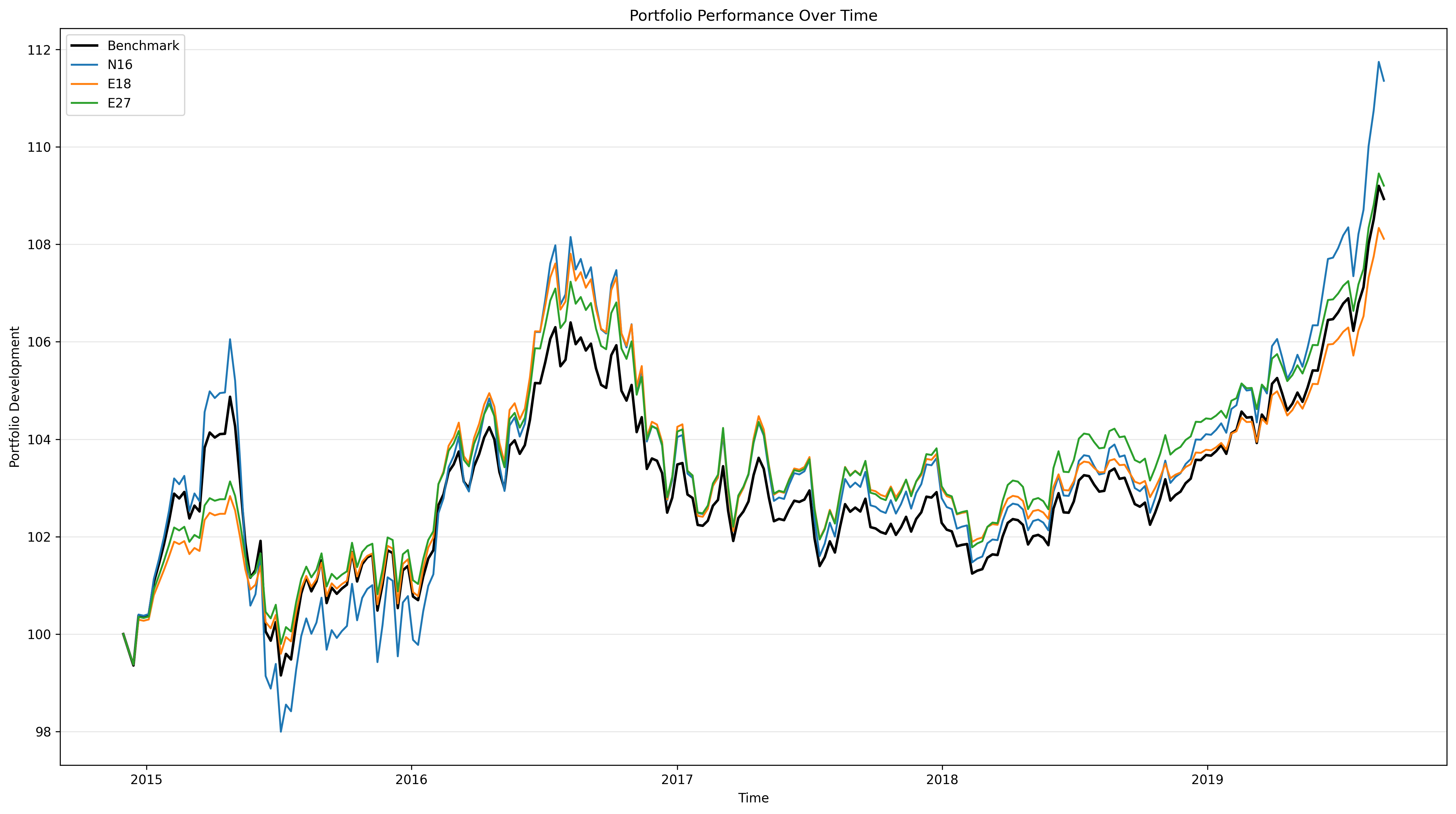}
\captionof{figure}{Performance of the different portfolios between December 2014 and September 2019. } \label{fig:result1_eu}
\end{center}

\begin{center}
\includegraphics[width=\textwidth]{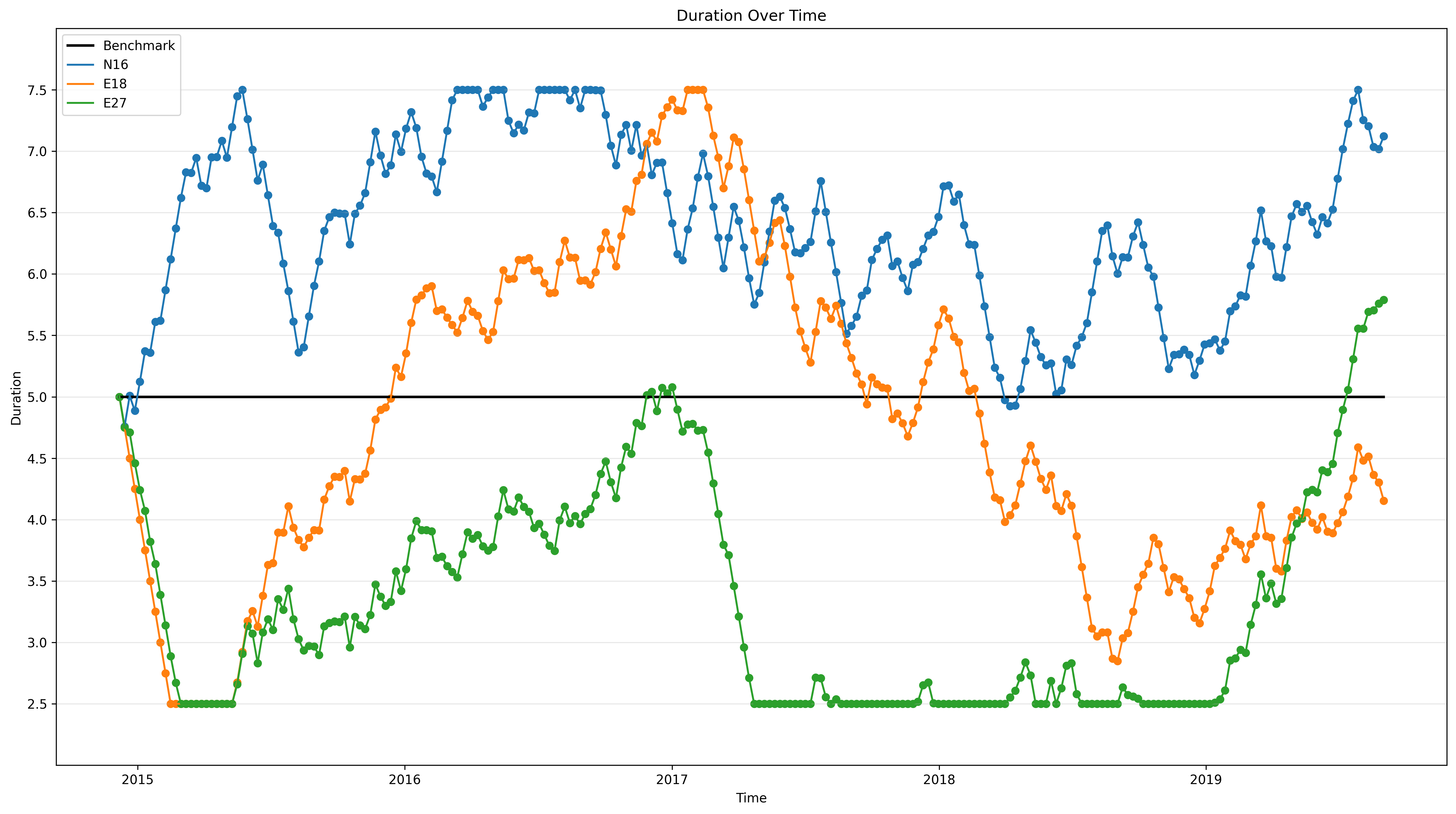}
\captionof{figure}{Duration of the different portfolios between December 2014 and September 2019. } \label{fig:duration1_eu}
\end{center}

\subsection{Period 2 ($\text{P}_2$)}

\begin{table}[ht]
\caption{Results for the different models in the period between September 2019 and November 2021. }\label{tab:result2_eu}
\centering
\begin{tabular}{|p{1.5cm}|p{1.4cm}|p{1.4cm}|p{1.4cm}|p{1.4cm}|p{1.4cm}|p{1.4cm}|p{1.4cm}|}
\hline
\textbf{Model} &\textbf{Ret.} & \textbf{$\Omega$} & \textbf{MDD}  & \textbf{Rank. $\Omega$} & \textbf{Rank. MDD} & \textbf{Sum Rank.} & \textbf{Rank. $\text{P}_2$} \\
\hline

\textbf{N16} & -1.27 & 0.76 & -3.55 & 3 & 2 & 5 & 3\\ \hline
\textbf{E18} & -0.93 & 0.95 & -2.96 & 1 & 1 & 2 & 1\\ \hline
\textbf{E27} & -0.91 & 0.95 & -4.74 & 1 & 3 & 4 & 2\\ \hline
\textbf{Bench.} & -0.78 & - & -3.73 & - & - & - & - \\ \hline
\end{tabular}
\end{table}

\begin{center}
\includegraphics[width=\textwidth]{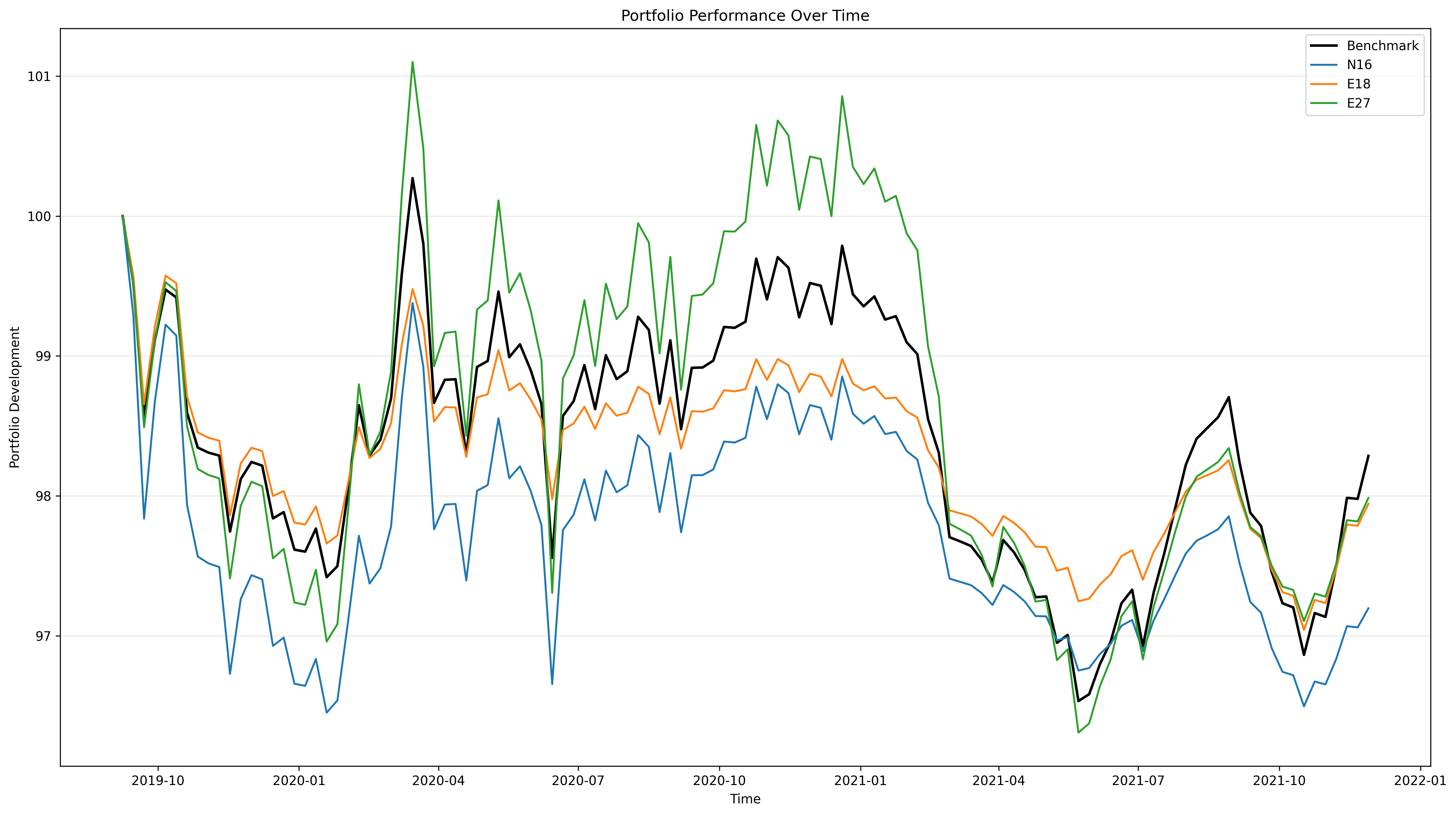}
\captionof{figure}{Performance of the different portfolios between September 2019 and November 2021. } \label{fig:result2_eu}
\end{center}

\begin{center}
\includegraphics[width=\textwidth]{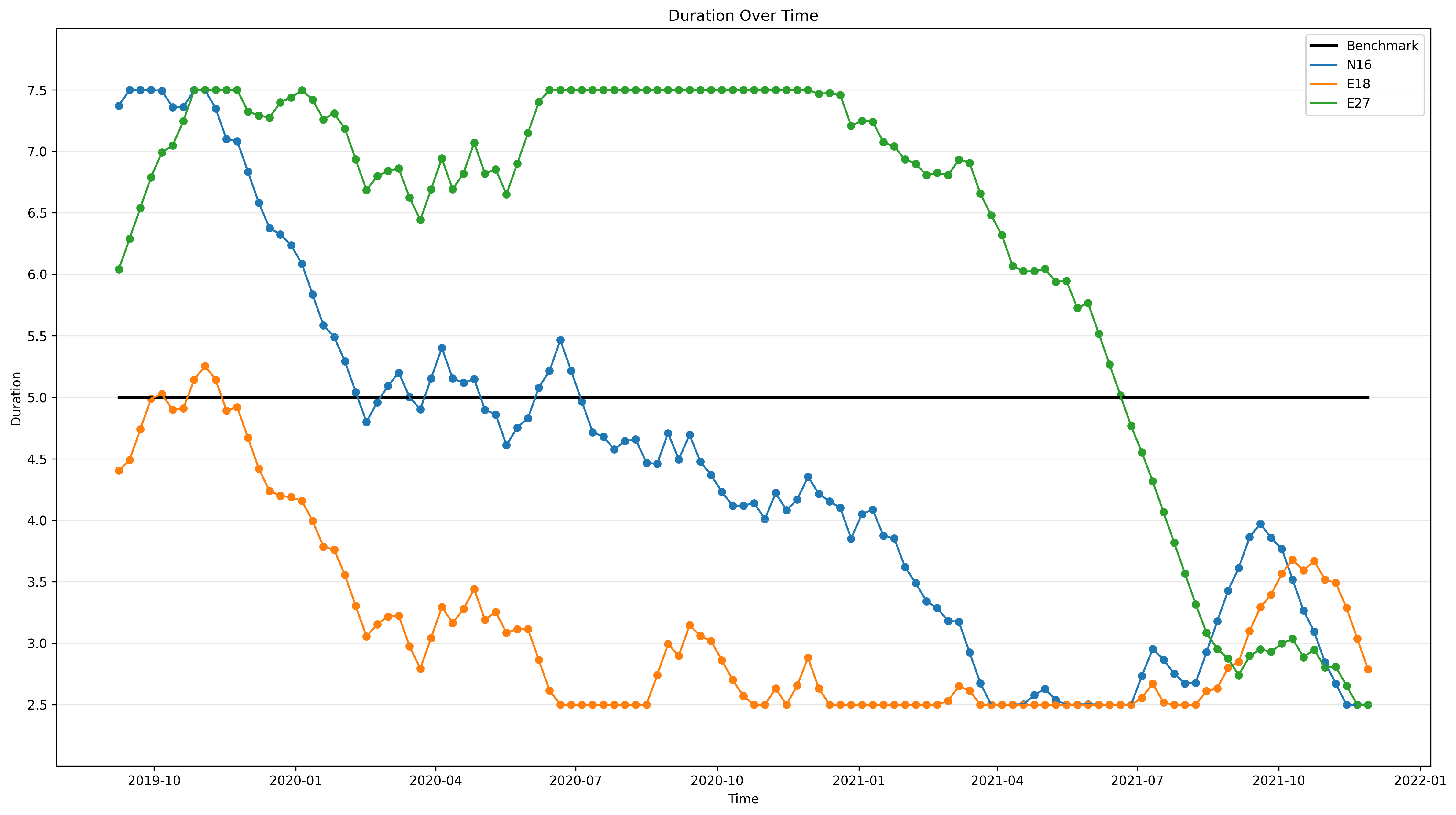}
\captionof{figure}{Duration of the different portfolios between September 2019 and November 2021. } \label{fig:duration2_eu}
\end{center}

\subsection{Period 3 ($\text{P}_3$)}

\begin{table}[ht]
\caption{Results for the different models in the period between December 2021 and October 2023. }\label{tab:result3_eu}
\centering
\begin{tabular}{|p{1.5cm}|p{1.4cm}|p{1.4cm}|p{1.4cm}|p{1.4cm}|p{1.4cm}|p{1.4cm}|p{1.4cm}|}
\hline
\textbf{Model} &\textbf{Ret.} & \textbf{$\Omega$} & \textbf{MDD}  & \textbf{Rank. $\Omega$} & \textbf{Rank. MDD} & \textbf{Sum Rank.} & \textbf{Rank. $\text{P}_3$} \\
\hline

\textbf{N16} & -3.69 & 1.89 & -7.57 & 1 & 3 & 4 & 1\\ \hline
\textbf{E18} & -3.14 & 1.88 & -7.47 & 2 & 2 & 4 & 1\\ \hline
\textbf{E27} & -2.50 & 1.78 & -6.53  & 3 & 1 & 4 & 1\\ \hline
\textbf{Bench.} & -6.55 & - & -12.52 & - & - & - & - \\ \hline
\end{tabular}
\end{table}

\begin{center}
\includegraphics[width=\textwidth]{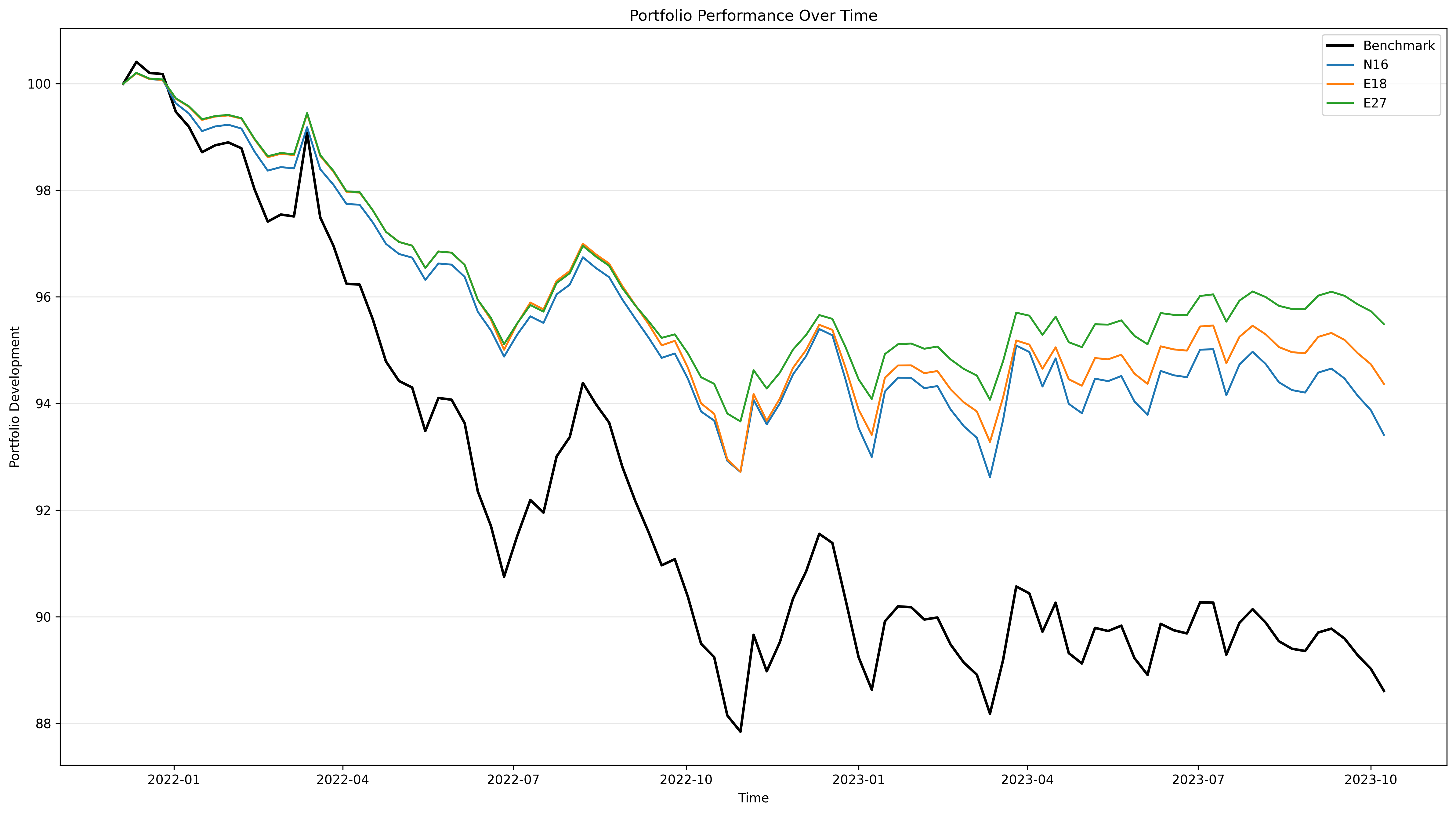}
\captionof{figure}{Performance of the different portfolios between December 2021 and October 2023. } \label{fig:result3_eu}
\end{center}

\begin{center}
\includegraphics[width=\textwidth]{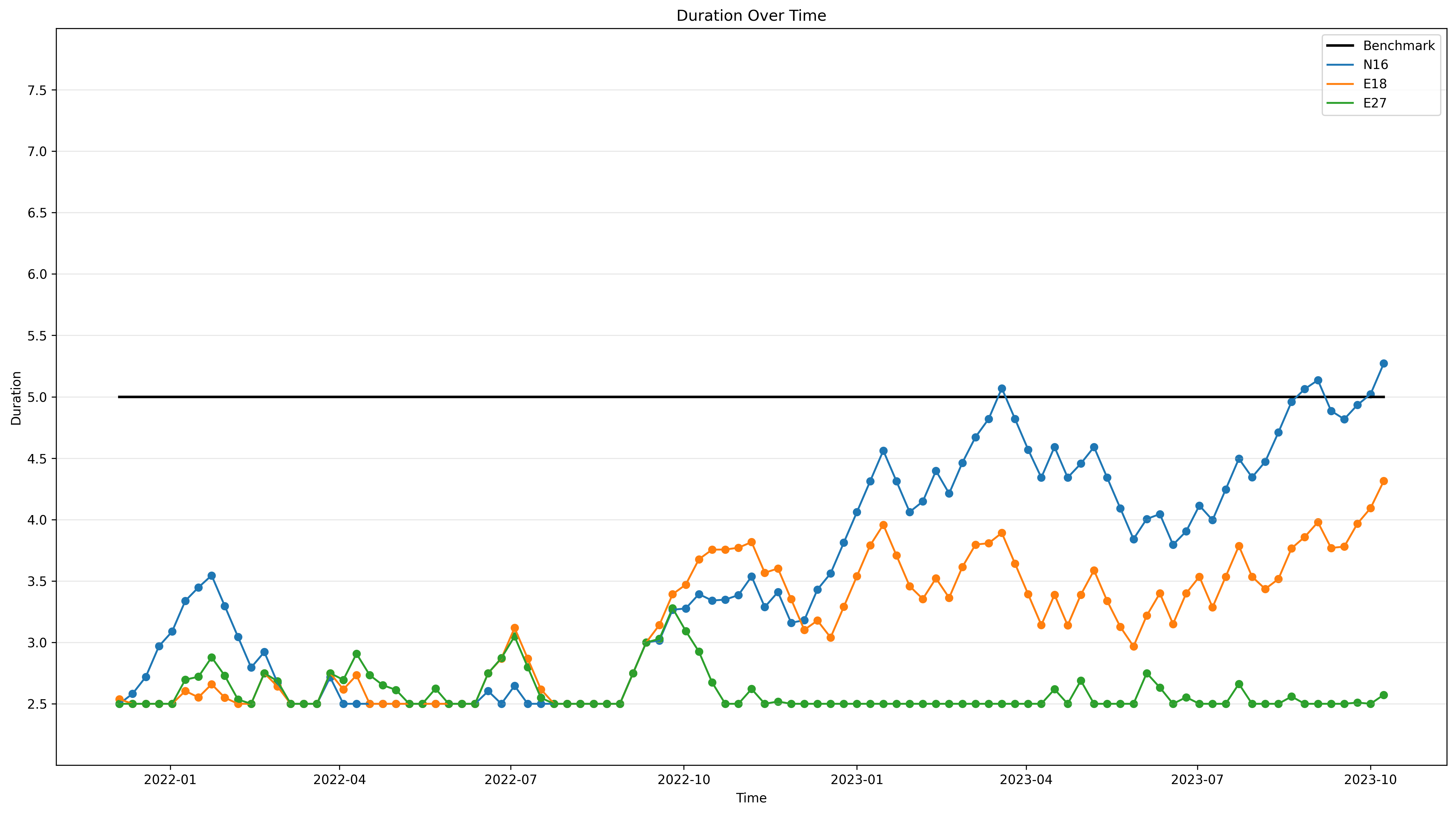}
\captionof{figure}{Duration of the different portfolios between December 2021 and October 2023. } \label{fig:duration3_eu}
\end{center}

\subsection{Period 4 ($\text{P}_4$)}

\begin{table}[ht]
\caption{Results for the different models in the period between October 2023 and February 2025. }\label{tab:result4_eu}
\centering
\begin{tabular}{|p{1.5cm}|p{1.4cm}|p{1.4cm}|p{1.4cm}|p{1.4cm}|p{1.4cm}|p{1.4cm}|p{1.4cm}|}
\hline
\textbf{Model} &\textbf{Ret.} & \textbf{$\Omega$} & \textbf{MDD}  & \textbf{Rank. $\Omega$} & \textbf{Rank. MDD} & \textbf{Sum Rank.} & \textbf{Rank. $\text{P}_4$} \\
\hline

\textbf{N16} & 5.34 & 1.29 & -0.90 & 1 & 1 & 2 & 1\\ \hline
\textbf{E18} & 3.98 & 0.63 & -2.46 & 3 & 2 & 5 & 2\\ \hline
\textbf{E27} & 3.40 & 0.64 & -2.92 & 2 & 3 & 5 & 2\\ \hline
\textbf{Bench.} & 4.67 & - & -2.00 & - & - & - & - \\ \hline
\end{tabular}
\end{table}

\begin{center}
\includegraphics[width=\textwidth]{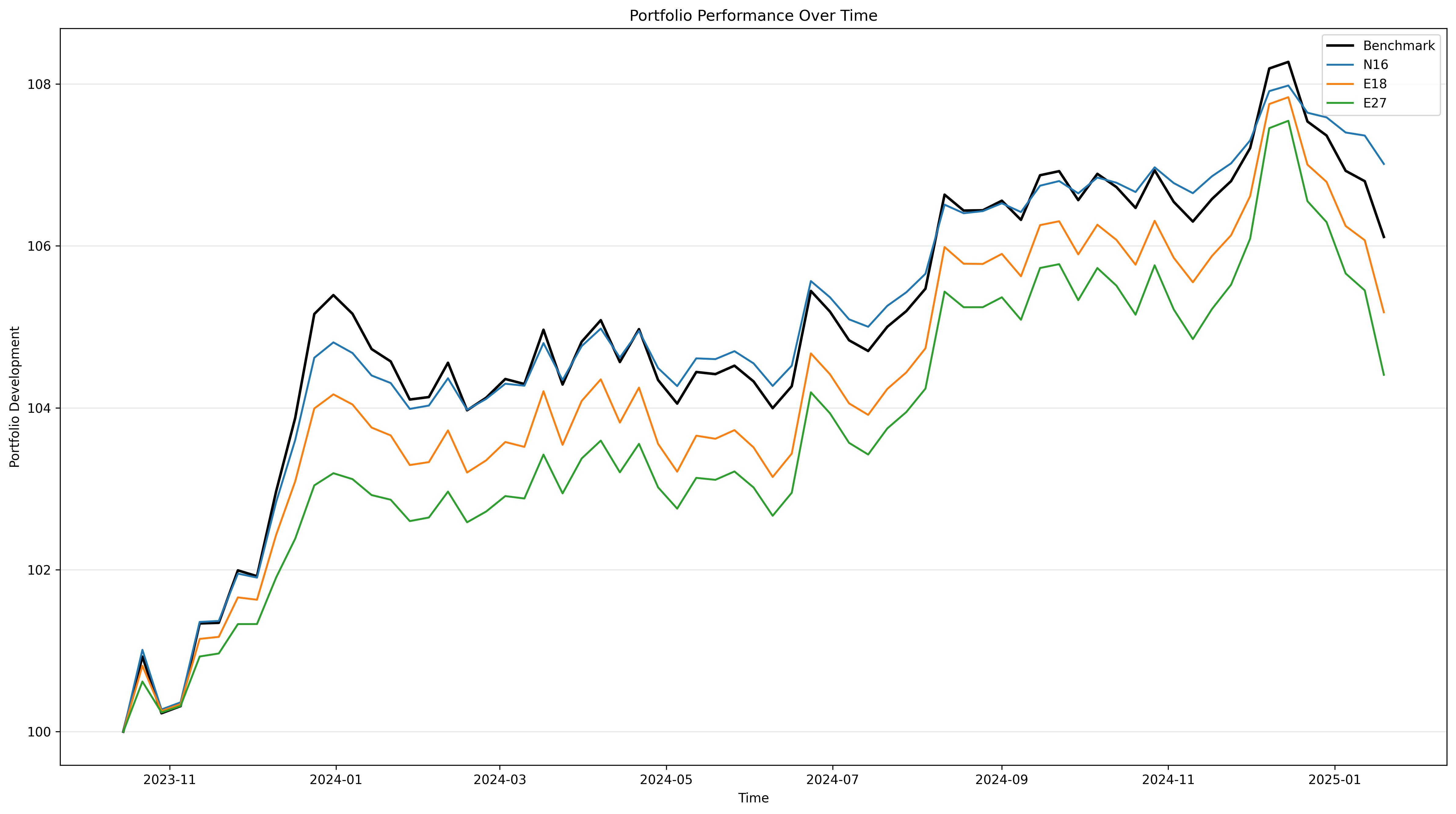}
\captionof{figure}{Performance of the different portfolios between October 2023 and February 2025. } \label{fig:result4_eu}
\end{center}

\begin{center}
\includegraphics[width=\textwidth]{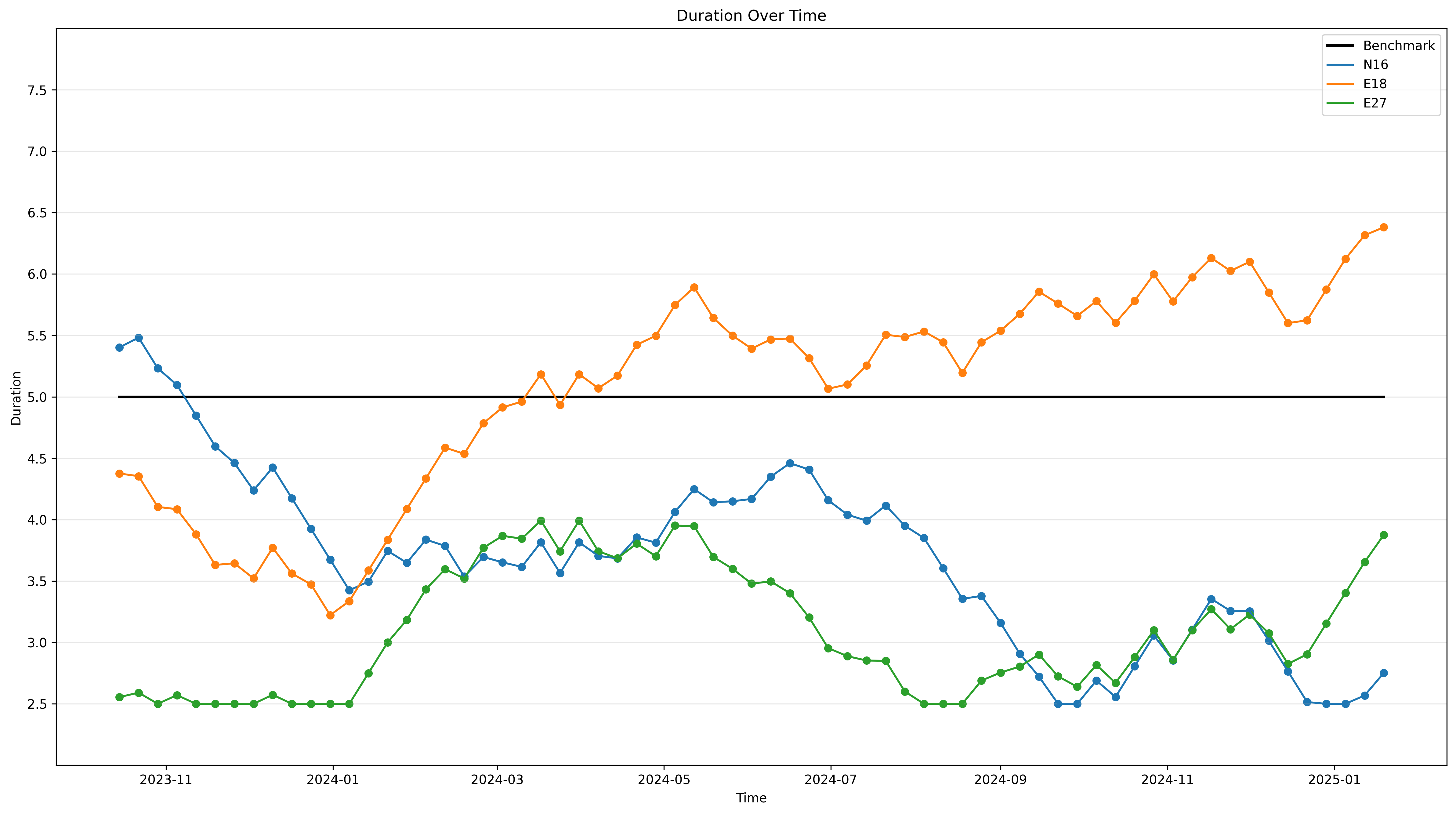}
\captionof{figure}{Duration of the different portfolios between October 2023 and February 2025.} \label{fig:duration4_eu}
\end{center}